\documentclass[twocolumn,amsmath]{aastex702}
\submitjournal{ApJ; 10/13/2026}

\begin{document}

\title{Chandra-JWST Probing the Effects of Moderate X-Ray Superflares on Protoplanetary Disks}

\correspondingauthor{Konstantin Getman}
\email{kug1@psu.edu}

\author[0000-0002-6137-8280]{Konstantin V. Getman}
\affiliation{Department of Astronomy \& Astrophysics \\
Pennsylvania State University \\ 
525 Davey Laboratory \\
University Park, PA 16802, USA}
\email{kug1@psu.edu}

\author[0000-0003-2822-2951]{Kevin L. Luhman}
\affiliation{Department of Astronomy \& Astrophysics \\
Pennsylvania State University \\ 
525 Davey Laboratory \\
University Park, PA 16802, USA}
\affiliation{Center for Exoplanets and Habitable Worlds \\ 525 Davey Laboratory \\ The Pennsylvania State University \\ University Park, PA 16802, USA}
\email{kll207@psu.edu}

\author[0000-0002-6278-9006]{Bayron Portilla-Revelo}
\affiliation{Department of Astronomy \& Astrophysics \\
Pennsylvania State University \\ 
525 Davey Laboratory \\
University Park, PA 16802, USA}
\affiliation{Center for Exoplanets and Habitable Worlds \\ 525 Davey Laboratory \\ The Pennsylvania State University \\ University Park, PA 16802, USA}
\email{bmp5924@psu.edu}

\author[0000-0003-1817-6576]{Christian Rab}
\affiliation{University Observatory, Faculty of Physics\\
Ludwig-Maximilians-Universit\"at M\"unchen\\
Scheinerstr. 1\\
D-81679 Munich, Germany}
\affiliation{Max-Planck-Institut f\"ur extraterrestrische Physik\\
Giessenbachstrasse 1\\
D-85748 Garching, Germany}
\email{rab@mpe.mpg.de}

\author[0009-0002-5380-403X]{Valentin Brunn}
\affiliation{INAF--Osservatorio Astrofisico di Arcetri\\
Largo E. Fermi 5\\
50125 Firenze, Italy}
\email{valentin.brunn@inaf.it}

\author[0000-0002-5077-6734]{Eric D. Feigelson}
\affiliation{Department of Astronomy \& Astrophysics \\
Pennsylvania State University \\ 
525 Davey Laboratory \\
University Park, PA 16802, USA}
\affiliation{Center for Exoplanets and Habitable Worlds \\ 525 Davey Laboratory \\ The Pennsylvania State University \\ University Park, PA 16802, USA}
\email{e5f@psu.edu}

\author[0000-0001-8068-0891]{Arjan Bik}
\affiliation{Department of Astronomy, Stockholm University\\
AlbaNova University Centre\\
106 91 Stockholm, Sweden}
\email{arjan.bik@astro.su.se}

\author[0000-0001-8694-4966]{Jan Forbrich}
\affiliation{Centre for Astrophysics Research\\
Department of Physics, Astronomy and Mathematics\\
University of Hertfordshire\\
Hatfield, AL10 9UW, UK}
\email{j.forbrich@herts.ac.uk}

\author[0009-0003-7663-5280]{Jenny Frediani}
\affiliation{Department of Astronomy, Stockholm University\\
AlbaNova University Centre\\
106 91 Stockholm, Sweden}
\email{jenny.frediani@astro.su.se}

\author[0000-0002-9593-7618]{Thomas J. Haworth}
\affiliation{Astronomy Unit, School of Physics and Astronomy, Queen Mary University of London, London E1 4NS, UK}
\email{t.haworth@qmul.ac.uk}

\author[0000-0003-1413-1776]{Charles J. Law}
\affiliation{Minnesota Institute for Astrophysics, University of Minnesota, 116 Church St. SE, Minneapolis, MN 55455, USA}
\email{law06855@umn.edu}

\author[0000-0003-2303-0096]{Marco Padovani}
\affiliation{INAF--Osservatorio Astrofisico di Arcetri\\
Largo E. Fermi 5\\
50125 Firenze, Italy}
\email{marco.padovani@inaf.it}

\author[0000-0001-9698-4080]{Mar\'ia Claudia Ram\'irez-Tannus}
\affiliation{Max-Planck-Institut f\"ur Astronomie\\
K\"onigstuhl 17\\
D-69117 Heidelberg, Germany}
\email{ramirez@mpia.de}

\author[0000-0002-7501-9801]{Andrew J. Winter}
\affiliation{Astronomy Unit, School of Physics and Astronomy, Queen Mary University of London, London E1 4NS, UK}
\email{andrew.winter@qmul.ac.uk}

\author[0000-0003-4452-0588]{Vladimir S. Airapetian}
\affiliation{American University\\ 4400 Massachusetts Avenue NW\\ Washington, DC 20016 USA}
\affiliation{NASA/GSFC/SEEC\\ Greenbelt, MD, 20771, USA}
\email{vladimir.airapetian-1@nasa.gov}

\author[0000-0003-2076-8001]{L. Ilsedore Cleeves}
\affiliation{University of Virginia, Charlottesville, VA 22904, USA}
\email{lic3f@virginia.edu}

\author[0000-0001-6010-6200]{Sergio A. Dzib}
\affiliation{Max-Planck-Institut fur Radioastronomie (MPIfR), Auf dem Hugel 69, 53121 Bonn, Germany}
\email{sdzib@mpifr.de}

\author[0000-0002-0631-7514]{Michael A. Kuhn}
\affiliation{Centre for Astrophysics Research\\
Department of Physics, Astronomy and Mathematics\\
University of Hertfordshire\\
Hatfield, AL10 9UW, UK}
\email{m.kuhn@herts.ac.uk}

\author[0000-0002-1566-389X]{Abygail R. Waggoner}
\affiliation{Department of Astronomy, University of Wisconsin-Madison, 475 N Charter St, Madison, WI 53706}
\email{awaggoner2@wisc.edu}

\author[0000-0002-5462-9387]{Rens Waters}
\affiliation{Department of Astrophysics/IMAPP, Radboud University, PO Box 9010, 6500 GL Nijmegen, The Netherlands}
\affiliation{SRON Netherlands Institute for Space Research, Niels Bohrweg 4, NL-2333 CA Leiden, the Netherlands}
\email{rens.waters@ru.nl}

\begin{abstract}
Powerful X-ray flares from young stars may rapidly alter the physical and chemical conditions of their planet-forming disks, but direct observational tests remain scarce. We present the first coordinated Chandra and JWST/NIRSpec study of infrared spectral variability associated with pre-main-sequence (PMS) X-ray flares in the Orion Nebula. Combining six Chandra ACIS-I observations with three NIRSpec-MOS visits, we analyze two-epoch spectra of 10 X-ray Flare stars and 12 matched non-flaring Controls among 84 observed disk-bearing PMS stars.

We investigate three diagnostics: narrow 2.8--4.3~$\mu$m features, H~I~Br$\alpha$, and the CO fundamental band. Flare stars show tentative evidence for relatively fewer variable narrow features at longer wavelengths than Controls, while feature incidence decreases strongly with projected distance from the dominant ionizing O star $\Theta^{1}$~Ori~C. Most classifiable features appear in absorption. These properties suggest that at least some features may arise in externally irradiated disk atmospheres and/or associated photoevaporative flows, with their apparent epoch-to-epoch variability potentially reflecting spatial sampling of such locally structured material rather than intrinsic temporal changes.

Br$\alpha$ behaves differently: flare stars show systematically stronger variability and uniquely exhibit absorption-core profiles, suggesting control by magnetospheric accretion and stellar magnetic processes rather than disk irradiation. The CO fundamental band shows no significant flare-associated variability, consistent with thermochemical models predicting weak responses to moderate X-ray superflares.

This pilot study demonstrates the potential of coordinated Chandra--JWST observations to probe how young stellar magnetic activity affects planet-forming environments, motivating future monitoring to build larger superflare samples and extend them into the rare X-ray mega-flare regime.
\end{abstract}



\section{Introduction} \label{sec_introduction} 
Pre-main sequence (PMS) stars emit intense X-ray and ultraviolet (UV) radiation that is predicted to strongly influence their surrounding environments, particularly protoplanetary disks and the processes governing planet formation. X-ray and UV irradiation is expected to ionize and heat disk gas and drive photoevaporative mass loss. X-rays, in particular, can penetrate deep into disks, with sufficiently hard X-rays potentially reaching the midplane at terrestrial-planet-forming radii and enhancing ionization that may enable magnetorotational instability (MRI), thereby influencing gas turbulence, viscosity, accretion, planetesimal formation through the streaming instability, and planetary migration. X-ray ionization may also alter ion-molecular chemistry and contribute to the production of disk winds and jets. Most of these predicted X-ray-driven effects remain poorly constrained empirically and are based primarily on astrophysical calculations \citep{Glassgold2000,Glassgold2007,Johansen2011,Alexander2014,Owen2019,Ercolano2021,Woitke2024,Lim2024}. Observational evidence is considerably more limited, with tentative links between high-energy irradiation and [Ne~II] emission, the [Ne~III]/[Ne~II] line ratio, and variable HCO$^{+}$ in some disks \citep{Espaillat2013,Espaillat2023,Cleeves2017,Waggoner2023}, with recent JWST surveys greatly expanding the available mid-IR disk diagnostics \citep{Arulanantham2025}.

Traditional theoretical models often assume that disks experience steady X-ray and UV irradiation, neglecting the substantial flux and spectral variations driven by large magnetic reconnection-induced X-ray flares and accretion-powered UV outbursts \citep{Gorti2009,Owen2019}. However, recent time-dependent simulations suggest that coronal mass ejections (CMEs) and stellar energetic particles associated with powerful PMS flares may accelerate disk dispersal and contribute to planetary atmospheric erosion \citep{Airapetian2020,AlvaradoGomez2022,Hazra2022,Airapetian2023,Hu2022}. Furthermore, disk ionization levels and ion-molecular chemistry can rapidly respond to sudden large X-ray flares and flare-generated suprathermal particles (STPs), i.e., energetic protons and electrons accelerated during magnetic reconnection events \citep{Ilgner2006,Rab2017,Waggoner2022,Brunn2024,Washinoue2024}.

Despite these theoretical advances, the empirical impact of large X-ray flares on protoplanetary disks remains largely unexplored due to the stochastic nature of magnetic reconnection-driven flares and the age- and mass-dependent flare occurrence rates in PMS stars. Coronal magnetic reconnection is most efficient during the early PMS evolutionary stages, before substantial radiative cores develop. For instance, a young ($t < 10$~Myr), fully convective, gravitationally contracting solar-mass PMS star can have a volume up to 30 times that of the contemporary Sun. A convection-driven magnetic dynamo operating throughout the deep convective interior of a PMS star is significantly more efficient at generating powerful X-ray flares than the solar tachocline dynamo, operating at the interface between the Sun's radiative and convective zones. As a result, a solar-mass PMS star can produce a very strong superflare ($E_{\rm X,fl} > 10^{35}$~erg) roughly every two weeks and an extreme mega-flare ($E_{\rm X,fl} > 10^{36}$~erg) roughly every four months --- events no longer magnetically feasible for the modern Sun, where $E_{\rm X,fl,\odot} < 10^{33}$~erg \citep{GetmanFeigelson2021,Getman2026b}. Lower-mass ($<0.5$~M$_{\odot}$) PMS stars, being smaller, generate weaker flares than their solar-mass siblings. For these stars, flares with $E_{\rm X,fl} \geq 10^{35}$~erg are already rare but can still significantly influence their surrounding environments. Over the PMS lifetime, protoplanetary disks are therefore exposed to millions of energetic magnetic reconnection events, potentially accompanied by powerful CMEs; after disk dispersal, the same hostile magnetic environment irradiates newly exposed primordial planetary atmospheres \citep{Getman2024,Getman2026a}.

Studying the effects of an $E_{\rm X,fl} > 10^{35}$~erg superflare on a single solar-mass PMS star and its protoplanetary disk over a two-week period would require a simultaneous X-ray photometric and IR spectroscopic monitoring campaign --- an endeavor that is both logistically challenging and prohibitively expensive. The challenge becomes even greater for rarer, more powerful mega-flares ($E_{\rm X,fl} > 10^{36}$~erg), leaving a critical gap in empirical data needed to understand their influence on disk chemistry and planet formation.

A promising strategy for overcoming these challenges is to leverage multi-telescope observations of large statistical samples of coeval PMS stars in densely populated star-forming regions. For example, a recent 70-ksec Chandra observation --- comprising multiple snapshots over 5.5 days --- of the Orion Nebula region, including the Orion Nebula Cluster (ONC) and the Orion Molecular Cloud~1 (OMC-1) filament, captured numerous powerful PMS X-ray flares. Among these, \citet{Getman2024} identified a hundred flares with $E_{\rm X,fl} > 10^{34}$~erg from a selected subset of the brightest X-ray stars, representing only 10\% of the underlying $>2000$ PMS stellar population within the Chandra ACIS-I $17\arcmin \times 17\arcmin$ field of view. These flares were rapidly identified within a day of observation, enabling prompt multi-telescope follow-up with HET, VLBA, and ALMA. This approach facilitates the investigation of various flare-driven phenomena, including measuring stellar magnetic fields during intense flares (HET; \citealt{Getman2025}), detecting flare-associated CMEs (VLBA), and probing flare-induced HCO$^{+}$ chemistry in the outer regions of protoplanetary disks (ALMA). Notably, among two flaring PMS stars in the Orion Nebula with gas-rich outer disks that were followed up with ALMA after Chandra-detected flares, both exhibited the expected flare-driven changes in HCO$^{+}$ chemistry (private communication).

The James Webb Space Telescope (JWST) has the unique capability to perform infrared spectroscopy of numerous objects simultaneously using the Near-Infrared Spectrograph (NIRSpec) and Microshutter Assembly (MSA), providing an even more efficient means of empirically exploring the effects of large stellar flares on the chemistry of inner protoplanetary disks.

We carried out nearly simultaneous Chandra ACIS-I and JWST/NIRSpec-MSA observations of the Orion Nebula region. These observations were designed to identify and characterize PMS X-ray flares in the 0.5--8~keV band, obtain near-simultaneous infrared (2.87--5.14~$\mu$m) spectra of protoplanetary disks around Orion PMS stars spanning a broad range of stellar masses ($0.1$--$1.5$~M$_{\odot}$), and compare empirical disk properties between two stellar subsamples (X-ray flaring and non-flaring stars). In addition, we compare the observational results with predictions from thermo-chemical disk models.

This study represents an initial exploratory effort, providing a unique Chandra--JWST/NIRSpec dataset for a limited number of PMS systems with captured mild X-ray superflares. Future longer-duration Chandra--JWST campaigns will expand these samples, including a larger number of rare and powerful X-ray mega-flares, providing critical constraints for improving theoretical models of protoplanetary disk evolution and advancing our understanding of planet formation.

The present study focuses on the variability of narrow spectral features at 2.8--4.3~\micron, H~I Br$\alpha$, and CO fundamental-band emission in the two-epoch NIRSpec spectra of 10 X-ray-flaring (Flare) and 12 matched non-flaring (Control) star--disk systems. A complementary study (Paper~II; \citealt{Getman2026c}) employs all available NIRSpec spectra with wavelength coverage encompassing the Br$\alpha$ and CO diagnostics for more than 70 PMS stars, regardless of X-ray flare activity or the availability of two-epoch spectroscopy, to investigate relationships of Br$\alpha$ and CO emission with stellar properties, accretion, and the local environment.

The current paper is organized as follows. Section~\ref{sec_modeling} presents model predictions for the effects of X-ray flares on disk chemistry. Section~\ref{sec_methods} describes our coordinated Chandra/JWST observations of the Orion Nebula and data reduction, summarizes the resulting X-ray and infrared datasets, and defines the Flare and Control samples. Section~\ref{sec_analyses_results} presents analyses of NIRSpec spectral variability in three diagnostics: (1) numerous narrow spectral features across 2.8--4.3~\micron; (2) the H~I Br$\alpha$ line at 4.05~\micron; and (3) the CO fundamental rovibrational band forest spanning 4.3--5.2~\micron. In Section~\ref{sec_discussion}, we discuss the plausible origins of the narrow NIRSpec spectral features and the responses of Br$\alpha$ and the CO fundamental band to X-ray flares. Supplementary tables are provided in Appendix~\ref{sec_appendix_tables}.

\begin{figure*}[!t]
\centering
\includegraphics[width=0.85\textwidth]{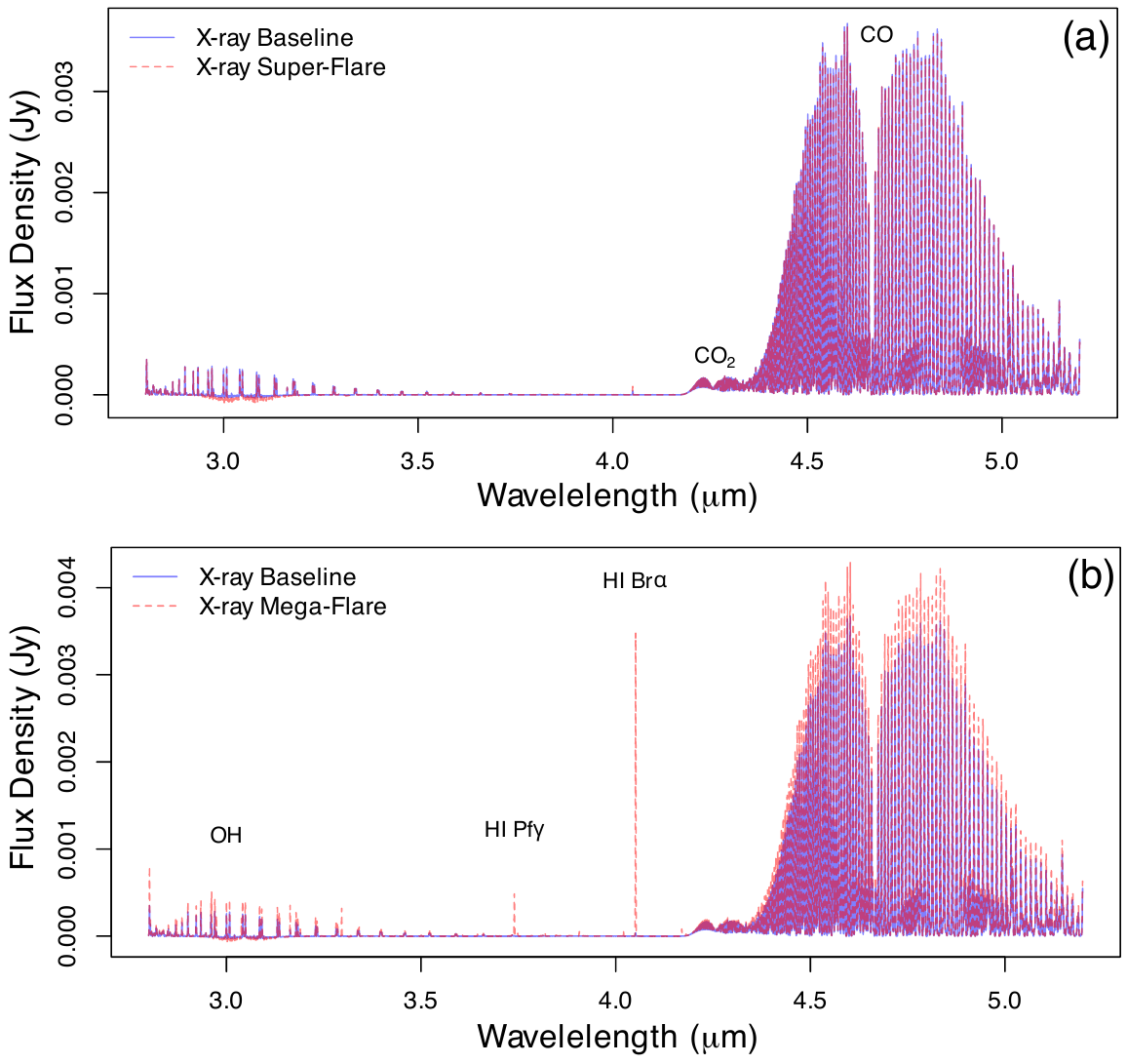}
\caption{ProDiMo predictions of flare-induced changes in the continuum-subtracted H~I, OH, CO$_2$, and CO emission-line spectrum of the protoplanetary disk surrounding a 0.7~$M_{\odot}$ Orion PMS star, driven by enhanced X-ray/EUV ionization and gas heating. The upper and lower panels compare the baseline model (blue) with models of a mild X-ray superflare ($E_{\rm X,fl}\sim10^{34.5}$~erg) and an X-ray
mega-flare ($E_{\rm X,fl}\sim10^{36.5}$~erg), respectively (red). Labels identify representative spectral features from the modeled atomic and molecular species.} \label{fig_prodimo_outcome}
\end{figure*}

\section{Model Predictions for the Effects of X-ray Flares on Disk Chemistry}\label{sec_modeling}

ProDiMo (Protoplanetary Disk Model) is a sophisticated radiation thermo-chemical code that self-consistently solves the continuum radiative transfer, gas heating--cooling balance, and chemical composition of protoplanetary disks \citep{Woitke2009,Woitke2024,Kamp2010,Thi2011,Rab2018}. The code predicts the physical and chemical structure of disks under irradiation by stellar optical, UV, and X-ray photons, as well as cosmic rays and suprathermal particles.

Our simulations adopt the DIANA standard T~Tauri disk model of \citet{Woitke2016}; see their Table~3. The model represents a primordial disk of mass $0.01~M_{\odot}$ surrounding a $0.7~M_{\odot}$ star and assumes a uniform gas-to-dust mass ratio of 100. The calculations are static and neglect X-ray scattering, while X-ray absorption and its coupling to the disk chemistry and thermal balance are included. Neglecting scattering provides a computationally efficient approximation appropriate for the warm disk surface layers probed by the JWST spectra considered here, since X-ray scattering becomes most important for harder photons penetrating toward the disk midplane \citep{Aresu2011,Rab2018}. The model includes detailed X-ray chemistry following \citet{Aresu2011}, including Ne and Ar and their singly and doubly ionized states, while the broader chemical network is based on \citet{Kamp2017}.

Three models were computed that differ only in the stellar X-ray luminosity and coronal plasma temperature: (1) a baseline model with $L_{\rm X}=10^{30}$~erg~s$^{-1}$ and $T_{\rm X}=10$~MK; (2) a mild-superflare model with $L_{\rm X}=3\times10^{30}$~erg~s$^{-1}$ and $T_{\rm X}=40$~MK; and a mega-flare model with $L_{\rm X}=10^{32}$~erg~s$^{-1}$ and $T_{\rm X}=120$~MK. According to the flare scaling relations of \citet{Getman2024,Getman2025}, the adopted X-ray luminosities and plasma temperatures approximately correspond to flare energies of $E_{\rm X,fl}\sim10^{34.5}$~erg and $E_{\rm X,fl}\sim10^{36.5}$~erg for the modeled  mild-superflare and mega-flare cases, respectively.

Synthetic spectra were generated using ProDiMo's escape-probability formalism \citep{Woitke2009,Woitke2024}. The CO rovibrational spectrum is calculated with a detailed non-LTE molecular model \citep{Thi2013}, whereas most other molecular lines considered here are treated in LTE using molecular data from HITRAN \citep{Gordon2022}.

For the warm inner disk regions ($\lesssim5$--10~AU) probed by JWST/NIRSpec G395H/F290LP (2.87--5.14~\micron), the models predict flare-induced enhancements of several atomic and molecular emission lines (Figure~\ref{fig_prodimo_outcome}). X-ray and extreme-ultraviolet (XEUV) photons produced during magnetic reconnection events photoionize hydrogen and heavier elements, generating free electrons and ions (e.g., C$^+$ and H$_3^+$) that drive a cascade of chemical and thermal changes. In the modeled mega-flare case, the predicted line fluxes increase by more than 500\% for H~I recombination lines, by approximately 60\% for OH, and by about 20\% for both CO$_2$ and CO (Figure~\ref{fig_prodimo_outcome}b). In contrast, the mild-superflare model predicts a significant enhancement only for the hydrogen recombination lines, whose fluxes increase by more than 100\% (Figure~\ref{fig_prodimo_outcome}a). The weak response of molecular emission to moderate X-ray flares is consistent with recent ProDiMo calculations exploring X-ray and irradiation effects in planet-forming disks \citep{PortillaRevelo2026,Frediani2026}.

An important limitation is that the modeled H~I emission originates exclusively from the disk. The models do not include magnetospheric accretion columns, chromospheric emission, stellar winds, or intervening circumstellar gas between the star and disk, which could both contribute substantially to the observed hydrogen recombination spectrum and attenuate ionizing radiation before it reaches the disk. Consequently, the predicted H~I enhancements should be interpreted as the response of the irradiated disk rather than as predictions for the total H~I recombination-line emission, such as Br$\alpha$, observed from a PMS star--disk system.

Our calculations also do not include energetic suprathermal particles accelerated during magnetic reconnection events. Such particles are expected to produce additional collisional ionization and heating \citep{Brunn2023,Brunn2024}, potentially leading to stronger chemical and line-emission responses than predicted here.

Finally, PMS X-ray superflares and mega-flares typically persist for only several hours to a few days, whereas the present calculations assume a steady-state disk structure. We approximate a flare by imposing an enhanced stellar X-ray luminosity and harder incident spectrum and solving for the corresponding steady-state thermo-chemical solution. From these ProDiMo solutions, gas-mass-weighted averages over the CO rovibrational line-emitting regions yield characteristic cooling timescales of $\sim2$~days and chemical timescales of $\sim2$--4~days for the baseline and mild-superflare models, increasing to $\sim4$--5~days in the mega-flare model. Because these timescales are comparable to or longer than typical flare durations and the delays between flare peaks and NIRSpec observations (see below), two potentially counteracting effects may arise: (1) an incomplete response, whereby a short flare ends before the disk develops the full thermal or chemical perturbation predicted by the flare-enhanced steady-state model; and (2) delayed relaxation, whereby a flare-induced perturbation persists after the X-ray emission subsides. Nevertheless, our steady-state models provide useful first-order estimates of the disk response to sustained X-ray irradiation, although they cannot predict the amplitude or timing of CO variability during and after individual flares. Future coordinated Chandra--JWST campaigns that enlarge the superflare sample, capture rare mega-flares, and improve temporal sampling will enable more meaningful comparisons with time-dependent ProDiMo flare calculations.

\begin{figure*}[!t]
\centering
\includegraphics[width=0.85\textwidth]{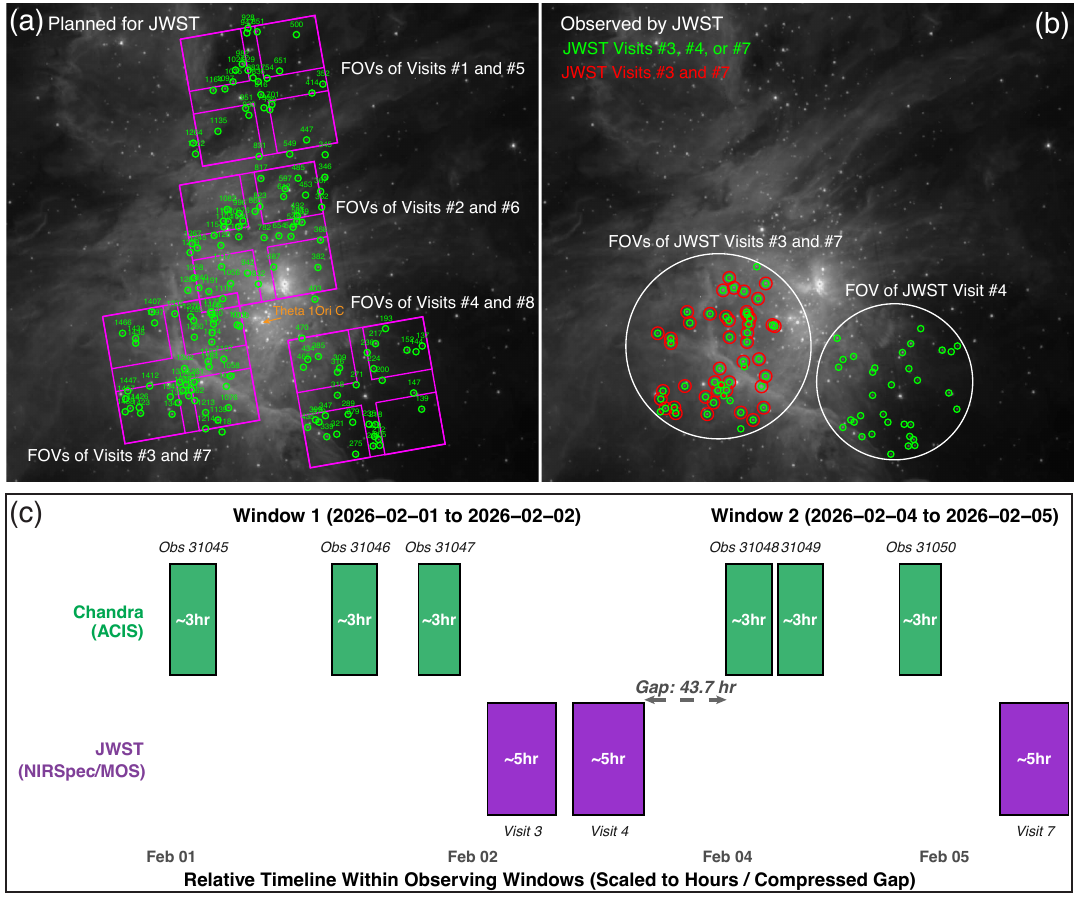}
\caption{(a) Four JWST/NIRSpec-MOS fields (purple rectangles) corresponding to eight planned visits (two visits per field) targeting protoplanetary disk-bearing PMS stars (green points) in the Orion Nebula, scheduled for early February 2026. (b) Only three of the eight planned visits were successfully completed. Visits~3 and 7 observed PMS systems in the same region southeast of $\theta^1$~Ori~C at two epochs, while Visit~4 targeted PMS systems southwest of $\theta^1$~Ori~C. JWST spectra were obtained for more than 80 PMS systems (green points), including $>40$ sources observed at both epochs in Visits~3 and 7 (red circles). (c) Timeline of the six successful Chandra (green) and three successful JWST (purple) observations of the Orion Nebula. In each Chandra observation, the $17\arcmin \times 17\arcmin$ ACIS-I field of view covered a much larger area than the JWST observations, encompassing all JWST fields.} \label{fig_planned_observations}
\end{figure*}

Looking ahead to the observational results presented in the following sections, the currently available coordinated Chandra--JWST observations of Orion PMS stars contain no clear X-ray mega-flares and only a handful of powerful superflares. Consequently, the absence of statistically significant flare-induced changes in the observed disk CO emission is not unexpected and is broadly consistent with the ProDiMo predictions for the mild-superflare model (Figure~\ref{fig_prodimo_outcome}a). However, despite the limited number of strong flare events ($E_{\rm X,fl} > 10^{35}$~erg), our observations reveal spectral variability not captured by the present models, including changes in H~I Br$\alpha$ likely associated with magnetospheric accretion and other stellar magnetic processes, as well as numerous narrow features tentatively associated with molecular material in externally irradiated disk/proplyd environments. The latter may reflect intrinsic temporal variability, spatially dependent sampling of extended circumstellar material, and/or residuals from the structured nebular background.

\section{Methods}\label{sec_methods}

\subsection{Observing campaign overview}\label{sec_campaign_overview}

As part of our coordinated Chandra ACIS-I and JWST/NIRSpec-MOS campaign targeting numerous disk-bearing PMS systems in the Orion Nebula, eight JWST visits were scheduled (Figure~\ref{fig_planned_observations}a). These observations were designed to obtain NIRSpec spectra for approximately 150 disk-bearing stars at two epochs: an earlier epoch (February 1--2, 2026; Visits 1--4) and a later epoch (February 4--5; Visits 5--8). Unfortunately, only three of the eight visits (Visits 3, 4, and 7) were successfully executed.

The failure of the remaining five visits was due to two independent causes. First, Visits 1 and 5 (the northernmost pointing) failed because JWST was unable to acquire a suitable guide star. Although a guide star had initially been selected, it was subsequently found to be a binary system, which likely prevented successful guide-star acquisition. Second, Visits 2, 6, and 8 failed because an insufficient number of MSATA reference stars were identified. Many of the selected target acquisition (TA) stars were saturated, primarily because of the bright nebular background combined with our choice of the F140X MSATA filter. The F110W filter would have been a more appropriate choice, as it is less sensitive to nebular background and is therefore better suited for target acquisition in the Orion Nebula. These lessons have been incorporated into the planning of future JWST/NIRSpec-MOS observations of this region.

Although two-epoch NIRSpec spectroscopy was not obtained for the full sample of 150 disk-bearing stars, the successful Visits 3, 4, and 7 yielded NIRSpec spectra for 84 stars (green points in Figure~\ref{fig_planned_observations}b) in at least one epoch, with 44 stars having observations at both epochs (red points in Figure~\ref{fig_planned_observations}b). Table~\ref{tab_stellar_properties} lists the properties of these 84 stars. 

Among the 84 JWST targets previously identified in the Chandra Orion Ultradeep Project (COUP; \citealt{Getman2005}) with at least one NIRSpec spectrum, approximately 80\% are M-type stars and 20\% are K-type stars. Their stellar masses span $0.1$--$1.5~M_{\odot}$, with the 25th, 50th (median), and 75th percentiles equal to 0.2, 0.3, and $0.6~M_{\odot}$, respectively. The visual extinctions range from $A_V=0$ to 13.8~mag, with corresponding 25th, 50th, and 75th percentiles of 1.6, 2.7, and 5.0~mag. The stellar properties adopted throughout this paper are taken from Table~6 of \citet{Getman2022}.

All six planned Chandra ACIS-I snapshot observations were executed successfully. Figure~\ref{fig_planned_observations}c shows the timeline of the resulting Chandra--JWST observing campaign. The first observing window (February 1--2, 2026) consisted of three Chandra observations (3~hr each), immediately followed by two JWST observations (Visits 3 and 4; 5~hr each). After a 44~hr gap, the second observing window (February 4--5, 2026) included the remaining three Chandra observations, immediately followed by one JWST observation (Visit 7).

\subsection{NIRSpec Data}\label{sec_NIRSpec_data}

The James Webb Space Telescope (JWST; \citealt{Gardner2006,Rigby2023}) NIRSpec Multi-Object Spectroscopy (MOS; \citealt{Jakobsen2022}) observations (Proposal ID: 9292) were obtained using the G395H/F290LP high-resolution grating/filter combination ($R \approx 2700$), covering the 2.87--5.14~$\mu$m wavelength range. The three completed NIRSpec-MOS visits, each with a total exposure duration of approximately 5~hours, utilized five Micro-Shutter Array (MSA) configurations. For each MSA configuration, observations were performed using the NRSRAPID readout pattern with four groups per integration (to mitigate detector saturation) and ten integrations at each of the three nod positions. Dedicated NIRCam pre-imaging was not required because the target fields were fully covered by archival NIRCam and HST/WFC3 imaging \citep{Robberto2020,McCaughrean2023,Luhman2024a}.

Data reduction was performed using the official JWST calibration pipeline (version 1.20.2; \citealt{Bushouse2025}), with CRDS version 13.1.1 and calibration context \texttt{jwst\_1464.pmap}, following the standard STScI
JWST NIRSpec MOS pipeline workflow\footnote{STScI JWST Pipeline Notebook for
NIRSpec MOS: \url{https://github.com/spacetelescope/jwst-pipeline-notebooks/blob/main/notebooks/NIRSPEC/MOS/JWPipeNB-NIRSpec-MOS.ipynb}.},
together with a customized reduction pipeline incorporating a cross-dispersion aperture correction to account for source-position-dependent slit losses caused by intra-shutter positioning. Both processing approaches yield broadly consistent spectral results.

\begin{figure*}[!t]
\centering
\includegraphics[width=0.85\textwidth]{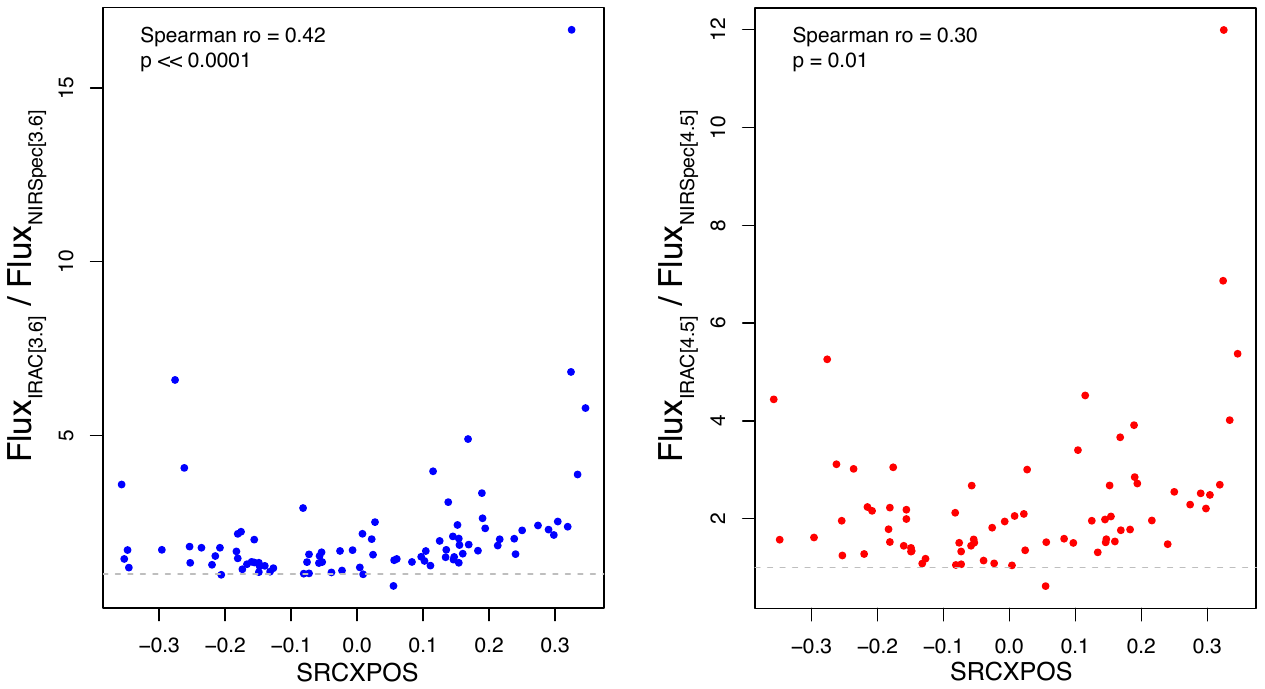}
\caption{IRAC-to-NIRSpec flux ratio as a function of source position along the dispersion direction within the MSA shutter, at wavelengths near 3.6~\micron\ (left) and 4.5~\micron\ (right), for Visit~3. Source position (SRCXPOS) is given in dimensionless relative shutter units, with zero corresponding to the shutter center. The Spitzer/IRAC fluxes are systematically higher than the median JWST/NIRSpec fluxes measured over the corresponding wavelength ranges, and the IRAC-to-NIRSpec flux ratio increases with SRCXPOS. These trends indicate that the JWST pipeline pathloss correction does not fully recover the absolute fluxes of PMS systems in our dataset.} \label{fig_IRAC_vs_NIRSPEC}
\end{figure*}

Because the observations used three nod positions (A, B, and C) for each MSA configuration, each nodded exposure was treated in turn as the science exposure, while the other two nods were treated as background exposures. The standard reduction therefore produced $(A-(B+C)/2)$, $(B-(A+C)/2)$, and $(C-(A+B)/2)$ spectra, which were subsequently combined. As a sanity check on two effects discussed below --- the unusual Br$\alpha$ absorption profiles and the sensitivity of the narrow spectral features at 2.8--4.3~\micron\ to science-nod position and background selection --- we also reprocessed each science nod independently without combining the three nods. For each science nod, we considered both the average of the other two nods and each individual alternative nod as the background --- for example, $(A-(B+C)/2)$, $(A-B)$, and $(A-C)$ for science nod A. 

A major challenge identified during the JWST data reduction is that the calibrated NIRSpec spectra exhibit significant flux-level and spectral-slope inconsistencies for the same sources observed in different MSA configurations. In addition, substantial discrepancies are present between Spitzer Space Telescope IRAC photometry (3.6 and 4.5~\micron; \citet{Megeath2012}) and the corresponding NIRSpec fluxes at similar wavelengths (Figure~\ref{fig_IRAC_vs_NIRSPEC}). For example, in Visit 3, the IRAC-to-NIRSpec flux ratio shows a statistically significant dependence on the source position along the dispersion direction within the MSA shutters (SRCXPOS), with Spearman correlation probabilities ranging from $<0.0001$ to 0.01. These spatially dependent flux trends suggest that the current NIRSpec pathloss correction\footnote{See the following pathloss memo: \url{https://www.stsci.edu/contents/news/jwst/2025/new-pathloss-corrections-for-nirspec-multi-object-and-fixed-slit-spectroscopy-are-available}.} does not fully remove position-dependent flux calibration systematics for these observations.

A detailed assessment of these effects, including NIRSpec and IRAC flux comparisons across multiple bands, source positions within MSA shutters, flux discrepancy metrics and their dependence on spatial offsets within shutters, and NRS Y-pixel distributions of extracted spectra, was communicated to the JWST NIRSpec team. Although we have not yet received feedback regarding possible mitigation strategies, we proceeded with the scientific analysis using two complementary approaches designed to minimize the impact of these systematic flux uncertainties, depending on the scientific objective.

For analyses focused on variability of numerous narrow spectral features between X-ray flare and non-flare epochs, the two NIRSpec spectra of each COUP-JWST source obtained at different epochs were normalized by aligning their continua. The continua were estimated using a rolling median filter \citep{Tukey1977}, implemented with the \texttt{rollapply} routine from the R package \texttt{zoo} \citep{Zeileis2005}. This approach provides a robust local continuum estimate that is minimally affected by narrow emission or absorption features. This procedure preserves relative changes in narrow spectral features while minimizing the impact of epoch-to-epoch flux calibration offsets.

For analyses involving integrated line fluxes, including H~I Br$\alpha$ and CO emission variability, as well as correlations of H~I Br$\alpha$ and CO luminosities with stellar and environmental properties (Paper~II), we adopted a different approach. The NIRSpec continua were first estimated (via \texttt{ctool} \citep{Pontoppidan2026} or  rolling-quantile), compared with the corresponding Spitzer/IRAC fluxes, and used to derive empirical flux scaling corrections. The corrected NIRSpec spectra were then adjusted for extinction. This approach provides an approximate absolute flux calibration while accounting for the apparent NIRSpec-to-IRAC flux offsets.

\begin{figure*}[!t]
\centering
\includegraphics[width=0.85\textwidth]{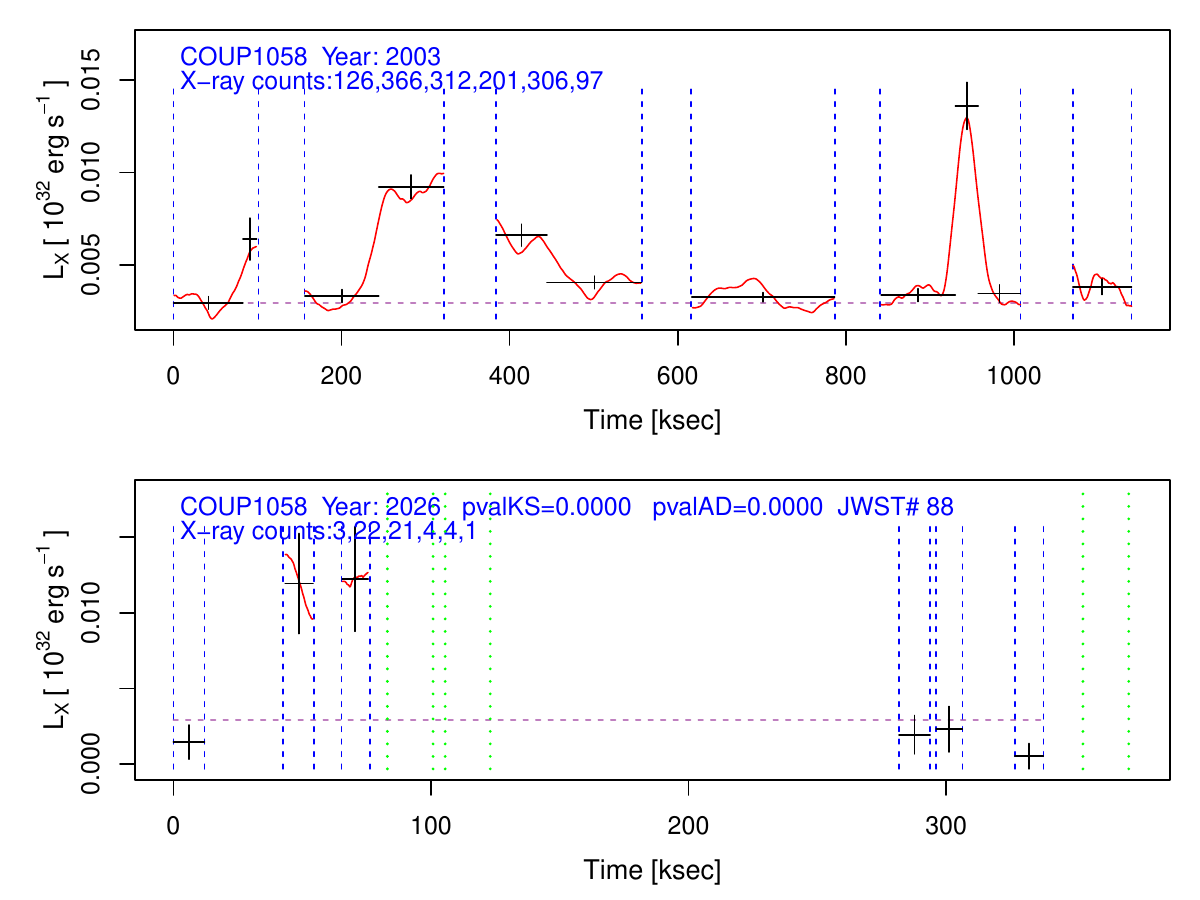}
\caption{Example Chandra X-ray light curves in units of intrinsic stellar X-ray luminosity for COUP1058--JWST88. The two panels correspond to the two-epoch observations: the deep 2003 COUP exposure (top) and the six shallow Chandra snapshots obtained in early February 2026 contemporaneously with JWST visits (bottom). The time origins ($t=0$) are 2003 January 8 20:58:18.8 UT and 2026 February 1 02:03:19 UT for the 2003 and 2026 panels, respectively. Blue dashed lines mark the start and end times of Chandra observations; green dashed lines mark the JWST visits (\#3, \#4, and \#7). Red curves show kernel density estimates (KDEs). Black segments with $1\sigma$ uncertainties \citep{Gehrels1986} represent Bayesian Blocks. Dashed magenta lines indicate the characteristic emission level during the 2003 COUP epoch. Legends provide COUP and JWST identifiers, observation epochs, Kolmogorov--Smirnov (KS) and Anderson--Darling (AD) test $p$-values comparing photon arrival times with a constant source model, and total detected X-ray photons per observation. Figure Set atlases of Chandra light curves and JWST spectra for all COUP--JWST PMS systems are available online, organized into four source categories: strong X-ray variability, mild X-ray variability, X-ray constant, and X-ray undetected in February 2026.} \label{fig_Chandra_Lcs}
\end{figure*}

Both approaches assume that the continuum flux level does not undergo significant astrophysical variability between the two NIRSpec epochs. In the presence of the identified pathloss-related systematics, genuine continuum variability cannot be robustly recovered without independent contemporaneous photometric observations. Consequently, epoch-to-epoch changes in the integrated Br$\alpha$ and CO fluxes are conditional on this stable-continuum assumption, whereas changes in line-profile morphology are less sensitive to the absolute continuum scaling.

\subsection{Chandra Data}\label{sec_Chandra_data}

The six Chandra observations (ObsIDs 31045--31050) obtained during our coordinated February 1--5, 2026 campaign (Figure~\ref{fig_planned_observations}c) were performed with the ACIS-I imaging array \citep{Garmire2003} on board the Chandra X-ray Observatory \citep{Weisskopf2002}. The data were reduced using CIAO v4.18 \citep{Fruscione2006}, CALDB v4.12.3, HEASOFT v6.33.2, and ACIS Extract v5658 2022-01-25 (AE; \citealt{Broos2010,Broos2012}), following the procedures described by \citet{Getman2024}. Standard processing included reprocessing the Level~1 event files to Level~2, applying the latest calibration products, identifying and filtering background flares, generating exposure maps, and performing point-source detection using maximum-likelihood image reconstruction \citep{Broos2010}. Source and background events were then extracted with AE, and source properties, including net counts, median photon energies, apparent fluxes, and light curves, were derived.

\begin{figure*}[!t]
\centering
\includegraphics[width=0.85\textwidth]{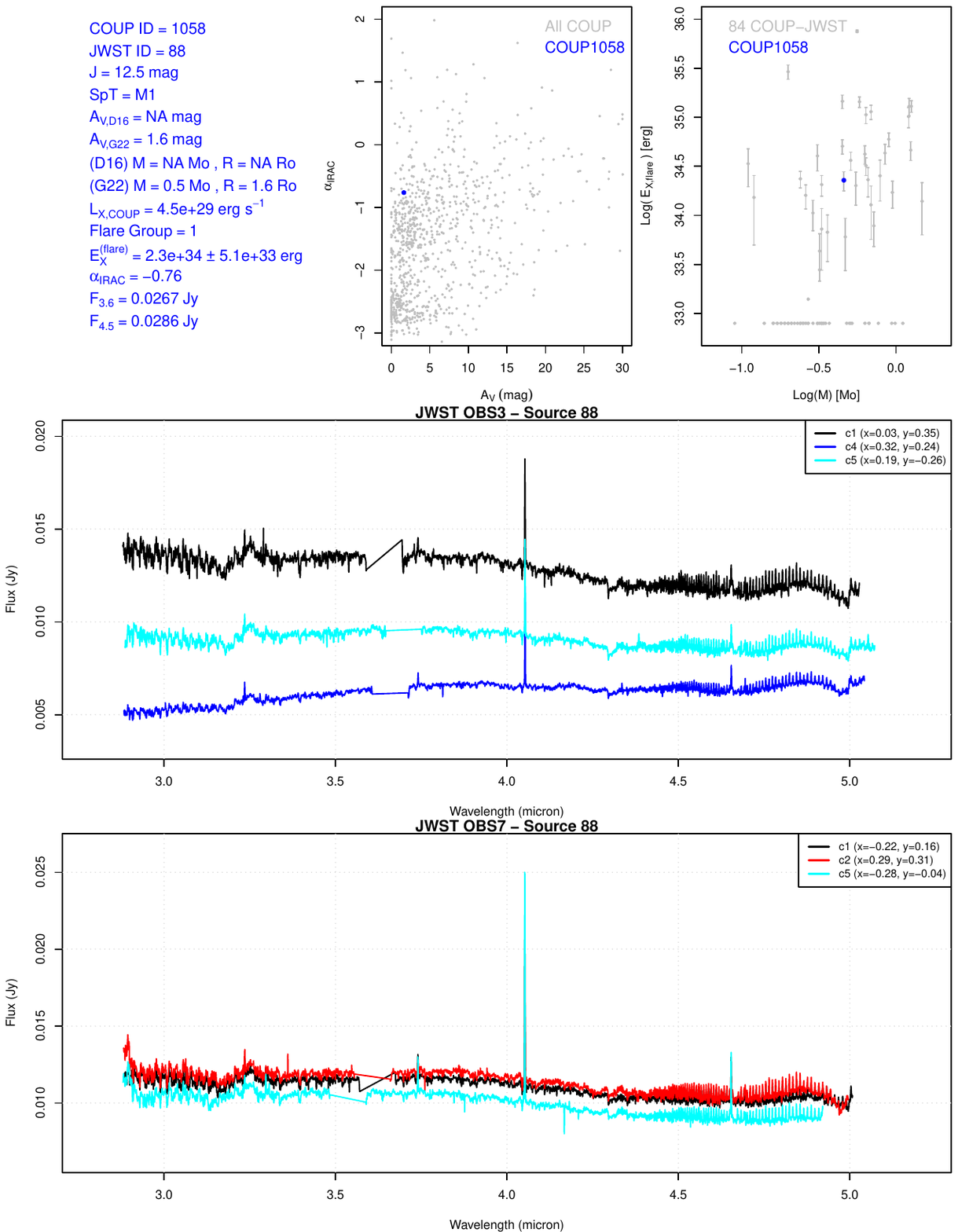}
\caption{Example of COUP1058--JWST88. The upper panels summarize stellar and environmental properties and show the source position (blue point) on the Spitzer/IRAC SED slope ($\alpha_{IRAC}$) versus visual extinction ($A_V$) and X-ray flare energy ($E_{\rm X,fl}$) versus stellar mass ($M$) diagrams. The $\alpha_{IRAC}$--$A_V$ diagram includes all COUP stars, while the $E_{\rm X,fl}$--$M$ diagram includes COUP--JWST systems from this project. Points plotted at $\log(E_{\rm X,fl}/{\rm erg})<33$ represent upper limits. The middle and bottom panels show all available JWST/NIRSpec spectra of COUP1058--JWST88 from all JWST visits and MSA configurations, color-coded by configuration. Legends provide the source X and Y positions within the corresponding MSA shutters. Figure Set atlases of Chandra light curves and JWST spectra for all COUP--JWST PMS systems are available online, organized into four source categories: strong X-ray variability, mild X-ray variability, X-ray constant, and X-ray undetected in February 2026.} \label{fig_NIRSPec_spectra}
\end{figure*}

The X-ray photon arrival times were segmented using the Bayesian Blocks algorithm \citep{Scargle1998,Scargle2013}, adopting the event-data Cash fitness function and the calibrated change-point prior $ncp_{\rm prior}=4-\ln(73.53\,p_0N^{-0.478})$, with $p_0=0.01$. We used the R implementation\footnote{Available at \url{https://github.com/AlicePagano/ASPA-Project-Bayesian-Blocks-Algorithm-in-R}.}, previously employed by \citet{Getman2024}. The algorithm partitions the photon arrival sequence into statistically homogeneous intervals separated by change points. Bayesian Blocks have become a standard tool for identifying X-ray variability and flaring activity in stellar light curves (e.g., \citealt{GetmanFeigelson2021}).

The resulting block representation was used to identify characteristic (a.k.a. baseline) emission intervals and elevated X-ray emission associated with major flares, and to determine the characteristic X-ray luminosity and energetics of each flare. For every Bayesian Block, the AE-derived point-spread-function-corrected apparent photon flux, $F_{\rm X,phot}$ (photons~s$^{-1}$~cm$^{-2}$), was converted to the intrinsic X-ray luminosity, $L_{\rm X}$, in the 0.5--8.0~keV energy band. For each source, a single conversion factor between $F_{\rm X,phot}$ and $L_{\rm X}$ was derived by scaling the observed photon flux to the time-averaged intrinsic X-ray luminosity measured from COUP. Details of this source-specific calibration procedure are given in \citet{Getman2024}.

Bayesian Blocks with luminosities exceeding three times the characteristic luminosity were tentatively identified as candidate large X-ray flares. For each candidate event, the flare energy, $E_{\rm X,fl}$, was estimated by subtracting the characteristic luminosity and integrating the excess luminosity over the flare duration. Because of the discontinuous Chandra coverage, these energies often represent lower limits to the total X-ray flare energies.

Based on the Bayesian Blocks analysis, visual inspection of the X-ray light curves, comparison with the 2003 COUP light curves, and AD and KS tests for departures from a constant photon arrival rate (Figure~\ref{fig_Chandra_Lcs}), the 84 COUP-JWST targets were classified into four groups. Group~1 consists of 18 stars exhibiting convincing large X-ray flares during the 2026 Chandra observations, typically showing highly significant departures from a constant photon arrival rate in the AD and KS tests ($p\lesssim10^{-3}$). Group~2 contains 20 stars showing weaker X-ray variability or possible flaring, generally with less significant variability statistics (typically at least one AD or KS probability below 0.05). Group~3 comprises 26 stars whose photon arrival distributions are consistent with little or no variability ($p>0.05$), and therefore represent predominantly constant X-ray emitters. Group~4 includes 20 stars that were not detected in the 2026 Chandra observations.

\subsection{Chandra-JWST Source Atlases and Overview of the Dataset}\label{sec_source_atlases}

To facilitate visualization of the combined Chandra and JWST data products, and to guide subsequent sample selection, we assembled a summary sheet for each COUP-JWST target containing the Chandra X-ray light curve, the corresponding JWST/NIRSpec spectra, and key stellar properties. These summaries consist of two pages for sources in Groups~1--3 and one page for Group~4 (Figures~\ref{fig_Chandra_Lcs} and \ref{fig_NIRSPec_spectra}). The individual source summaries were combined into four PDF atlases corresponding to the four X-ray variability groups.

Inspection of these atlases provides an initial overview of the X-ray flare properties and infrared spectroscopic diversity of the COUP-JWST sample. Given the limited number of successful JWST visits (three out of eight; Section~\ref{sec_campaign_overview}), the current dataset does not contain any captured X-ray mega-flares with energies $E_{\rm X,fl} > 10^{36}$~erg. Only a small number of events reach the regime of powerful superflares, whereas most detected flares have moderate superflare energies in the range $10^{33} \lesssim E_{\rm X,fl} \lesssim 10^{35}$~erg (Figure~\ref{fig_NIRSPec_spectra}, top-right panel). Only two sources, COUP152-JWST5 and COUP385-JWST35, exhibit the most energetic detected flares, with $\log(E_{\rm X,fl}/{\rm erg}) > 35.2$. However, these systems lack second-epoch NIRSpec observations and therefore cannot be used to assess flare-associated spectral variability. Consequently, the two-epoch analysis presented below does not probe the response of disk spectra to the most extreme X-ray flares.

The NIRSpec spectra reveal a rich variety of disk and stellar spectral signatures. Many spectra exhibit molecular absorption features, including H$_2$O bands primarily near 2.8--3.0~\micron\ and OH absorption features distributed across the 3.0--4.3~\micron\ region, with morphologies resembling stellar photospheric molecular absorption patterns. To assess the stellar contribution to these spectra, we compare the observations with synthetic stellar atmosphere models from the BT-Settl CIFIST grid \citep{Allard2012}. The models adopt the solar abundances of \citet{Caffau2011}, are reddened using the individual visual extinctions ($A_V$) derived for the COUP-JWST sources and the wavelength-dependent extinction law of \citet{Gordon2023}, and are scaled to reproduce the observed $J$-band magnitudes. These comparisons provide a reference for identifying photospheric molecular absorption and separating stellar atmospheric signatures from disk-related emission and absorption features.

\begin{figure*}[!t]
\centering
\includegraphics[width=0.85\textwidth]{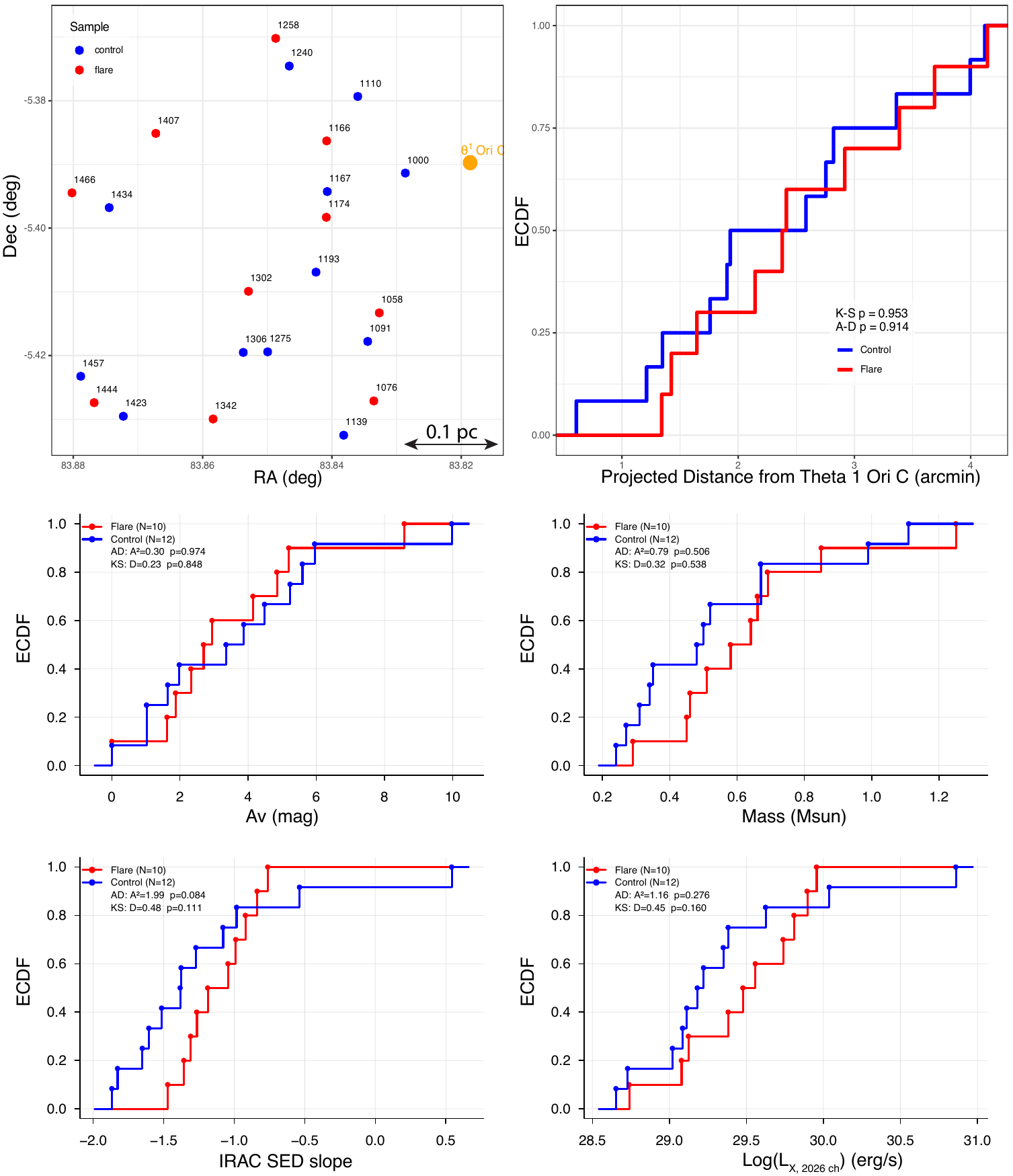}
\caption{Comparison of stellar and environmental properties for the 10 COUP--JWST Flare sample members (red) and 12 Control sample members (blue). The compared properties include celestial positions, projected distance from $\theta^1$~Ori~C (orange), visual extinction, stellar mass, Spitzer/IRAC SED slope, and baseline (characteristic) stellar X-ray luminosity during the 2026 epoch. KS and AD test $p$-values are provided in the figure legends.} \label{fig_samples_Flare_vs_Control}
\end{figure*}

Hydrogen recombination lines are commonly detected, predominantly in emission, with H~I Br$\alpha$ at 4.05~\micron\ being the strongest feature. Several weaker H~I lines are also seen in many spectra, including Br~9 at 3.30~\micron, Pf$\gamma$ at 3.74~\micron, Hu~13 at 4.17~\micron, Pf$\beta$ at 4.65~\micron, and Hu~11 at 4.67~\micron. A substantial fraction of disk-bearing systems display the CO fundamental emission-band forest between 4.3 and 5.2~\micron, while others show CO absorption, likely associated with high disk inclinations, absorption by foreground molecular material, or proplyd and photospheric contributions.

Several spectra also exhibit broad continuum depressions shortward of 3.3~\micron, potentially arising from ice absorption, particularly the broad H$_2$O-ice band near 3~\micron, with possible contributions from weaker ice features such as NH$_3$ near 2.93~\micron, as recently observed with JWST/NIRSpec toward a highly inclined protoplanetary disk \citep{Sturm2023,Bergner2024}.

\subsection{Selection of the Flare and Control Samples}\label{sec_samples}

The aforementioned atlases (Figures~\ref{fig_Chandra_Lcs} and \ref{fig_NIRSPec_spectra}) also served as the basis for selecting the two samples used to investigate the impact of stellar X-ray flares on protoplanetary disk chemistry. Candidate Flare sample members were required to satisfy the following criteria: (1) NIRSpec spectra available at both epochs (Visits 3 and 7; Figure~\ref{fig_planned_observations}c) and free of significant instrumental artifacts, such as artificial ``wiggles'' caused by PSF undersampling and/or sub-pixel spectral-trace tilt on the detector; and (2) clear evidence for a strong or moderate X-ray flare during one or both observing epochs. These criteria yielded a Flare sample of 10 COUP-JWST systems (Table~\ref{tab_stellar_properties}), comprising seven sources from Group~1 and three from Group~2.

The Control sample consists of 12 COUP-JWST systems with high-quality two-epoch NIRSpec spectra and no evidence for significant X-ray variability during the 2026 Chandra observations. All Control sample members were selected from Group~3 (Table~\ref{tab_stellar_properties}).

As shown in Figure~\ref{fig_samples_Flare_vs_Control}, the Flare and Control samples have statistically indistinguishable distributions of projected distance from $\theta^1$~Ori~C, visual extinction, stellar mass, IRAC spectral energy distribution slope, and characteristic X-ray luminosity. Their spatial distributions across the observed Orion fields are also similar. This close matching minimizes potential biases arising from environmental or stellar properties and enables a more robust assessment of the effects of stellar X-ray flares on disk chemistry.

\section{Science Analyses and Results}\label{sec_analyses_results}

Employing these Flare and Control COUP-JWST samples, this paper investigates the impact of stellar flares on NIRSpec spectral variability through three complementary analyses. Specifically, we examine flare-associated changes in: (1) numerous narrow spectral features across the 2.8--4.3~\micron\ wavelength range; (2) the hydrogen recombination Br$\alpha$ line at 4.05~\micron; and (3) the CO fundamental ro-vibrational band forest spanning 4.3--5.2~\micron.

\subsection{Variability of Narrow NIRSpec Spectral Features (2.8--4.3~\micron)}\label{sec_narrow_features_analyses}

\begin{figure*}[!t]
\centering
\includegraphics[width=0.85\textwidth]{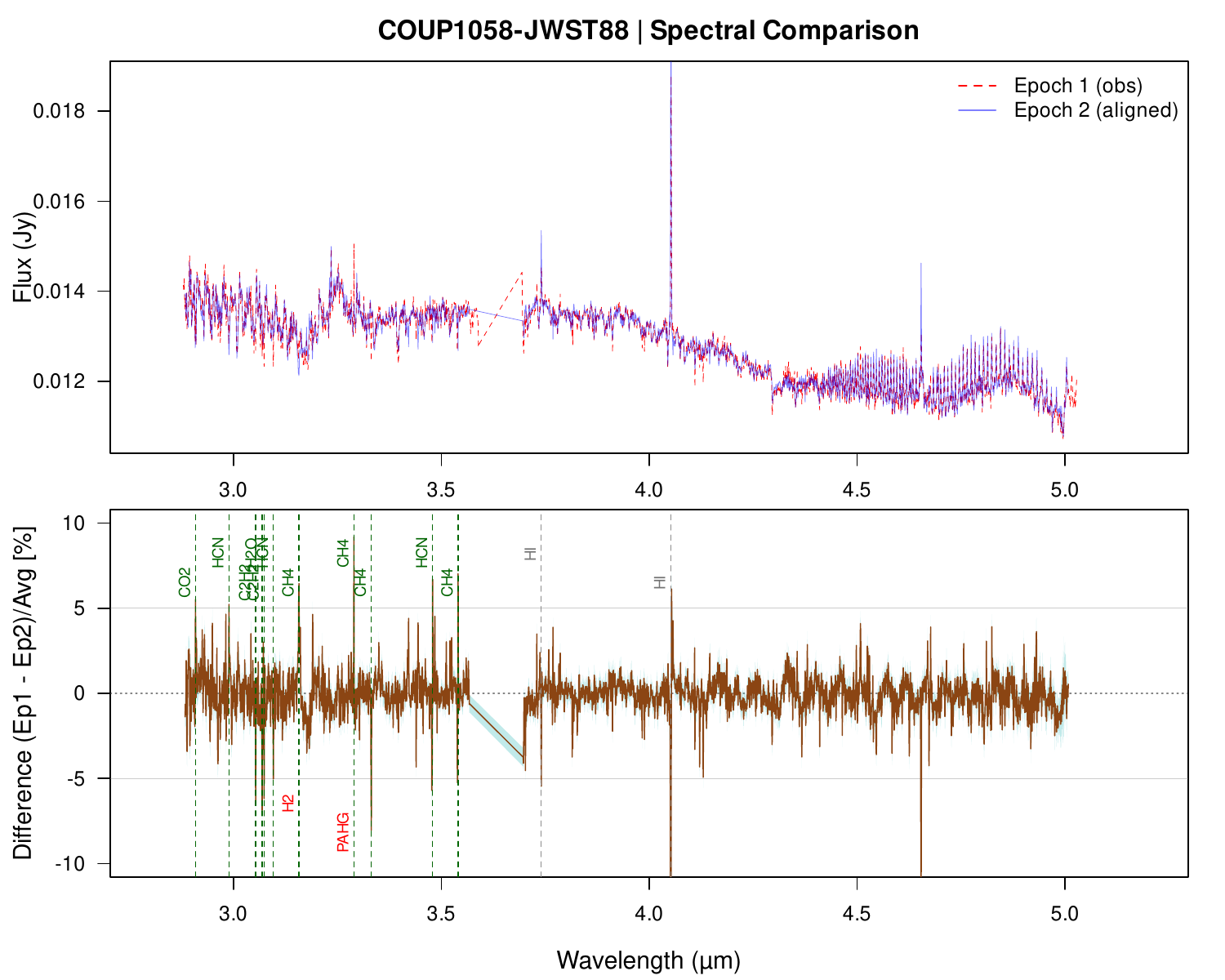}
\caption{Example of COUP1058--JWST88 showing the comparison of NIRSpec spectra between two epochs: one near an X-ray flare (red dashed) and one away from the flare (solid blue). For Flare sources, Epoch~1 denotes the flare-associated epoch and Epoch~2 the comparison epoch, irrespective of chronological order; for Controls, the epoch assignment is random. The spectra were first aligned by scaling their continua using a rolling-median method. The bottom panel shows the relative spectral difference, $(F_1-F_2)/[(|F_1|+|F_2|)/2]\times100\%$, as a function of wavelength. The propagated $1\sigma$ uncertainty envelope (cyan) is typically too narrow to be visible. For features exhibiting $>5\%$ differences at wavelengths below 4.3~\micron, dark-green dashed lines mark the narrow spectral features included in our analysis; upper dark-green labels give their candidate molecular identifications based on wavelength matches to HITRAN transitions. Lower red labels give candidate species associations from wavelength matches to the PDRs4All Orion Bar line list. Gray dashed lines labeled H~I mark prominent hydrogen recombination lines with $>5\%$ differences at wavelengths below 4.3~\micron\ that were excluded during the initial narrow-feature screening. All HITRAN and PDRs4All associations are tentative. A complete Figure Set of NIRSpec spectral comparisons for the 10 Flare and 12 Control COUP--JWST systems is available online.} \label{fig_narrow_features_Ep1_vs_Ep2}
\end{figure*}

For each of the 10 Flare and 12 Control sample members, the two-epoch NIRSpec spectra were aligned by matching their rolling-median continua. Figure~\ref{fig_narrow_features_Ep1_vs_Ep2} and the accompanying online Figure Set compare the aligned spectra for all targets. We identified 281 narrow spectral features (excluding the prominent hydrogen recombination lines considered in our initial line screening) within the 2.8--4.3~\micron\ wavelength range by subtracting the two continuum-aligned epoch spectra and examining their relative differences, $(F_1-F_2)/[(|F_1|+|F_2|)/2]\times100\%$. Features exhibiting absolute relative differences $>5\%$ were identified as strongly variable. Such flux changes are statistically significant, typically exceeding the propagated uncertainty envelopes (cyan) by factors of several or more (Figure~\ref{fig_narrow_features_Ep1_vs_Ep2}). The 281 highly variable narrow features are listed in Table~\ref{tab_narrow_features}.

For all identified high-variability spectral features, quality-control diagnostics indicate nominal spectral extraction: the NIRSpec data-quality flags are clean (DQ = 0), and the spatial extraction footprint is unchanged (NPIXELS = 5.0). In addition, the ratio of read-noise variance to photon Poisson variance ($\mathrm{FLUX\_VAR\_RNOISE}/\mathrm{FLUX\_VAR\_POISSON}$) is consistently low ($<0.25$), indicating that the affected spectral bins are photon-noise dominated rather than read-noise dominated, with no indication from the pipeline diagnostics that the affected spectral bins coincide with detector gaps or other low-throughput regions.

Visual inspection of the individual-epoch spectra indicates that 32 of the 281 features cannot be reliably classified as either absorption or emission because they coincide with rapidly changing local pseudo-continuum structure. Among the remaining 249 features, 186 (75\%) are classified as absorption, with no evidence for emission in the other epoch; 43 (17\%) are classified as emission, with no evidence for absorption in the other epoch; and 20 (8\%) switch between absorption and emission between the two epochs. 
\begin{figure*}[!t]
\centering
\includegraphics[width=0.85\textwidth]{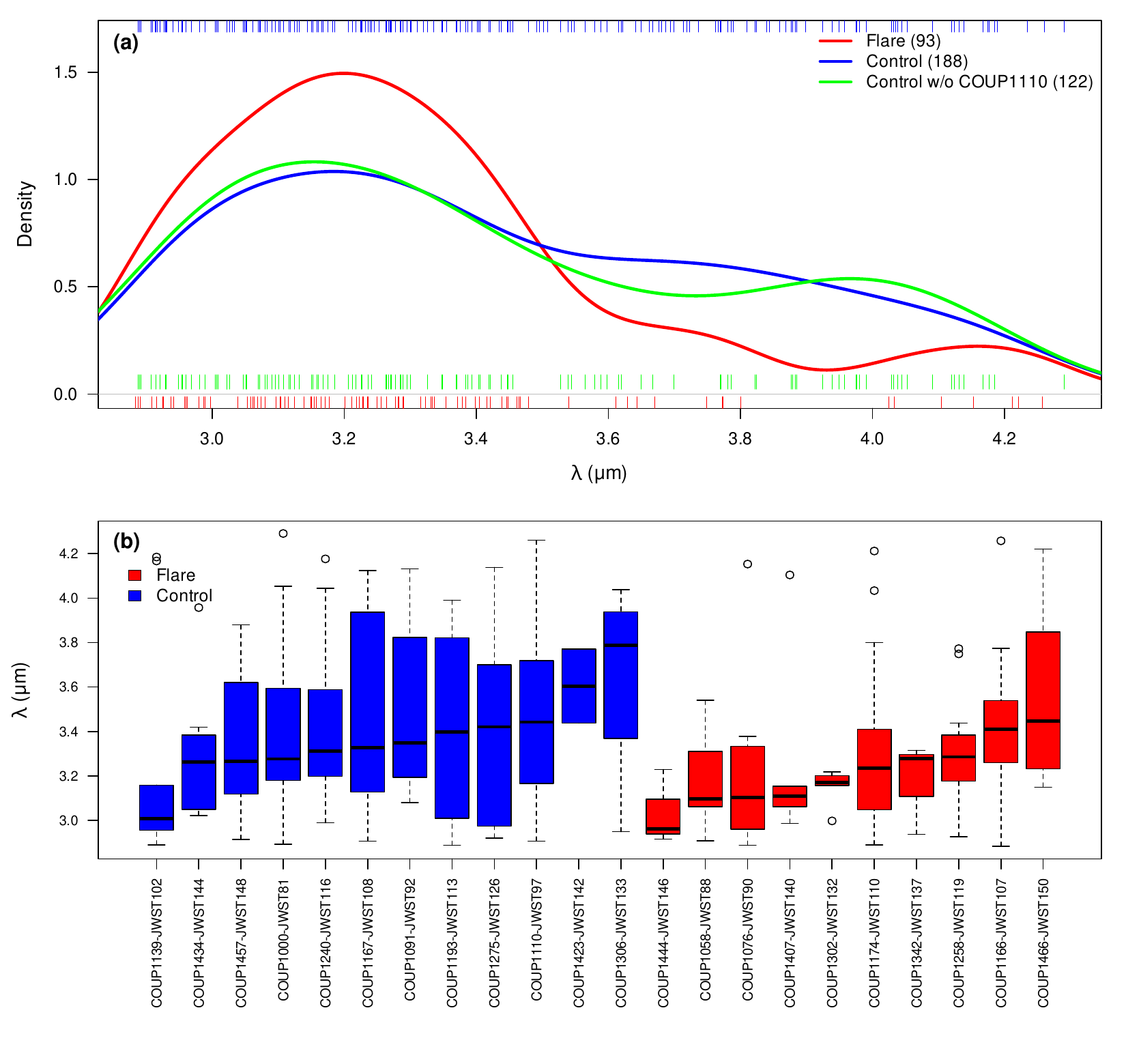}
\caption{(a) Kernel density distributions of wavelengths for 281 narrow spectral features showing strong two-epoch NIRSpec flux variability in the 10 Flare (red; 93 features) and 12 Control (blue; 188 features) COUP--JWST systems. The distribution for 11 Control systems (green; 122 features), excluding COUP1110--JWST97 with an unusually large number of variable features, is also shown. The Flare and full Control distributions differ significantly (KS: $p=0.0024$; AD: $p=0.0069$), whereas after excluding COUP1110--JWST97 the difference is marginally significant (KS: $p=0.034$; AD: $p=0.042$). (b) Source-level comparison of the wavelength distributions of variable features using box plots for individual Flare (red) and Control (blue) systems. Unlike panel (a), this analysis treats each star as a single statistical unit by comparing q25, median (q50), and q75 wavelengths. After Benjamini--Hochberg FDR correction for these three comparisons, the q75 Flare--Control difference is significant according to the KS test ($q_{\rm FDR}=0.0077$) and marginally significant according to the AD test ($q_{\rm FDR}=0.022$), whereas q25 (KS: $q_{\rm FDR}=0.71$; AD: $q_{\rm FDR}=0.83$) and q50 (KS: $q_{\rm FDR}=0.11$; AD: $q_{\rm FDR}=0.10$) show no significant differences.} \label{fig_narrow_features_KDE_source_level_comparison}
\end{figure*}

Our primary science question is whether the strongest spectral changes observed in stars experiencing X-ray flares differ statistically from those in stars without detected flares. Figure~\ref{fig_narrow_features_KDE_source_level_comparison}a compares the kernel density estimates (KDEs) of the wavelengths of the 281 highly variable narrow spectral features for the Flare (red; 93 features) and Control (blue; 188 features) samples, together with the Control sample excluding COUP1110--JWST97 (green; 122 features), which contains substantially more highly variable narrow features than any other target. The feature-level wavelength distributions differ significantly between the Flare and Control samples (KS: $p=0.0024$; AD: $p=0.0069$). After excluding COUP1110--JWST97 from the Control sample, the difference becomes marginally significant (KS: $p=0.034$; AD: $p=0.042$). Because individual spectral features from the same star are not statistically independent, we regard the complementary source-level analysis in Figure~\ref{fig_narrow_features_KDE_source_level_comparison}b as the more conservative test. This analysis treats each star as a single statistical unit by comparing the q25, median (q50), and q75 wavelengths of its highly variable features. To account for the three quantile comparisons, we applied the Benjamini--Hochberg false-discovery-rate (FDR) correction \citep{BenjaminiHochberg1995,Miller2001} separately to each statistical test. For q75, the Flare--Control difference is significant according to the KS test ($q_{\rm FDR}=0.0077$) and marginally significant according to the AD test ($q_{\rm FDR}=0.022$). The q25 comparison is not significant (KS: $q_{\rm FDR}=0.71$; AD: $q_{\rm FDR}=0.83$), nor is the q50 comparison (KS: $q_{\rm FDR}=0.11$; AD: $q_{\rm FDR}=0.10$).

\begin{figure*}[!t]
\centering
\includegraphics[width=0.85\textwidth]{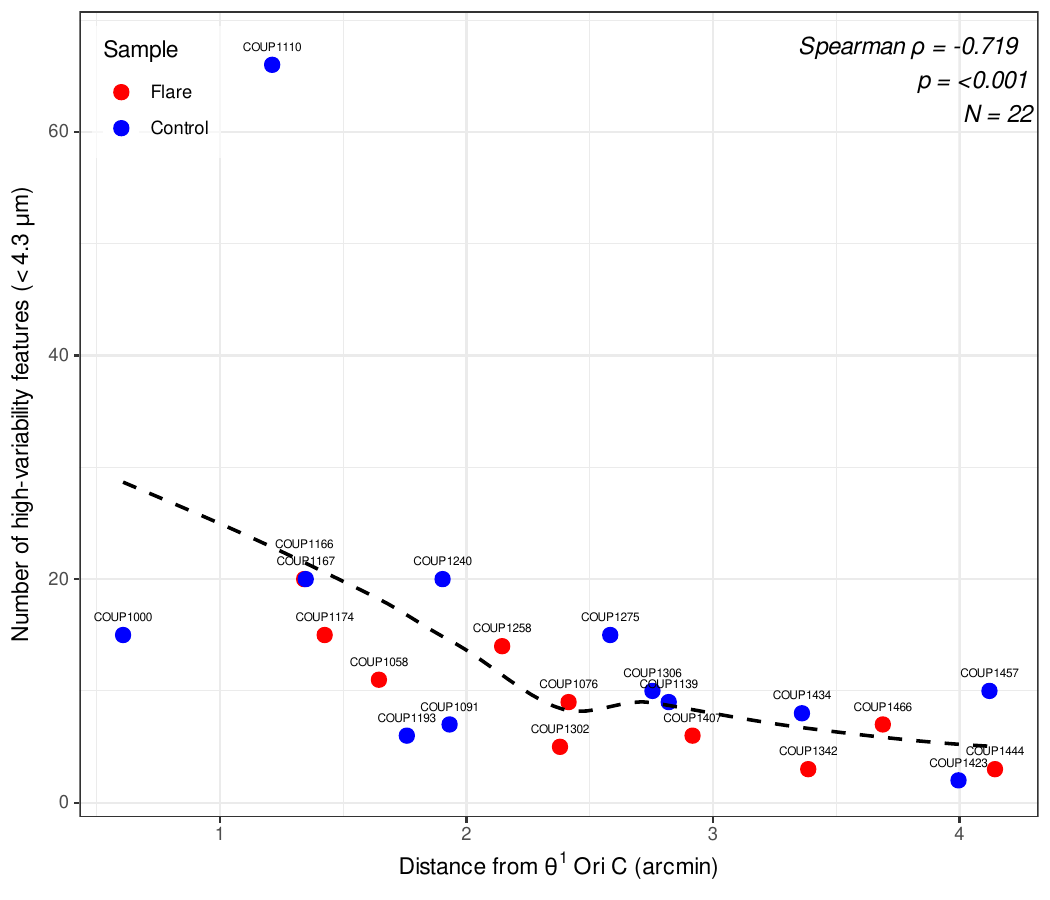}
\caption{The number of highly variable NIRSpec spectral features as a function of projected distance from $\theta^1$~Ori~C. Spearman correlation results are listed in the figure legends. COUP source identifiers for the 10 Flare (red) and 12 Control (blue) COUP--JWST systems are shown as point labels. The black dashed curve represents a locally estimated scatterplot smoothing (LOESS) fit \citep{Cleveland1979}.} \label{fig_narrow_features_distance_to_Theta1OriC}
\end{figure*}

Overall, the results presented in Figure~\ref{fig_narrow_features_KDE_source_level_comparison} provide tentative evidence for a Flare--Control difference concentrated primarily toward the long-wavelength tail of the distribution rather than a uniform shift across the entire 2.8--4.3~\micron\ interval. At the source level, q25 and q50 do not differ significantly between the samples, whereas the q75 difference is significant according to the KS test and marginally significant according to the AD test. Specifically, the Flare sample appears to contain relatively fewer highly variable narrow spectral features at longer wavelengths (approximately 3.5--4.3~\micron) than the Control sample. Given the small source samples (10 Flare and 12 Control stars), this possible difference requires confirmation with larger samples.

The possible Flare--Control difference in the long-wavelength tail cannot be attributed to NIRSpec-MOS spectral truncation or occasional spectral gaps. The distributions of spectral truncation wavelengths and spectral gap-edge wavelengths are statistically indistinguishable between the Flare and Control samples, indicating that differences in wavelength coverage do not bias the comparison.

To investigate the nature of these highly variable narrow spectral features, their wavelengths were cross-matched against the empirical PDRs4All NIRSpec line list for the Orion Bar \citep{Peeters2024}, providing a direct comparison with transitions observed in externally irradiated PDR gas in the Orion Nebula; the PDRs4All NIRSpec mosaic lies within the area covered by our NIRSpec-MOS field. Using one nominal NIRSpec resolution element ($R\simeq2700$) as the matching tolerance, 67 of the 281 features (24\%) coincide with at least one PDRs4All transition. These candidate associations include several high-order H~I recombination transitions not included among the prominent hydrogen lines removed during our initial feature screening; because wavelength coincidence alone does not establish an H~I identification, these features are retained in the narrow-feature sample. Circularly shifting the complete observed feature-wavelength pattern relative to the PDRs4All line list yields a mean chance-coincidence rate of 23\%, statistically indistinguishable from the observed rate ($p=0.39$). A complementary comparison of the distributions of the nearest-line separations, expressed in units of the NIRSpec spectral resolution, likewise shows no excess of unusually close matches ($p=0.35$). Thus, individual PDRs4All associations listed in Table~\ref{tab_narrow_features} should be regarded as tentative wavelength coincidences rather than secure line identifications; at the population level, their frequency and wavelength proximity are consistent with chance superposition on the dense PDR spectrum.

Nevertheless, the composition of these tentative PDRs4All associations provides useful context for the types of transitions present near the observed feature wavelengths. Among the 67 matched features, 31 (46\%) have H$_2$ or tentative H$_2$ associations, 17 (25\%) H~I, eight (12\%) PAH-related, four (6\%) He~I, and four (6\%) CH$^+$ associations; the remaining three are associated with O~I or [Fe~III]. Thus, the candidate PDRs4All associations are dominated by molecular hydrogen and hydrogen recombination transitions, although the chance-coincidence tests above preclude interpreting these frequencies as secure identifications.

We additionally cross-matched the feature wavelengths against selected molecular transitions from the HITRAN2020 database \citep{Gordon2022}, representing species commonly observed or predicted in protoplanetary disks and their immediate environments (C$_2$H$_2$, C$_2$H$_4$, CH$_4$, CO$_2$, CO, H$_2$O, H$_2$, HCN, NH$_3$, OH, and SO), and against the Vienna Atomic Line Database (VALD; \citealt{Ryabchikova2015}), containing atomic and molecular transitions expected in the photospheres of late-K and M-type pre-main-sequence stars. For each observed NIRSpec feature, the 20 nearest HITRAN or VALD transitions within a wavelength window of $\pm0.002~\micron$ of the measured feature centroid were identified. Among these candidates, the transition with the strongest predicted line strength was adopted as the preferred wavelength association.

The wavelength-offset analysis shows that the VALD associations are systematically farther from the observed feature centroids than the corresponding HITRAN associations. This result suggests that the highly variable narrow NIRSpec features are more closely associated in wavelength with molecular transitions cataloged in HITRAN than with atomic or photospheric transitions listed in VALD. However, because the matching procedure is based on wavelength proximity and transition strength, without radiative-transfer modeling or verification against predicted molecular spectra, these associations should be regarded as plausible candidate identifications rather than definitive molecular assignments.

The tentative HITRAN associations are dominated by a small number of molecular species. Of the 281 highly variable narrow spectral features, 39.9\% are associated with CH$_4$, 19.2\% with HCN, 18.9\% with C$_2$H$_2$, and 10.0\% with NH$_3$, while CO$_2$, H$_2$O, OH, and SO together account for the remaining 12.1\%. The relative distributions of the candidate molecular species are broadly similar between the Flare and Control samples. CH$_4$, HCN, and C$_2$H$_2$ together account for 78.5\% of the Flare associations and 77.7\% of the Control associations, with CH$_4$ the most frequent candidate association in both samples (41.9\% and 38.8\%, respectively).

Finally, Figure~\ref{fig_narrow_features_distance_to_Theta1OriC} shows the number of highly variable narrow NIRSpec spectral features detected in each of the 22 Flare and Control COUP--JWST systems as a function of projected distance from $\theta^1$~Ori~C. A statistically significant anti-correlation is found ($\rho=\mathrm{-0.72}$, $p<\mathrm{0.001}$), indicating that the number of highly variable features decreases with increasing projected distance from the Trapezium's dominant ionizing star. This trend suggests that the incidence of narrow spectral variability is influenced by environmental conditions associated with the intense UV radiation field of $\theta^1$~Ori~C.

\begin{figure*}[!t]
\centering
\includegraphics[width=0.85\textwidth]{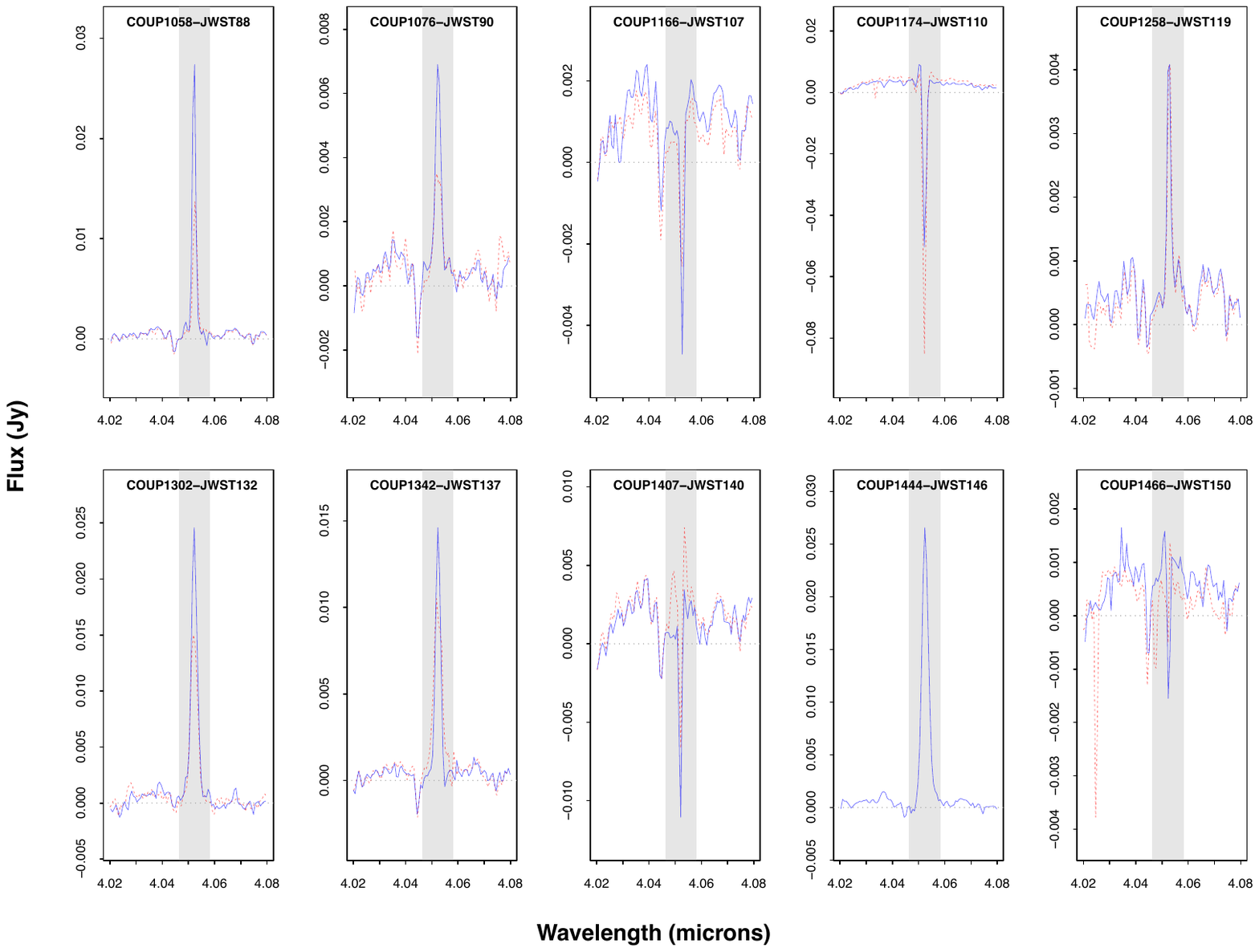}
\caption{Examples of NIRSpec spectra of the H~I Br$\alpha$ line for the 10 Flare sample members, scaled to the Spitzer/IRAC flux level and corrected for extinction. Spectra obtained during X-ray flare and non-flare epochs are shown as dashed red and solid blue lines, respectively. The vertical gray bar marks the wavelength interval used for flux integration. A Figure Set containing the H~I Br$\alpha$ spectra of the 12 Control sample members is available online.} \label{fig_Bralpha_spectra}
\end{figure*}

\begin{figure*}[!t]
\centering
\includegraphics[width=0.85\textwidth]{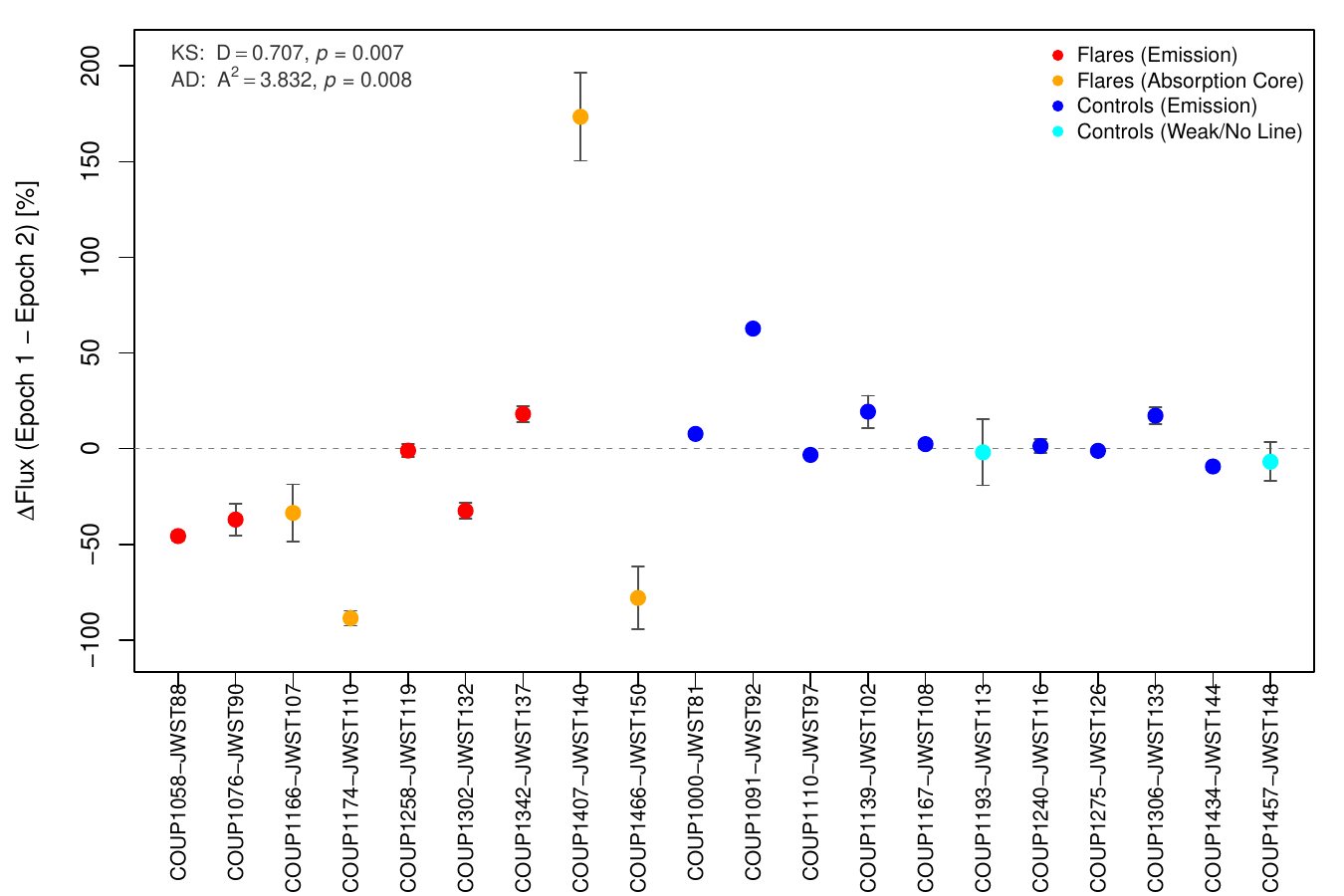}
\caption{Difference in the NIRSpec flux integrated over the H~I Br$\alpha$ waveband between the two observing epochs, defined as $\Delta \mathrm{Flux} = 100\% \times \frac{F_{\rm epoch1} - F_{\rm epoch2}}{(|F_{\rm epoch1}| + |F_{\rm epoch2}|)/2}$. For the Flare sample members, Epoch~1 corresponds to the X-ray flare epoch; for the Control sample members, Epoch~1 is chosen randomly from the two epochs. The x-axis lists the COUP--JWST source IDs. COUP1444--JWST146 and COUP1423--JWST142 are omitted because H~I Br$\alpha$ was not covered in at least one epoch. Flare sample members are color-coded red (H~I Br$\alpha$ in emission) and orange (H~I Br$\alpha$ core in absorption); Control sample members are shown in blue (H~I Br$\alpha$ in emission) and cyan (H~I Br$\alpha$ weak or undetected). Results of the KS and AD tests comparing the $\Delta \mathrm{Flux}$ distributions of the Flare and Control samples are given in the figure legend.} \label{fig_Bralpha_analysis}
\end{figure*}

Overall, the analyses presented in this section reveal two observational trends in the variability of narrow NIRSpec spectral features within the 2.8--4.3~\micron\ wavelength range. First, source-level wavelength distributions provide tentative evidence that stars with detected X-ray flares have relatively fewer highly variable features at longer wavelengths than Control stars, although this q75 difference requires confirmation with larger samples. Second, the incidence of highly variable narrow spectral features decreases significantly with increasing projected distance from $\theta^1$~Ori~C. Comparison with the empirical PDRs4All spectrum yields a number and wavelength proximity of candidate matches consistent with chance, while the HITRAN/VALD comparison favors molecular over stellar photospheric wavelength associations, although individual line assignments from both comparisons remain tentative. Thus, the possible deficit of long-wavelength variable features among Flare stars and the increased incidence of variable features toward $\theta^1$~Ori~C point, respectively, to potential flare-related and external-environmental influences whose physical origins are considered in the Discussion (\S~\ref{sec_origins_of_narrow_features}).

\subsection{Variability of the Hydrogen Recombination Br$\alpha$ Line}\label{sec_Bralpha_analyses}

Hydrogen recombination Br$\alpha$ and the CO fundamental ro-vibrational band are among the brightest spectral features in the COUP-JWST NIRSpec sample (Figure~\ref{fig_NIRSPec_spectra}). Our ProDiMo simulations (Figure~\ref{fig_prodimo_outcome}) predict that, assuming both tracers originate predominantly in the disk surface layers, X-ray mega-flares ($E_{\rm X,fl} \gtrsim 10^{36}$~erg) should enhance the emission of both Br$\alpha$ and CO. For the more common mild X-ray superflares ($E_{\rm X,fl} \lesssim 10^{35}$~erg), however, only Br$\alpha$ is predicted to exhibit a measurable increase in emission. Although no X-ray mega-flares were captured during our coordinated Chandra--JWST campaign, the observed mild superflares provide the first opportunity to test these predictions empirically.

Figure~\ref{fig_Bralpha_spectra} and the accompanying online Figure Set compare the Br$\alpha$ spectral region for the 10 Flare and 12 Control COUP-JWST systems. Among the 10 Flare stars, six exhibit Br$\alpha$ entirely in emission, whereas four display absorption cores superimposed on broader emission wings. Of the 12 Control stars, one lacks NIRSpec coverage in the Br$\alpha$ wavelength range, two show weak or undetected Br$\alpha$, and the remaining nine exhibit the line entirely in emission. The inferred Br$\alpha$ fluxes are listed in Table~\ref{tab_integrated_fluxes}.

The significance of the four Br$\alpha$ absorption systems becomes apparent when viewed in the context of the full COUP-JWST sample. Among the 73 stars with available NIRSpec coverage of the Br$\alpha$ region (Paper~II), only 11 show weak or undetected Br$\alpha$, four exhibit absorption cores, and the remaining 58 display the line in emission. Remarkably, all four absorption-core systems belong to the Flare sample. Paper~II further shows that Br$\alpha$ luminosity is strongly correlated with stellar accretion luminosity, indicating that the bulk of the Br$\alpha$ emission originates in the magnetospheric accretion region rather than the disk surface. The independent-nod reductions, employing various background subtraction schemes (\S~\ref{sec_NIRSpec_data}), confirm that these four Flare sources are the only systems among the 73 stars with Br$\alpha$ coverage for which the absorption-core morphology persists systematically across the different science-nod and background-subtraction combinations. Several additional stars (COUP 1019, 1107, 1110, 1193, 1260, 1265, 1276, 1324, and 1335) show clear Br$\alpha$ emission in most reductions, although a small number of nod/background combinations produce noisy profiles, P-Cygni-like morphology, or apparent central absorption. We therefore do not classify these several cases as robust absorption-core systems.

To quantify the spectral variability, we measured the fractional change in the integrated Br$\alpha$-band flux between the two observing epochs,
\begin{equation}
\Delta {\rm Flux} =
100\% \times
\frac{F_{\rm epoch1}-F_{\rm epoch2}}
{(|F_{\rm epoch1}|+|F_{\rm epoch2}|)/2}.
\end{equation}

For each Flare-sample source, Epoch~1 is defined as the NIRSpec epoch associated with the stronger/relevant X-ray flare, independent of chronological order, and Epoch~2 as the comparison epoch. Thus, the Epoch~1/Epoch~2 designation refers to flare status rather than chronological order. For the Control sample, which by definition has no detected flare in either epoch, the Epoch~1/Epoch~2 assignment is made randomly.

Figure~\ref{fig_Bralpha_analysis} presents $\Delta {\rm Flux}$ for the Flare and Control samples. Red and orange symbols denote Flare stars with Br$\alpha$ emission and absorption-core profiles, respectively, while blue and cyan symbols represent Control stars with Br$\alpha$ emission and weak/non-detections. The distributions of $\Delta {\rm Flux}$ differ significantly between the Flare and Control samples, with both the Kolmogorov--Smirnov and Anderson--Darling tests yielding $p$-values of 0.007--0.008.

The observed behavior differs qualitatively from the ProDiMo predictions. Rather than exhibiting enhanced Br$\alpha$ emission during the flare epoch, the Flare sample generally shows a reduction in integrated Br$\alpha$ flux. Moreover, the presence of Br$\alpha$ absorption cores in four Flare systems is not reproduced by the current disk-based simulations. These observational results suggest that the dominant source of Br$\alpha$ variability is not fully captured by the present models.

\begin{figure*}[!t]
\centering
\includegraphics[width=0.85\textwidth]{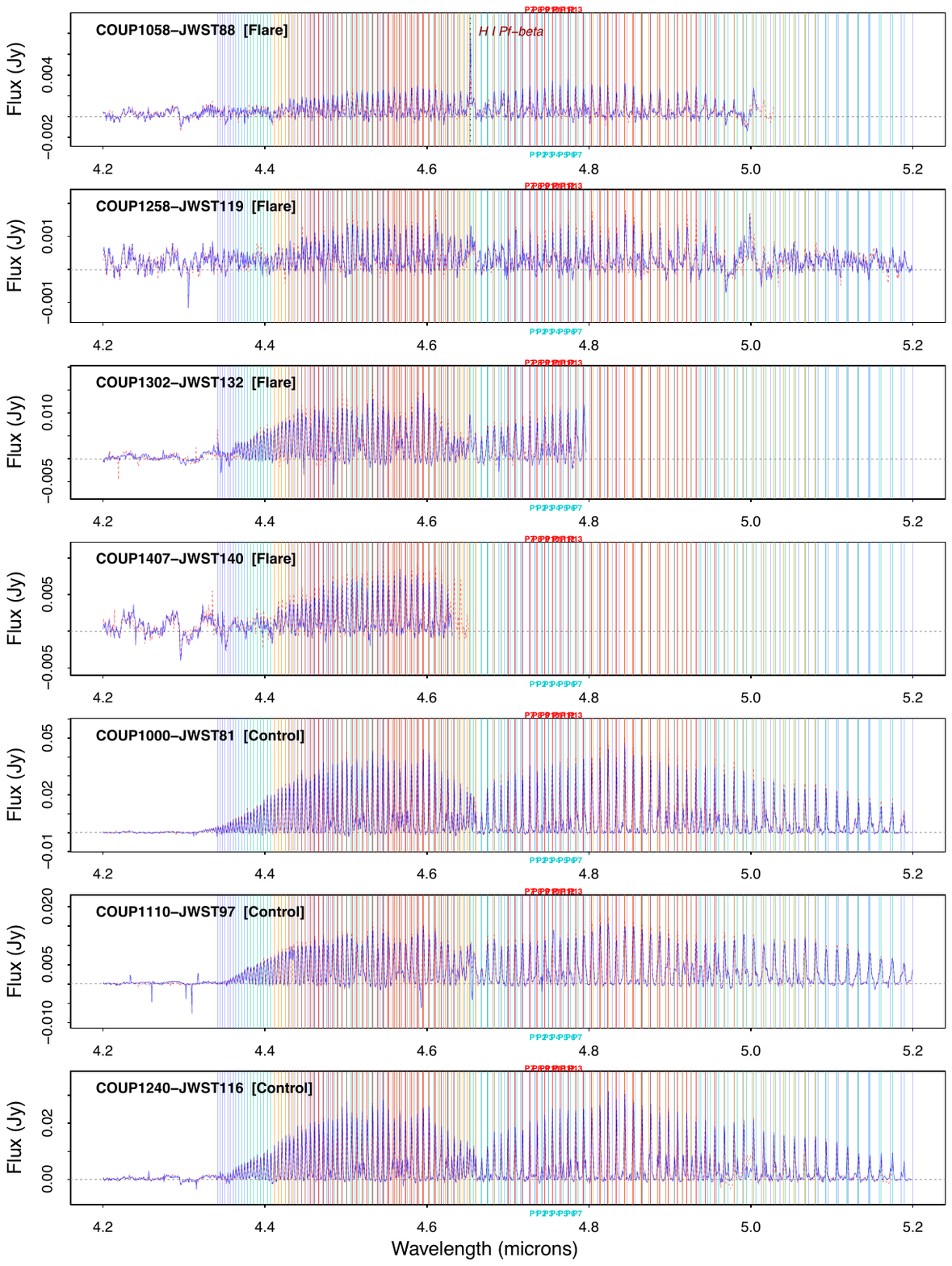}
\caption{Examples of NIRSpec spectra of the CO fundamental line forest over the 4.3--5.2~\micron\ band for the seven COUP--JWST sources (4 Flare and 3 Control sample members) exhibiting strong CO emission. The spectra are scaled to the Spitzer/IRAC flux level and corrected for extinction. Spectra obtained during X-ray flare and non-flare epochs are shown as dashed red and solid blue lines, respectively. Vertical lines mark theoretical CO transitions from the HITRAN database, color-coded by line strength at $T_{\rm gas}=1200$~K across five quantiles ranging from strongest to weakest as red, orange, teal, light blue, and blue. For illustration, the red and teal tick marks label selected $^{12}$CO $v=1\rightarrow0$ fundamental ($P7$--$P13$) and $v=2\rightarrow1$ hot-band ($P1$--$P7$) transitions, respectively. Following \citet{Banzatti2015}, the fundamental line cores predominantly trace slow molecular disk winds, whereas the vibrationally excited hot-band transitions probe hot gas in the inner Keplerian disk. An online Figure Set presents the NIRSpec spectra of the remaining Flare and Control sample members, which exhibit weak CO emission, CO absorption, or incomplete CO spectral coverage.} \label{fig_CO_spectra}
\end{figure*}

\begin{figure*}[!t]
\centering
\includegraphics[width=0.85\textwidth]{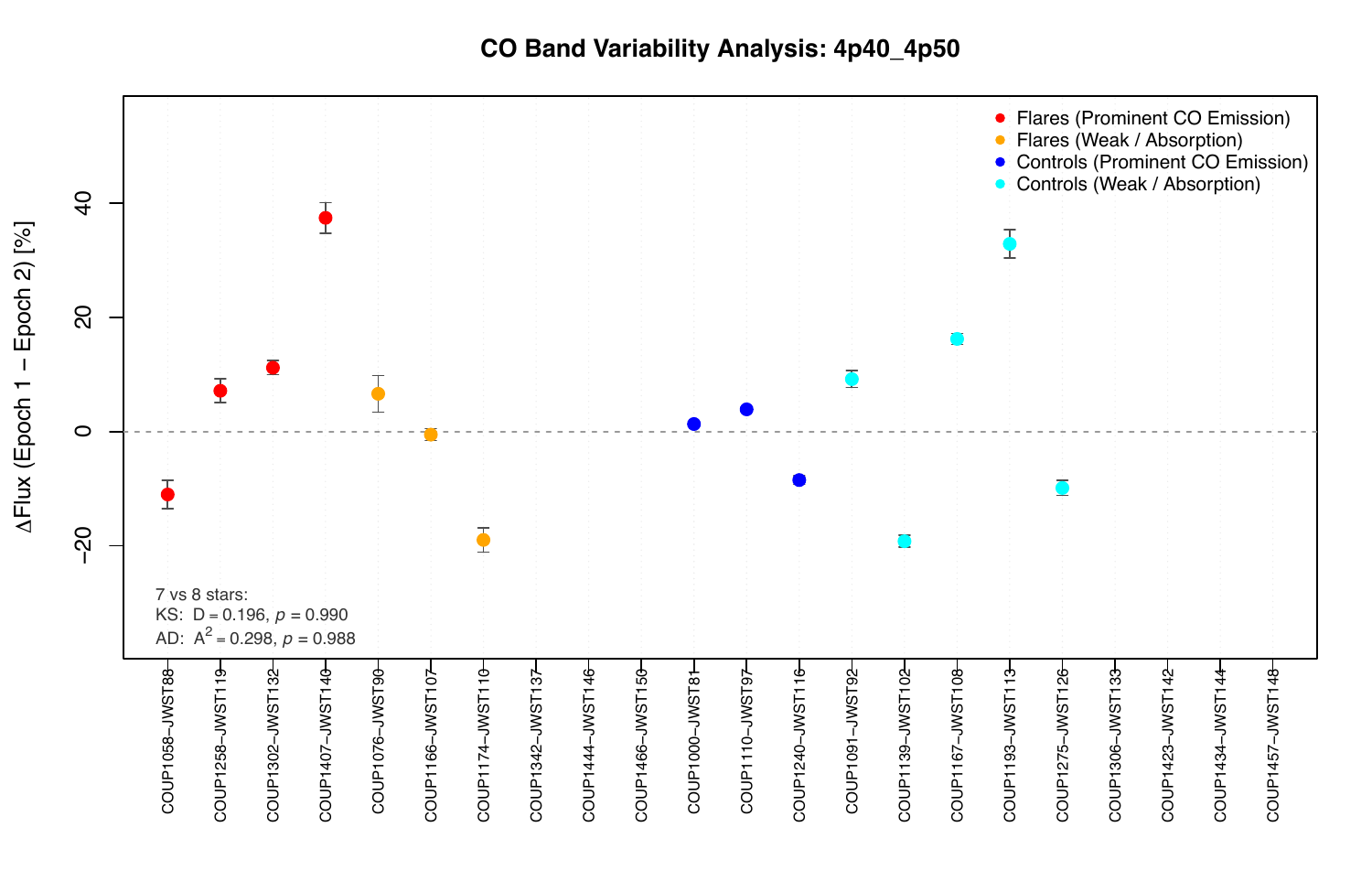}
\caption{Analog of Figure~\ref{fig_Bralpha_analysis}, but for NIRSpec spectral flux variability in the 4.4--4.5~\micron\ CO sub-band. Flare sample members are color-coded red (strong CO emission) and orange (weak CO emission or CO absorption); Control sample members are shown in blue (strong CO emission) and cyan (weak CO emission or CO absorption). The absence of plotted points for some sources reflects incomplete CO spectral coverage. Results of the KS and AD tests comparing the $\Delta\mathrm{Flux}$ distributions of the Flare and Control samples are given in the figure legend. An online Figure Set presents analogous variability plots for several additional CO sub-bands.} \label{fig_CO_analysis}
\end{figure*}

\subsection{Variability of the CO Fundamental Band Forest}\label{sec_CO_analyses}

Figure~\ref{fig_CO_spectra} and the accompanying online Figure Set compare the CO spectral region for the 10 Flare and 12 Control COUP-JWST systems. Among the 10 Flare stars, four exhibit strong CO emission, five show weak CO emission or absorption, and one lacks spectral coverage of the CO region. Among the 12 Control stars, three exhibit strong CO emission, four show strong CO absorption, two display weak CO emission or absorption, and three lack spectral coverage of the CO region.

For spectra with strong CO emission, we estimated the continuum using the \texttt{ctool} method \citep{Pontoppidan2026}, which is designed to enhance the visibility of CO emission lines. For the remaining spectra, we adopted a rolling-quantile continuum with $q=0.9$, which suppresses emission peaks and thereby emphasizes absorption features. The two approaches were selected because they provide more appropriate continuum estimates for emission- and absorption-dominated CO spectra, respectively. The inferred CO fluxes are listed in Table~\ref{tab_integrated_fluxes}.

The CO fundamental rovibrational band at 4.7~\micron\ is a well-established tracer of warm molecular gas in young stellar systems and is commonly observed in emission from the inner regions of protoplanetary disks (e.g., \citealt{Bast2011,Brown2013}). CO absorption is also observed toward young stars and can arise from stellar photospheres as well as warm circumstellar disk material or cooler molecular gas along the line of sight, with the observed morphology depending strongly on source geometry and disk inclination (e.g., \citealt{Boogert2002,Horne2012}). Recent JWST/NIRSpec observations of young protostellar systems likewise reveal the CO fundamental series in both emission and absorption \citep{Federman2024}. 

Approximately half of the Flare and Control stars in our sample exhibit weak or strong CO absorption rather than emission. Comparison with synthetic stellar photospheric spectra of similar spectral type, reddened by the corresponding extinction and degraded to the NIRSpec spectral resolution, reveals two distinct absorption morphologies. Weak absorption spectra (e.g., COUP1076--JWST90 and COUP1166--JWST107) broadly resemble stellar photospheric CO absorption, whereas strong absorption spectra (e.g., COUP1139--JWST102 and COUP1167--JWST108) display well-separated R- and P-branch structures that resemble CO emission-band spectra inverted in intensity. The latter morphology may therefore trace circumstellar molecular gas seen in absorption against the continuum, potentially associated with inclined disks or disk/proplyd environments, although absorption by foreground molecular material cannot be excluded.

As in the analysis of the H~I Br$\alpha$ line, we quantified the CO spectral variability by measuring $\Delta {\rm Flux}$ between the flare-associated and comparison epochs, using the same Epoch~1/Epoch~2 convention defined above. The CO fundamental band was divided into seven contiguous wavelength intervals extending from the R-branch toward the P-branch, and $\Delta {\rm Flux}$ was measured independently for each interval. Figure~\ref{fig_CO_analysis} presents the results for the 4.4--4.5~\micron\ interval, while the accompanying online figure-set provides analogous comparisons for the remaining six intervals. Red and orange symbols denote Flare stars with strong CO emission and weak-emission-or-absorption spectra, respectively, whereas blue and cyan symbols represent the corresponding Control stars.

None of the seven wavelength intervals exhibits a statistically significant difference between the Flare and Control samples. For every interval, both the Kolmogorov--Smirnov and Anderson--Darling tests yield $p$-values exceeding 0.5, indicating statistically indistinguishable distributions of $\Delta {\rm Flux}$.

Among the seven systems exhibiting strong CO emission, six show remarkably stable CO-band fluxes over the three-day interval, with $|\Delta {\rm Flux}| \lesssim 10$--15\% across all seven wavelength intervals. The only exception is COUP1407--JWST140, a member of the Flare sample, for which $|\Delta {\rm Flux}|$ ranges from approximately 25\% to 40\%, depending on the wavelength interval considered.

Unlike the H~I Br$\alpha$ line, the CO fundamental band exhibits little evidence for enhanced variability associated with mild X-ray superflares.

\section{Discussion}\label{sec_discussion}

\subsection{Plausible Origins of the Narrow NIRSpec Spectral Features
}\label{sec_origins_of_narrow_features}

The source-level analysis provides tentative evidence that stars experiencing X-ray flares have relatively fewer highly variable narrow spectral features at longer wavelengths (approximately 3.5--4.3~\micron) than stars in the Control sample (\S~\ref{sec_narrow_features_analyses}). This difference is confined primarily to the upper portion of the wavelength distributions, as traced by q75, and remains significant after FDR correction according to the KS test ($q_{\rm FDR}=0.0077$) and marginally significant according to the AD test ($q_{\rm FDR}=0.022$). The small source samples (10 Flare and 12 Control stars) and the differing statistical significance of the two tests warrant caution. If this possible difference is confirmed as intrinsic by future coordinated Chandra--JWST observations with larger source samples, it would raise the possibility that moderate X-ray superflares, and perhaps accompanying CMEs, suppress rather than enhance the occurrence or detectability of some narrow spectral variations. One speculative possibility is that flare-associated X-rays and energetic particles accelerated by accompanying CMEs alter the ionization, chemistry, or excitation of molecular material in disk surface layers or associated flows, thereby reducing the detectability of some features during the 1--4 days between flare onset and the NIRSpec observations. CME-driven energetic particles have been proposed to affect circumstellar disk chemistry \citep{Hu2022}. However, there is presently no theoretical basis for associating the tentative deficit specifically at 3.5--4.3~\micron\ with material at particular disk radii. In particular, the ProDiMo calculations presented in this work model the response of the disk to optical, UV, and X-ray irradiation, but do not include CME-associated energetic-particle irradiation or externally driven photoevaporative flows that may contribute to the narrow features discussed below.

The majority of the morphologically classifiable features appear in absorption against the local NIRSpec pseudo-continuum. Among the 249 features that can be classified in at least one epoch, 186 (75\%) show absorption with no evidence for emission in the other epoch, while an additional 20 (8\%) switch between absorption and emission. This prevalence of absorption suggests that at least some of the features may arise in relatively cool material along the line of sight rather than in hot emitting gas. Their wavelength distributions are more closely matched by molecular transitions in the HITRAN2020 database than by atomic or molecular transitions expected from stellar photospheres listed in VALD. However, because the wavelength matching is based solely on transition proximity and relative line strength, without radiative-transfer modeling or spectral synthesis, the resulting molecular associations should be regarded as plausible candidates rather than definitive identifications.

The direct comparison with the empirical PDRs4All Orion Bar spectrum provides a complementary constraint. Although 67 of the 281 features (24\%) coincide within one nominal NIRSpec resolution element with a PDRs4All transition, both the number of matches and their wavelength separations are statistically consistent with chance superposition on the dense PDR spectrum. The candidate matches are dominated by H$_2$ and H~I transitions, with smaller contributions from PAH-related features, He~I, CH$^+$, O~I, and [Fe~III], but these associations cannot presently be interpreted as secure identifications. Thus, the PDRs4All comparison neither establishes nor excludes an origin of individual narrow features in externally irradiated material; rather, the strongest evidence for an environmental connection comes from the spatial dependence of the feature incidence discussed below.

The Orion protoplanetary disks studied here reside in an unusually harsh radiation environment. Unlike isolated T~Tauri disks, they are continuously exposed to the intense far-ultraviolet (FUV) radiation field of $\theta^1$~Ori~C, which heats disk surface layers, establishes photodissociation regions, and drives external photoevaporative flows \citep{Winter2022}. JWST and ALMA observations of the nearly edge-on Orion proplyd d203--506, located within the same JWST/NIRSpec-MOS field sampled by our targets, directly reveal an FUV-driven molecular photoevaporative flow and demonstrate how external irradiation reshapes the thermal and chemical structure of protoplanetary disks \citep{Berne2024}. JWST observations of the same system detected CH$_3^+$, providing direct evidence that UV irradiation can activate hydrocarbon chemistry in disk atmospheres \citep{Berne2023}, while subsequent studies revealed neutral carbon, CH$^+$, CH, and likely vibrationally excited H$_3^+$, further demonstrating the rich FUV-driven chemistry of the irradiated disk and wind \citep{Goicoechea2024,Zannese2025,Schroetter2025,Goicoechea2025}. Recent thermo-chemical models likewise predict that external irradiation substantially modifies disk thermal structure, gas-phase abundances, and the observability of volatile molecules \citep{PortillaRevelo2025,Gross2025,Keyte2025,Frediani2026}.

Our observations reveal a strong association between the incidence of apparent spectral variability and the external radiation environment. The number of highly variable narrow spectral features decreases significantly with projected distance from $\theta^1$~Ori~C ( Figure~\ref{fig_narrow_features_distance_to_Theta1OriC}), consistent with stronger UV-driven heating, photochemistry, and photoevaporative activity in the most strongly irradiated disks. Interestingly, much of this trend occurs within $2$--$2.5\arcmin$ (0.2--0.3~pc) of $\theta^1$~Ori~C, comparable to the projected scale of the Orion Bright Bar, across which the geometry and physical conditions of the nebular ionization front change substantially \citep{ODell2017}. Thus, the observed distance dependence may reflect a combination of the declining external radiation field and large-scale structural variations within the Orion Nebula. Although projection effects and intrinsic disk-to-disk differences undoubtedly contribute to the observed scatter, this correlation is consistent with an origin of at least some of the highly variable absorption features in externally irradiated disk atmospheres and/or associated photoevaporative flows.

An additional contribution to the observed distance dependence may arise from the known spatial--age structure of the ONC. Previous studies have found evidence for an age gradient in which younger stellar populations are preferentially concentrated toward the central Orion Nebula \citep{Getman2014b,Getman2018}. This possibility is also consistent with independent trends found in the larger COUP--JWST samples analyzed in Paper~II, irrespective of X-ray flare status or the availability of two-epoch NIRSpec spectra. In those samples, both Br$\alpha$ and CO luminosities correlate positively with visual extinction, while analysis of the much larger parent COUP population shows that $A_V$, X-ray median energy, IRAC spectral slope, and accretion luminosity are mutually correlated and can serve, in part, as tracers of circumstellar evolutionary state. Thus, the larger incidence of highly variable narrow features toward $\theta^1$~Ori~C could partly reflect a greater prevalence of younger, more embedded, and actively accreting star--disk systems toward the ONC center, rather than being driven solely by the stronger external FUV radiation field. The present small sample does not allow these spatially covariant environmental and evolutionary effects to be disentangled.

\begin{figure*}[!t]
\centering
\includegraphics[width=0.95\textwidth]{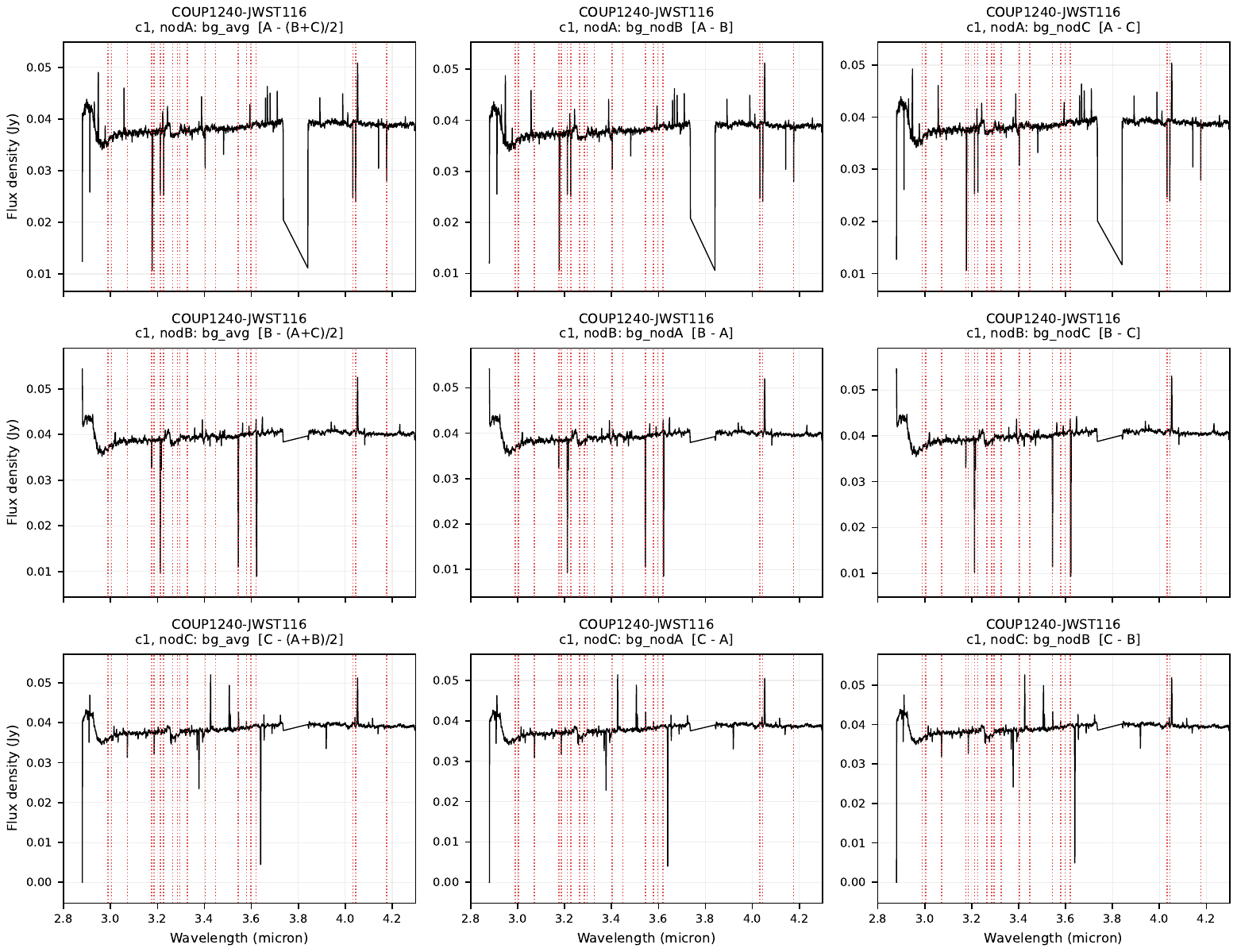}
\caption{Example of the sensitivity of narrow spectral features to science-nod position and background selection for COUP~1240 in Observation~3 (MSA configuration c1). Rows show spectra extracted with nods A, B, and C, respectively, as the science exposure. Within each row, the background is constructed from the average of the other two nods (left), the first individual alternative nod (middle), or the second individual alternative nod (right). Red dashed vertical lines mark the wavelengths of the highly variable narrow features identified from the two-epoch spectral comparison (\S~\ref{sec_narrow_features_analyses}). For a fixed science nod, the narrow-feature morphology is generally similar among the three background choices, whereas more substantial differences are apparent among the different science nod positions. This behavior suggests that spatial sampling of the source and its immediate environment contributes more strongly to the narrow-feature morphology than the specific choice among the alternative background nods.} \label{fig_nod_bg_analysis}
\end{figure*}

Although the individual molecular identifications remain highly uncertain, the tentative HITRAN associations are strongly concentrated among a few species: CH$_4$ (39.9\%), HCN (19.2\%), C$_2$H$_2$ (18.9\%), and NH$_3$ (10.0\%) together account for 88\% of the preferred associations. The relative distributions of these candidate species are broadly similar between the Flare and Control samples. Among these, NH$_3$ and CH$_4$ are particularly intriguing because both molecules have historically proven difficult to detect in the gas phase of protoplanetary disks despite being predicted to constitute important volatile reservoirs \citep{Oberg2011,Oberg2023}. Observations, laboratory studies, and thermo-chemical models indicate that these species reside predominantly in icy grain mantles in dense molecular clouds, protostellar envelopes, and disk midplanes \citep{Boogert2015,Ballering2021,McClure2023,Sturm2023}. Nevertheless, warm gaseous CH$_4$ has been detected in absorption toward the highly inclined inner disk of GV~Tau~N \citep{Gibb2013}, while warm NH$_3$ was subsequently detected in the same system \citep{Najita2021}. More recently, JWST spectroscopy has revealed rich hydrocarbon chemistry, including CH$_4$, C$_2$H$_2$, and larger hydrocarbons, in the warm inner atmospheres of disks around very low-mass stars \citep{Tabone2023,Arabhavi2024}, while recent thermo-chemical calculations predict that NH$_3$ should be present in many disks despite producing intrinsically weak infrared signatures \citep{Vlasblom2026}. Thus, if some of our tentative wavelength associations ultimately prove correct, the apparent scarcity of previous NH$_3$ and CH$_4$ detections may reflect observational limitations rather than the absence of these molecules from irradiated disk atmospheres.

Maintaining gas-phase abundances of such volatile molecules does not necessarily require wholesale thermal sublimation of icy grains. Laboratory experiments demonstrate efficient ultraviolet photodesorption of NH$_3$- and CH$_4$-bearing ice analogs \citep{MartinDomenech2018}, while X-ray irradiation also efficiently desorbs volatile species from astrophysical ices \citep{Dupuy2018}; NH$_3$-rich ices themselves undergo X-ray-induced photodesorption \citep{TorresDiaz2024}.

We emphasize that this astrophysical interpretation remains speculative because the molecular carriers of the highly variable absorption features have not been securely identified. In particular, although the feature wavelengths show closer agreement with HITRAN molecular transitions than with VALD photospheric transitions, the direct PDRs4All comparison provides no population-level evidence for preferential coincidence with known Orion Bar transitions. Nevertheless, the combination of (i) predominantly absorption-line morphology, (ii) closer wavelength agreement with HITRAN molecular transitions than with VALD photospheric transitions, and (iii) the pronounced dependence of feature incidence on projected distance from $\theta^1$~Ori~C is consistent with an origin of at least some features in externally irradiated disk atmospheres and/or photoevaporative flows. However, these observations do not establish that the differences between the two NIRSpec epochs represent intrinsic temporal variability. If the apparent spectral differences are genuinely temporal and some of the tentative molecular identifications are correct, episodic stellar X-ray flares could temporarily alter gas-phase molecular abundances in disk surface layers whose underlying thermal and chemical structure is set in part by continuous external FUV irradiation.

An alternative possibility is that some of the apparent two-epoch variability is spatial rather than temporal in origin. The two NIRSpec epochs (Visit 3 and Visit 7) were obtained at slightly different aperture position angles (APA = 203.2 and 205.5 degrees). Although this difference is small, the Orion targets reside in a spatially complex environment containing proplyd ionization fronts, photoevaporative flows, and highly structured nebular emission. Importantly, while the external FUV field from $\theta^1$~Ori~C may help establish such irradiated structures and their molecular content, it should remain essentially constant at a given system over the few-day interval between our NIRSpec epochs and therefore cannot by itself account for rapid intrinsic spectral variability. Our alternative nod-by-nod reductions (\S~\ref{sec_NIRSpec_data}) provide qualitative support for spatially dependent spectral structure. As illustrated by the representative case of COUP~1240 in Figure~\ref{fig_nod_bg_analysis}, the narrow-feature morphology can vary substantially among science nod positions and/or MSA configurations, whereas, for a fixed science nod, it is generally much less sensitive to whether the background is constructed from the average of the other two nods or from either individual nod. Thus, some of the apparent two-epoch ``variability'' between Visits~3 and 7 illustrated in Figure~\ref{fig_narrow_features_Ep1_vs_Ep2} could instead trace spatial--chemical differences across the immediate environments of the star--disk systems sampled by the different science-nod positions and MSA configurations. This behavior suggests that at least some of the spectral differences arise on relatively local spatial scales sampled by the source apertures, rather than being produced solely by variations among the more spatially separated background positions. Such locally structured material could include irradiated disk atmospheres, proplyd ionization fronts, or photoevaporative flows, although small-scale nebular structure and position-dependent instrumental or extraction effects cannot be excluded.

Thus, some features identified here as ``variable'' could reflect spatial-sampling differences rather than intrinsic temporal changes. The relative stability of the narrow-feature morphology to alternative background choices for a fixed science nod argues against simple background-selection effects as the dominant explanation, but does not exclude residual contamination from spatially structured nebular or molecular-cloud emission. These possibilities do not exclude an origin of the features in irradiated disk atmospheres or photoevaporative flows; indeed, the apparent sensitivity to local source-aperture sampling is qualitatively consistent with such an origin. Nevertheless, the present data cannot uniquely distinguish intrinsic temporal variability from spatially dependent circumstellar/nebular structure or position-dependent instrumental effects. Future work could test these alternatives by examining the two-dimensional NIRSpec data and source extendedness, proplyd morphology, local nebular surface brightness, and aperture orientation. Such analyses are beyond the scope of the present study.

\subsection{Magnetospheric Origin of the Br$\alpha$ Variability}\label{sec_origin_of_Bralpha}

The Br$\alpha$ results (\S~\ref{sec_Bralpha_analyses}) differ fundamentally from those obtained for the narrow molecular spectral features. Whereas the latter plausibly trace warm molecular gas in the disk atmosphere, the observed Br$\alpha$ variability appears inconsistent with a predominantly disk-surface origin. Our ProDiMo simulations predict that mild X-ray superflares should produce enhanced Br$\alpha$ emission if the line forms primarily within the irradiated surface layers of the disk. Instead, the Flare sample exhibits systematically weaker integrated Br$\alpha$ fluxes during the flare epoch, and all four Br$\alpha$ absorption-core systems identified in the entire sample of 73 COUP-JWST stars belong exclusively to the Flare sample. Together with the strong correlation between Br$\alpha$ luminosity and stellar accretion luminosity reported in Paper~II, these observations suggest that the Br$\alpha$ variability is governed primarily by processes operating close to the central star rather than by direct heating of the disk surface.

An important caveat is that the empirical NIRSpec flux correction assumes negligible astrophysical continuum variability between epochs. If the near-IR continuum brightened during a flare-associated epoch while the intrinsic Br$\alpha$ line flux remained unchanged, the continuum-based scaling could produce an apparent reduction in the recovered Br$\alpha$ flux; continuum fading could have the opposite effect. Thus, the inferred systematic Br$\alpha$ weakening should be interpreted as a change relative to an assumed stable continuum until contemporaneous photometry becomes available. This limitation does not readily explain the observed changes in Br$\alpha$ line morphology, particularly the absorption-core profiles independently recovered from individual nod spectra.

A plausible explanation is that the large-scale magnetic structures responsible for the observed X-ray superflares also perturb the stellar magnetosphere. Magnetospheric accretion models predict that the geometry and efficiency of funnel flows are highly sensitive to the configuration of the stellar magnetic field (e.g., \citealt{Bouvier2007,Romanova2008,Romanova2015}). Observational monitoring of AA~Tau has further shown that magnetospheric inflation and restructuring can accompany substantial changes in accretion-funnel geometry and episodes of reduced accretion \citep{Bouvier2007b}. Magnetic reconnection associated with powerful flares could therefore temporarily modify the topology of the accretion funnels, reducing the volume or density of ionized gas contributing to magnetospheric line emission. Because Br$\alpha$ emission is thought to originate predominantly within these accretion structures, such magnetic restructuring could weaken the observed Br$\alpha$ emission, consistent with the systematic decrease in integrated Br$\alpha$ flux observed during the flare-associated epochs. To our knowledge, however, a systematic decrease in accretion-line emission following PMS X-ray superflares has not previously been established observationally, making the Br$\alpha$ behavior found here a potentially new manifestation of flare--magnetosphere coupling.

The occurrence of Br$\alpha$ absorption cores exclusively among Flare stars provides an additional clue. One possibility is that flare-active systems contain larger columns of relatively cool neutral hydrogen located along the line of sight to the Br$\alpha$-emitting region. Such absorbing material could reside within magnetospheric condensations, prominences, post-eruptive plasma, or coronal mass ejections associated with enhanced magnetic activity. Solar and stellar observations demonstrate that major magnetic eruptions can produce cool absorbing plasma embedded within otherwise hot coronal environments (e.g., \citealt{Mackay2010,Argiroffi2019,Moschou2019,Fuhrmeister2020,Kowalski2024,Namekata2025}). In particular, the flare observed in the young active M~dwarf 2MASS~J07471385+5020386 exhibited a substantially deepened He~I 1.083~\micron\ absorption core together with blueshifted He~I and H$\alpha$ emission wings tracing outward-moving material \citep{Fuhrmeister2020}. Although the atomic formation mechanism of He~I differs from that of Br$\alpha$, these observations illustrate that powerful stellar magnetic events can simultaneously modify chromospheric line profiles and increase the column density of cool absorbing plasma along the line of sight. A similar geometric effect could contribute to the Br$\alpha$ absorption cores observed exclusively among our Flare stars. Partial occultation of the Br$\alpha$-emitting accretion columns by such material could reduce the apparent line flux or even produce absorption cores, qualitatively explaining why these unusual profiles are found only among stars exhibiting strong X-ray magnetic activity. Importantly, the absorbing gas need not be associated exclusively with the contemporaneous flare itself; repeated magnetic activity over many stellar rotation periods could maintain a population of cool circumstellar condensations capable of intermittently occulting the Br$\alpha$-emitting region.

The absence of enhanced Br$\alpha$ emission during the flare epoch may also reflect both the observing cadence and the radiative-transfer properties of the line. In most cases, the JWST observations sampled the late decay phases of the X-ray flares (roughly 1--4 days after the inferred impulsive phases, with the exact timing uncertain because of the discontinuous Chandra coverage; Figure~\ref{fig_Chandra_Lcs}), whereas chromospheric heating and hydrogen recombination emission likely reached their maximum during the earlier impulsive phase. Any flare-generated Br$\alpha$ enhancement may therefore have largely dissipated before the NIRSpec observations commenced. Furthermore, as discussed in Paper~II, Br$\alpha$ is expected to be substantially more optically thick than Br$\gamma$ and the higher-order mid-infrared hydrogen recombination lines, allowing it to probe a comparatively extended volume of ionized gas within the magnetospheric accretion flow. Consequently, relatively modest changes in the geometry or optical depth of the accretion columns may dominate the observed Br$\alpha$ variability, potentially overwhelming any short-lived chromospheric enhancement associated with the flare itself.

Taken together, these results suggest that coordinated X-ray and infrared observations can distinguish between flare responses arising in different physical components of young stellar systems. Unlike the narrow absorption features discussed in the previous section, whose origin may involve externally irradiated disk/proplyd material as well as spatial-sampling or background-subtraction effects, the Br$\alpha$ variability appears to be controlled primarily by magnetospheric accretion and other stellar magnetic processes operating close to the host star. Future coordinated Chandra--JWST monitoring with higher-temporal-cadence NIRSpec spectroscopy, particularly during the impulsive and early decay phases of powerful X-ray flares, together with detailed magnetohydrodynamic and radiative-transfer modeling of hydrogen recombination lines, will be required to determine whether a short-lived Br$\alpha$ enhancement precedes the subsequent decline observed here and whether magnetic restructuring of the accretion flow contributes to this unexpected behavior.

\subsection{CO Fundamental Band Response to X-ray Flares}\label{sec_CO_response}

In contrast to the numerous narrow spectral features and unlike the H~I Br$\alpha$ line, the CO fundamental band exhibits little evidence for enhanced variability associated with X-ray flares (\S~\ref{sec_CO_analyses}). Across all seven wavelength intervals spanning the 4.3--5.2~\micron\ CO band, the Flare and Control samples display statistically indistinguishable distributions of $\Delta {\rm Flux}$. Thus, within the sensitivity and temporal baseline of the present observations, we find no compelling observational evidence that mild PMS X-ray superflares ($\log(E_{\rm X,fl}/{\rm erg}) < 35$) produce measurable changes in the integrated CO fundamental emission.

This observational result is broadly consistent with the predictions of our ProDiMo thermo-chemical simulations. For the representative $0.7~M_{\odot}$ disk model, X-ray superflares with energies of $\log(E_{\rm X,fl}/{\rm erg})\sim34.5$ are predicted to produce little or no enhancement of the CO fundamental emission, whereas substantially more energetic mega-flares with $\log(E_{\rm X,fl}/{\rm erg})\sim36.5$ can increase the integrated CO emission by approximately 20\% (\S~\ref{sec_modeling}). The only possible observational counterpart is COUP1407--JWST140, which exhibits the largest CO variability among the seven strong CO emitters and also experienced the most energetic inferred flare in our disk CO-emitting sample ($\log(E_{\rm X,fl}/{\rm erg})=35.06$). Although this single object does not establish a statistical trend, it is qualitatively consistent with the expectation that stronger X-ray flares should produce larger perturbations of CO emission. 

The absence of a statistically significant flare signature should therefore not be interpreted as evidence that X-ray flares never influence CO emission in protoplanetary disks. The present experiment contains only seven systems with strong CO emission, of which only four belong to the Flare sample. Moreover, only one of these systems experienced a flare approaching the upper end of the sampled flare-energy distribution. The statistical power of the experiment was further reduced because only three of the eight planned JWST/NIRSpec-MOS visits were successfully executed, leaving only 44 disk-bearing stars with two-epoch spectroscopy instead of the approximately 150 systems originally envisioned. Consequently, the probability of capturing the rare but astrophysically important X-ray mega-flares expected to produce the strongest thermo-chemical response was substantially lower than anticipated.

Future coordinated Chandra--JWST monitoring campaigns can provide a much more stringent test of flare-driven CO variability. The experience gained from the present program --- including improved guide-star selection and optimization of the MSATA target-acquisition strategy for the bright Orion Nebula background --- should enable a substantially higher success rate for future NIRSpec-MOS observations. A larger sample of two-epoch spectra, combined with the detection of several X-ray mega-flares ($\log(E_{\rm X,fl}/{\rm erg})\gtrsim36$), would allow a direct observational test of the predicted flare-induced enhancement of the CO fundamental band and place significantly stronger constraints on the thermo-chemical response of planet-forming disks to extreme stellar magnetic activity.

The diverse morphologies of the CO absorption spectra also warrant further investigation. While weak absorption profiles are generally consistent with stellar photospheric absorption, the strongest absorption spectra exhibit well-developed R- and P-branch structures that are difficult to reconcile with a purely photospheric origin. Possible explanations include absorption by highly inclined disks, self-absorption within disk atmospheres, or intervening molecular material associated with disk winds or the surrounding molecular cloud. Detailed radiative-transfer modeling of these systems lies beyond the scope of the present study but offers a promising avenue for future work.

\section{Conclusions}\label{sec_conclusions}

We present the first coordinated Chandra--JWST/NIRSpec investigation of infrared spectral variability associated with X-ray flares from disk-bearing PMS stars. Six Chandra ACIS-I observations and three successful JWST/NIRSpec-MOS visits of the Orion Nebula yielded two-epoch spectroscopy for 10 X-ray Flare stars and 12 matched non-flaring Controls. We examined three complementary diagnostics of flare-associated variability: narrow spectral features at 2.8--4.3~\micron, H~I Br$\alpha$, and the CO fundamental band. Our principal findings are as follows.

\begin{enumerate}

\item Narrow spectral features show tentative flare-associated and stronger environmental trends (\S~\ref{sec_narrow_features_analyses}). Among 281 highly variable features, the majority of morphologically classifiable features appear in absorption against the local pseudo-continuum. Flare stars show tentative evidence for relatively fewer variable features at longer wavelengths (3.5--4.3~\micron) than Controls, with the source-level difference concentrated in the q75 wavelength statistic and requiring confirmation with larger samples. If confirmed, the physical origin of this tentative deficit remains unclear, but it could reflect flare-associated X-rays and/or CME-accelerated energetic particles perturbing the chemistry or excitation of molecular material in disk atmospheres and associated photoevaporative flows. The incidence of variable features decreases significantly with projected distance from $\theta^1$~Ori~C, providing stronger evidence for an environmental connection. Their wavelengths more closely match molecular transitions in HITRAN than stellar photospheric transitions in VALD, although individual molecular identifications remain tentative. A direct comparison with the empirical PDRs4All Orion Bar spectrum yields a number and wavelength proximity of candidate matches consistent with chance superposition, providing no population-level evidence for preferential coincidence with known PDR transitions. Taken together, the absorption-line morphology and strong dependence of feature incidence on projected distance from $\theta^1$~Ori~C are consistent with an origin of at least some features in externally irradiated disk atmospheres or photoevaporative flows, potentially modulated by the spatial--evolutionary structure of the ONC (\S~\ref{sec_origins_of_narrow_features}).  NIRSpec nod-by-nod reductions show that the narrow-feature morphology can vary substantially among science nod positions and/or MSA configurations while remaining comparatively insensitive to the choice of background for a fixed science nod. Thus, some of the apparent two-epoch ``variability'' may trace spatial--chemical structure across the immediate environments of the star--disk systems rather than intrinsic temporal changes. Residual contamination from small-scale nebular or molecular-cloud structure and position-dependent instrumental effects cannot be excluded.

\item Br$\alpha$ exhibits the clearest flare-associated spectral behavior, but in the opposite sense from the modeled disk response (\S~\ref{sec_Bralpha_analyses}). Flare stars show systematically stronger Br$\alpha$ variability and weaker integrated Br$\alpha$ fluxes during flare-associated epochs, while all four absorption-core profiles identified among the 73 stars with usable Br$\alpha$ spectra occur in the Flare sample. The integrated-flux result is conditional on the assumption of negligible astrophysical continuum variability, whereas the absorption-core morphology is less sensitive to this calibration limitation. Together with the strong Br$\alpha$--accretion relation found in Paper~II, these results favor an origin primarily in magnetospheric accretion and other stellar magnetic processes rather than direct X-ray heating of the disk (\S~\ref{sec_origin_of_Bralpha}). Magnetic restructuring of accretion flows and/or additional cool absorbing material associated with enhanced magnetic activity provide plausible explanations that require higher-cadence observations and detailed line-profile modeling.

\item The CO fundamental band shows no significant response to the moderate X-ray superflares sampled here (\S~\ref{sec_CO_analyses} and \ref{sec_CO_response}). Across all seven wavelength intervals spanning 4.3--5.2~\micron, the Flare and Control samples have statistically indistinguishable flux-change distributions. Six of the seven strong CO emitters remain stable to within 10--15\%, while COUP1407--JWST140, which experienced the strongest inferred flare among these systems, shows the largest CO changes (25--40\%). Although this individual case is suggestive, it does not establish a flare--CO relation. The overall absence of a detectable CO response is consistent with our ProDiMo calculations (\S~\ref{sec_modeling}), which predict little CO enhancement for a moderate $\log(E_{\rm X,fl}/{\rm erg})\sim34.5$ superflare but a 20\% enhancement for a $\log(E_{\rm X,fl}/{\rm erg})\sim36.5$ mega-flare.

\end{enumerate}

This pilot experiment demonstrates that coordinated X-ray and infrared spectroscopy can separate distinct responses within young star--disk systems, while also revealing the limitations imposed by small flare samples, sparse temporal coverage, and complex nebular backgrounds. Only three of eight planned NIRSpec-MOS visits succeeded, and no X-ray mega-flares were captured in systems with suitable two-epoch spectroscopy. Future coordinated Chandra--JWST campaigns with larger samples, higher temporal cadence, and several captured mega-flares will be required to test whether disk molecular emission responds measurably to the most energetic PMS flares, distinguish intrinsic temporal variability from local spatial--chemical structure and other spatial-sampling effects, and establish how stellar magnetic activity couples to both magnetospheric accretion and planet-forming disk gas.

\begin{acknowledgments} This work is supported by STScI grant \#JWST-GO-09292.001-A and SAO Chandra grant GO6-27010X. JF and AB acknowledge support from the Swedish National Space Agency (2022-00154). This paper
employs a list of Chandra data sets obtained by the Chandra
X-ray Observatory, contained in~\dataset[doi:10.25574/cdc.700]{https://doi.org/10.25574/cdc.700}. The JWST/NIRSpec observations analyzed in this work are available from MAST at \dataset[doi:10.17909/3hm9-rw51]{https://doi.org/10.17909/3hm9-rw51}. Support for program \#9292 was provided by NASA through a grant from the Space Telescope Science Institute, which is operated by the Association of Universities for Research in Astronomy, Inc., under NASA contract NAS 5-03127.  This work is based on observations made with the NASA/ESA/CSA James Webb Space Telescope. The data are associated with program 9292 and were obtained from the Mikulski Archive for Space Telescopes (MAST) at the Space Telescope Science Institute, which is operated by the Association of Universities for Research in Astronomy, Inc., under NASA contract NAS 5-03127.
\end{acknowledgments}





%
\facilities{CXO, JWST (NIRSpec)}

\software{R \citep{RCoreTeam2026}, JWST Calibration Pipeline \citep{Bushouse2025}, CTOOL \citep{Pontoppidan2026}, CIAO \citep{Fruscione2006}, Acis Extract \citep{Broos2010,Broos2012}}


\clearpage
\appendix
\section{Supplementary Tables}\label{sec_appendix_tables}

This Appendix presents three tables: (1) basic stellar properties of the 84 PMS stars observed in the coordinated Chandra--JWST campaign (Table~\ref{tab_stellar_properties}); (2) wavelengths, relative flux differences, spectral morphologies, and tentative HITRAN and PDRs4All identifications of the 281 highly variable narrow NIRSpec features at 2.8--4.3~\micron (Table~\ref{tab_narrow_features}); and (3) integrated NIRSpec Br$\alpha$ and CO fundamental-band fluxes (in seven CO sub-bands), with their statistical uncertainties, for the 10 Flare and 12 Control stars.

\begin{deluxetable*}{rrrrlrrrrcrrr}
\tabletypesize{\scriptsize}
\tablecaption{Properties of the COUP--JWST PMS Stars\label{tab_stellar_properties}}
\tablewidth{0pt}
\tablehead{
\colhead{COUP} & \colhead{JWST} & \colhead{R.A.} & \colhead{Decl.} & \colhead{SpT} & \colhead{$A_V$} & \colhead{$M_*$} & \colhead{$\alpha_{\rm IRAC}$} & \colhead{Group} & \colhead{Sample} & \colhead{$\log L_{\rm X,char}$} & \colhead{$\log E_{\rm X,fl}$} & \colhead{$\sigma_{\log E_{\rm X,fl}}$} \\
\colhead{} & \colhead{} & \colhead{deg} & \colhead{deg} & \colhead{} & \colhead{mag} & \colhead{$M_\odot$} & \colhead{} & \colhead{} & \colhead{} & \colhead{erg s$^{-1}$} & \colhead{erg} & \colhead{erg} \\
\colhead{(1)} & \colhead{(2)} & \colhead{(3)} & \colhead{(4)} & \colhead{(5)} & \colhead{(6)} & \colhead{(7)} & \colhead{(8)} & \colhead{(9)} & \colhead{(10)} & \colhead{(11)} & \colhead{(12)} & \colhead{(13)}
}
\startdata
1058 & 88 & 83.8326735 & -5.4133003 & M1 & 1.6 & 0.5 & -0.8 & 1 & F & 29.08 & 34.36 & 0.10 \\
1076 & 90 & 83.8335250 & -5.4271274 & M0.5 & 1.9 & 0.5 & -0.8 & 1 & F & 29.48 & 34.56 & 0.09 \\
1166 & 107 & 83.8408153 & -5.3862966 & M1.5 & 2.9 & 0.3 & -1.3 & 2 & F & 28.74 & 34.02 & 0.15 \\
1174 & 110 & 83.8408662 & -5.3982973 & G6/K3 & 0.0 & 0.5 & -1.5 & 1 & F & 29.74 & 34.70 & 0.07 \\
1258 & 119 & 83.8486980 & -5.3702009 & \nodata & 4.8 & 0.6 & -1.0 & 1 & F & 29.12 & 35.03 & 0.08 \\
1302 & 132 & 83.8529067 & -5.4099178 & M0e & 5.2 & 0.7 & -1.3 & 2 & F & 29.38 & 34.36 & 0.14 \\
1342 & 137 & 83.8583788 & -5.4299634 & K8 & 4.1 & 0.6 & -1.4 & 1 & F & 29.95 & 35.16 & 0.05 \\
1407 & 140 & 83.8672399 & -5.3851200 & M3e & 8.6 & 0.7 & -1.2 & 1 & F & 29.56 & 35.06 & 0.07 \\
1444 & 146 & 83.8767709 & -5.4274017 & K8e & 2.7 & 1.2 & -0.9 & 2 & F & 29.89 & 34.66 & 0.09 \\
1466 & 150 & 83.8802003 & -5.3944564 & K3/5 & 2.3 & 0.8 & -1.0 & 1 & F & 29.81 & 34.63 & 0.10 \\
1000 & 81 & 83.8286498 & -5.3913513 & M0e & 1.6 & 0.5 & -0.5 & 3 & C & 29.02 & \nodata & \nodata \\
1091 & 92 & 83.8344673 & -5.4177857 & \nodata & 5.6 & 0.3 & -1.5 & 3 & C & 29.38 & \nodata & \nodata \\
1110 & 97 & 83.8360057 & -5.3793205 & M1 & 5.2 & 0.7 & -1.4 & 3 & C & 29.22 & \nodata & \nodata \\
1139 & 102 & 83.8381809 & -5.4325099 & M0.5 & 6.0 & 1.0 & -1.4 & 3 & C & 30.86 & \nodata & \nodata \\
1167 & 108 & 83.8407240 & -5.3942534 & K5/7 & 3.4 & 0.7 & -1.3 & 3 & C & 29.35 & \nodata & \nodata \\
1193 & 113 & 83.8424630 & -5.4069044 & K2/3 & 4.5 & 1.1 & -1.8 & 3 & C & 30.04 & \nodata & \nodata \\
1240 & 116 & 83.8465979 & -5.3745462 & \nodata & 10.0 & 0.3 & -1.0 & 3 & C & 29.62 & \nodata & \nodata \\
1275 & 126 & 83.8499390 & -5.4194175 & M1 & 3.9 & 0.5 & 0.5 & 3 & C & 29.08 & \nodata & \nodata \\
1306 & 133 & 83.8537105 & -5.4195115 & M5 & 1.0 & 0.3 & -1.1 & 3 & C & 29.11 & \nodata & \nodata \\
1423 & 142 & 83.8722600 & -5.4295322 & M4 & 1.0 & 0.5 & -1.9 & 3 & C & 29.18 & \nodata & \nodata \\
1434 & 144 & 83.8744414 & -5.3967785 & M2e & 2.0 & 0.2 & -1.6 & 3 & C & 28.65 & \nodata & \nodata \\
1457 & 148 & 83.8788469 & -5.4232570 & M3.5 & 0.0 & 0.3 & -1.7 & 3 & C & 28.73 & \nodata & \nodata \\
\enddata
\tablecomments{The 22 stars in the two-epoch Flare and Control samples are listed here (10 Flare followed by 12 Control stars). The complete table of all 84 PMS stars observed in the coordinated Chandra--JWST program is available in machine-readable form in the online journal. Columns 1--2: COUP \citep{Getman2005} and JWST source identifiers. Columns 3--4: Gaia ICRS coordinates in degrees (at Epoch 2016.0). Column 5: Spectral type \citep{Skiff2014}. Columns 6--7: Visual extinction and stellar mass from \citet{Getman2022}. Column 8: Spitzer/IRAC spectral slope \citep{Megeath2012}. Column 9: Flare group classification used in this work. Column 10: Membership in the two-epoch Flare (F) or Control (C) sample; missing entries indicate stars outside these samples. Column 11: Logarithm of the characteristic X-ray luminosity (erg s$^{-1}$) measured in the 2026 campaign. Column 12: Logarithm of the 2026 X-ray flare energy (erg), often a lower limit to the total flare energy owing to incomplete temporal coverage by Chandra. Column 13: Propagated $1\sigma$ uncertainty in the logarithmic flare energy (erg).}
\end{deluxetable*}

\begin{deluxetable*}{rrcrrccll}
\tabletypesize{\footnotesize}
\tablecaption{Highly Variable Narrow NIRSpec Features at 2.8--4.3~\micron\label{tab_narrow_features}}
\tablewidth{0pt}
\tablehead{
\colhead{COUP} & \colhead{JWST} & \colhead{Sample} & \colhead{$\lambda$} & \colhead{$\Delta F$} & \colhead{Epoch 1} & \colhead{Epoch 2} & \colhead{HITRAN} & \colhead{PDRs4All} \\
\colhead{} & \colhead{} & \colhead{} & \colhead{\micron} & \colhead{\%} & \colhead{Type} & \colhead{Type} & \colhead{Species} & \colhead{Species} \\
\colhead{(1)} & \colhead{(2)} & \colhead{(3)} & \colhead{(4)} & \colhead{(5)} & \colhead{(6)} & \colhead{(7)} & \colhead{(8)} & \colhead{(9)}
}
\startdata
1058 & 88 & F & 2.90894 & 5.5 & e & p & CO2 & \nodata \\
1058 & 88 & F & 2.98945 & 5.2 & e & p & HCN & \nodata \\
1058 & 88 & F & 3.05370 & -6.3 & a & p & C2H2 & \nodata \\
1058 & 88 & F & 3.06923 & -6.9 & a & p & H2O & \nodata \\
1058 & 88 & F & 3.07396 & -6.2 & a & p & C2H2 & \nodata \\
1058 & 88 & F & 3.09624 & -5.0 & a & p & HCN & \nodata \\
1058 & 88 & F & 3.15773 & 6.5 & e & p & CH4 & H2 \\
1058 & 88 & F & 3.28996 & 9.2 & e & p & CH4 & PAHG \\
1058 & 88 & F & 3.33173 & -8.1 & a & p & CH4 & \nodata \\
1058 & 88 & F & 3.47917 & 6.7 & p & a & HCN & \nodata \\
\enddata
\tablecomments{The first 10 of 281 highly variable narrow NIRSpec features identified in the two-epoch spectra of the 10 Flare and 12 Control PMS stars are listed here. The complete 281-feature catalog is available in machine-readable form in the online journal. Columns 1--2: COUP and JWST source identifiers. Column 3: Flare (F) or Control (C) sample. Column 4: Measured feature-center wavelength in microns. Column 5: Relative spectral difference, $\Delta F=(F_1-F_2)/[(|F_1|+|F_2|)/2]\times100\%$, evaluated at the feature center. For Flare sources, Epoch 1 is the flare-associated epoch and Epoch 2 is the comparison epoch, regardless of chronological order; for Controls, the epoch assignment is random. Columns 6--7: Feature morphology in Epochs 1 and 2, respectively: e = emission, a = absorption, and p = local pseudo-continuum. Column 8: Tentative molecular species association with the HITRAN database \citep{Gordon2022}. Column 9: Tentative molecular or atomic species association with the empirical PDRs4All Orion Bar spectrum \citep{Peeters2024}. These associations are wavelength-based candidates rather than secure identifications.}
\end{deluxetable*}

\begin{deluxetable*}{ll}
\tabletypesize{\footnotesize}
\tablecaption{Extinction-Corrected Integrated Br$\alpha$ and CO Fluxes\label{tab_integrated_fluxes}}
\tablewidth{0pt}
\tablehead{
  \colhead{Column Label} & \colhead{Description}
}
\startdata
\texttt{COUP} & COUP source identifier \\
\texttt{JWST} & JWST source identifier \\
\texttt{Sample} & Sample: F = Flare; C = Control \\
\texttt{Br\_F1} & Br$\alpha$ integrated flux, Epoch 1 \\
\texttt{Br\_eF1} & 1$\sigma$ error in Br$\alpha$ flux, Epoch 1 \\
\texttt{Br\_F2} & Br$\alpha$ integrated flux, Epoch 2 \\
\texttt{Br\_eF2} & 1$\sigma$ error in Br$\alpha$ flux, Epoch 2 \\
\texttt{CO1\_F1} & CO integrated flux, 4.30--4.40~\micron, Epoch 1 \\
\texttt{CO1\_eF1} & 1$\sigma$ error in CO flux, 4.30--4.40~\micron, Epoch 1 \\
\texttt{CO1\_F2} & CO integrated flux, 4.30--4.40~\micron, Epoch 2 \\
\texttt{CO1\_eF2} & 1$\sigma$ error in CO flux, 4.30--4.40~\micron, Epoch 2 \\
\texttt{CO2\_F1} & CO integrated flux, 4.40--4.50~\micron, Epoch 1 \\
\texttt{CO2\_eF1} & 1$\sigma$ error in CO flux, 4.40--4.50~\micron, Epoch 1 \\
\texttt{CO2\_F2} & CO integrated flux, 4.40--4.50~\micron, Epoch 2 \\
\texttt{CO2\_eF2} & 1$\sigma$ error in CO flux, 4.40--4.50~\micron, Epoch 2 \\
\texttt{CO3\_F1} & CO integrated flux, 4.50--4.60~\micron, Epoch 1 \\
\texttt{CO3\_eF1} & 1$\sigma$ error in CO flux, 4.50--4.60~\micron, Epoch 1 \\
\texttt{CO3\_F2} & CO integrated flux, 4.50--4.60~\micron, Epoch 2 \\
\texttt{CO3\_eF2} & 1$\sigma$ error in CO flux, 4.50--4.60~\micron, Epoch 2 \\
\texttt{CO4\_F1} & CO integrated flux, 4.70--4.78~\micron, Epoch 1 \\
\texttt{CO4\_eF1} & 1$\sigma$ error in CO flux, 4.70--4.78~\micron, Epoch 1 \\
\texttt{CO4\_F2} & CO integrated flux, 4.70--4.78~\micron, Epoch 2 \\
\texttt{CO4\_eF2} & 1$\sigma$ error in CO flux, 4.70--4.78~\micron, Epoch 2 \\
\texttt{CO5\_F1} & CO integrated flux, 4.80--4.90~\micron, Epoch 1 \\
\texttt{CO5\_eF1} & 1$\sigma$ error in CO flux, 4.80--4.90~\micron, Epoch 1 \\
\texttt{CO5\_F2} & CO integrated flux, 4.80--4.90~\micron, Epoch 2 \\
\texttt{CO5\_eF2} & 1$\sigma$ error in CO flux, 4.80--4.90~\micron, Epoch 2 \\
\texttt{CO6\_F1} & CO integrated flux, 4.90--5.00~\micron, Epoch 1 \\
\texttt{CO6\_eF1} & 1$\sigma$ error in CO flux, 4.90--5.00~\micron, Epoch 1 \\
\texttt{CO6\_F2} & CO integrated flux, 4.90--5.00~\micron, Epoch 2 \\
\texttt{CO6\_eF2} & 1$\sigma$ error in CO flux, 4.90--5.00~\micron, Epoch 2 \\
\texttt{CO7\_F1} & CO integrated flux, 5.00--5.20~\micron, Epoch 1 \\
\texttt{CO7\_eF1} & 1$\sigma$ error in CO flux, 5.00--5.20~\micron, Epoch 1 \\
\texttt{CO7\_F2} & CO integrated flux, 5.00--5.20~\micron, Epoch 2 \\
\texttt{CO7\_eF2} & 1$\sigma$ error in CO flux, 5.00--5.20~\micron, Epoch 2 \\
\enddata
\tablecomments{The complete table contains measurements for 10 Flare and 12 Control stars and is available in machine-readable form. All fluxes and their 1$\sigma$ statistical uncertainties are extinction-corrected and expressed in erg~cm$^{-2}$~s$^{-1}$. Epoch 1 is the flare-associated epoch for Flare stars, and Epoch 2 is the comparison epoch, regardless of chronological order. For Controls, epoch assignments are random. Negative integrated fluxes indicate net absorption relative to the adopted continuum. Missing measurements are indicated by NA in the machine-readable table.}
\end{deluxetable*}

\clearpage

\bibliography{sample702}{}

@ARTICLE{Namekata2025,
       author = {{Namekata}, Kosuke and {Maehara}, Hiroyuki and {Notsu}, Yuta and {Honda}, Satoshi and {Ikuta}, Kai and {Nogami}, Daisaku and {Shibata}, Kazunari},
        title = "{Do Young Suns Produce Frequent, Massive CMEs? Results from Five-year Dedicated Optical Observations of EK Draconis and V889 Hercules}",
      journal = {\apj},
         year = 2025,
        month = nov,
       volume = {993},
       number = {1},
          eid = {80},
        pages = {80},
          doi = {10.3847/1538-4357/adfe70},
archivePrefix = {arXiv},
       eprint = {2510.22111},
 primaryClass = {astro-ph.SR},
       adsurl = {https://ui.adsabs.harvard.edu/abs/2025ApJ...993...80N}
}

@ARTICLE{Kowalski2024,
       author = {{Kowalski}, Adam F.},
        title = "{Stellar flares}",
      journal = {Living Reviews in Solar Physics},
         year = 2024,
        month = dec,
       volume = {21},
       number = {1},
          eid = {1},
        pages = {1},
          doi = {10.1007/s41116-024-00039-4},
archivePrefix = {arXiv},
       eprint = {2402.07885},
 primaryClass = {astro-ph.SR},
       adsurl = {https://ui.adsabs.harvard.edu/abs/2024LRSP...21....1K}
}

@ARTICLE{Fuhrmeister2020,
       author = {{Fuhrmeister}, B. and {Czesla}, S. and {Hildebrandt}, L. and {Nagel}, E. and {Schmitt}, J.~H.~M.~M. and {Jeffers}, S.~V. and {Caballero}, J.~A. and {Hintz}, D. and {Johnson}, E.~N. and {Sch{\"o}fer}, P. and {Zechmeister}, M. and {Reiners}, A. and {Ribas}, I. and {Amado}, P.~J. and {Quirrenbach}, A. and {Nortmann}, L. and {Bauer}, F.~F. and {B{\'e}jar}, V.~J.~S. and {Cort{\'e}s-Contreras}, M. and {Dreizler}, S. and {Galad{\'\i}-Enr{\'\i}quez}, D. and {Hatzes}, A.~P. and {Kaminski}, A. and {K{\"u}rster}, M. and {Lafarga}, M. and {Montes}, D.},
        title = "{The CARMENES search for exoplanets around M dwarfs. Variability of the He I line at 10 830 {\r{A}}}",
      journal = {\aap},
         year = 2020,
        month = aug,
       volume = {640},
          eid = {A52},
        pages = {A52},
          doi = {10.1051/0004-6361/202038279},
archivePrefix = {arXiv},
       eprint = {2006.09372},
 primaryClass = {astro-ph.SR},
       adsurl = {https://ui.adsabs.harvard.edu/abs/2020A&A...640A..52F}
}

@ARTICLE{Moschou2019,
       author = {{Moschou}, Sofia-Paraskevi and {Drake}, Jeremy J. and {Cohen}, Ofer and {Alvarado-G{\'o}mez}, Juli{\'a}n D. and {Garraffo}, Cecilia and {Fraschetti}, Federico},
        title = "{The Stellar CME-Flare Relation: What Do Historic Observations Reveal?}",
      journal = {\apj},
         year = 2019,
        month = jun,
       volume = {877},
       number = {2},
          eid = {105},
        pages = {105},
          doi = {10.3847/1538-4357/ab1b37},
archivePrefix = {arXiv},
       eprint = {1904.09598},
 primaryClass = {astro-ph.SR},
       adsurl = {https://ui.adsabs.harvard.edu/abs/2019ApJ...877..105M}
}

@ARTICLE{Argiroffi2019,
       author = {{Argiroffi}, C. and {Reale}, F. and {Drake}, J.~J. and {Ciaravella}, A. and {Testa}, P. and {Bonito}, R. and {Miceli}, M. and {Orlando}, S. and {Peres}, G.},
        title = "{A stellar flare-coronal mass ejection event revealed by X-ray plasma motions}",
      journal = {Nature Astronomy},
         year = 2019,
        month = may,
       volume = {3},
        pages = {742-748},
          doi = {10.1038/s41550-019-0781-4},
archivePrefix = {arXiv},
       eprint = {1905.11325},
 primaryClass = {astro-ph.SR},
       adsurl = {https://ui.adsabs.harvard.edu/abs/2019NatAs...3..742A}
}

@ARTICLE{Mackay2010,
       author = {{Mackay}, D.~H. and {Karpen}, J.~T. and {Ballester}, J.~L. and {Schmieder}, B. and {Aulanier}, G.},
        title = "{Physics of Solar Prominences: II{\textemdash}Magnetic Structure and Dynamics}",
      journal = {\ssr},
         year = 2010,
        month = apr,
       volume = {151},
       number = {4},
        pages = {333-399},
          doi = {10.1007/s11214-010-9628-0},
archivePrefix = {arXiv},
       eprint = {1001.1635},
 primaryClass = {astro-ph.SR},
       adsurl = {https://ui.adsabs.harvard.edu/abs/2010SSRv..151..333M}
}

@ARTICLE{Romanova2015,
       author = {{Romanova}, Marina M. and {Owocki}, Stanley P.},
        title = "{Accretion, Outflows, and Winds of Magnetized Stars}",
      journal = {\ssr},
         year = 2015,
        month = oct,
       volume = {191},
       number = {1-4},
        pages = {339-389},
          doi = {10.1007/s11214-015-0200-9},
archivePrefix = {arXiv},
       eprint = {1605.04979},
 primaryClass = {astro-ph.SR},
       adsurl = {https://ui.adsabs.harvard.edu/abs/2015SSRv..191..339R}
}

@ARTICLE{Romanova2008,
       author = {{Romanova}, Marina M. and {Kulkarni}, Akshay K. and {Lovelace}, Richard V.~E.},
        title = "{Unstable Disk Accretion onto Magnetized Stars: First Global Three-dimensional Magnetohydrodynamic Simulations}",
      journal = {\apjl},
         year = 2008,
        month = feb,
       volume = {673},
       number = {2},
        pages = {L171},
          doi = {10.1086/527298},
archivePrefix = {arXiv},
       eprint = {0711.0418},
 primaryClass = {astro-ph},
       adsurl = {https://ui.adsabs.harvard.edu/abs/2008ApJ...673L.171R}
}

@INPROCEEDINGS{Bouvier2007,
       author = {{Bouvier}, J. and {Alencar}, S.~H.~P. and {Harries}, T.~J. and {Johns-Krull}, C.~M. and {Romanova}, M.~M.},
        title = "{Magnetospheric Accretion in Classical T Tauri Stars}",
    booktitle = {Protostars and Planets V},
         year = 2007,
       editor = {{Reipurth}, Bo and {Jewitt}, David and {Keil}, Klaus},
        month = jan,
        pages = {479},
          doi = {10.48550/arXiv.astro-ph/0603498},
archivePrefix = {arXiv},
       eprint = {astro-ph/0603498},
 primaryClass = {astro-ph},
       adsurl = {https://ui.adsabs.harvard.edu/abs/2007prpl.conf..479B}
}

@ARTICLE{Bouvier2007b,
       author = {{Bouvier}, J. and {Alencar}, S.~H.~P. and {Boutelier}, T. and {Dougados}, C. and {Balog}, Z. and {Grankin}, K. and {Hodgkin}, S.~T. and {Ibrahimov}, M.~A. and {Kun}, M. and {Magakian}, T. Yu. and {Pinte}, C.},
        title = "{Magnetospheric accretion-ejection processes in the classical T Tauri star AA Tauri}",
      journal = {\aap},
         year = 2007,
        month = mar,
       volume = {463},
       number = {3},
        pages = {1017-1028},
          doi = {10.1051/0004-6361:20066021},
archivePrefix = {arXiv},
       eprint = {astro-ph/0611787},
 primaryClass = {astro-ph},
       adsurl = {https://ui.adsabs.harvard.edu/abs/2007A&A...463.1017B}
}

@ARTICLE{Dupuy2018,
       author = {{Dupuy}, R. and {Bertin}, M. and {F{\'e}raud}, G. and {Hassenfratz}, M. and {Michaut}, X. and {Putaud}, T. and {Philippe}, L. and {Jeseck}, P. and {Angelucci}, M. and {Cimino}, R. and {Baglin}, V. and {Romanzin}, C. and {Fillion}, J.-H.},
        title = "{X-ray photodesorption from water ice in protoplanetary disks and X-ray-dominated regions}",
      journal = {Nature Astronomy},
         year = 2018,
        month = jul,
       volume = {2},
        pages = {796-801},
          doi = {10.1038/s41550-018-0532-y},
archivePrefix = {arXiv},
       eprint = {1807.03725},
 primaryClass = {astro-ph.GA},
       adsurl = {https://ui.adsabs.harvard.edu/abs/2018NatAs...2..796D}
}

@ARTICLE{MartinDomenech2018,
       author = {{Mart{\'\i}n-Dom{\'e}nech}, R. and {Cruz-D{\'\i}az}, G.~A. and {Mu{\~n}oz Caro}, G.~M.},
        title = "{UV photoprocessing of NH$_{3}$ ice: photon-induced desorption mechanisms}",
      journal = {\mnras},
         year = 2018,
        month = jan,
       volume = {473},
       number = {2},
        pages = {2575-2582},
          doi = {10.1093/mnras/stx2510},
archivePrefix = {arXiv},
       eprint = {1710.06467},
 primaryClass = {astro-ph.GA},
       adsurl = {https://ui.adsabs.harvard.edu/abs/2018MNRAS.473.2575M}
}

@ARTICLE{Keyte2025,
       author = {{Keyte}, Luke and {Haworth}, Thomas J.},
        title = "{Impact of photoevaporative winds in chemical models of externally irradiated protoplanetary discs}",
      journal = {\mnras},
         year = 2025,
        month = feb,
       volume = {537},
       number = {1},
        pages = {598-616},
          doi = {10.1093/mnras/staf047},
archivePrefix = {arXiv},
       eprint = {2501.05172},
 primaryClass = {astro-ph.EP},
       adsurl = {https://ui.adsabs.harvard.edu/abs/2025MNRAS.537..598K}
}

@ARTICLE{Gross2025,
       author = {{Gross}, Rachel E. and {Cleeves}, L. Ilsedore},
        title = "{Modeling the Impact of Moderate External UV Irradiation on Disk Chemistry}",
      journal = {\apj},
         year = 2025,
        month = feb,
       volume = {980},
       number = {2},
          eid = {189},
        pages = {189},
          doi = {10.3847/1538-4357/ada939},
archivePrefix = {arXiv},
       eprint = {2501.06156},
 primaryClass = {astro-ph.EP},
       adsurl = {https://ui.adsabs.harvard.edu/abs/2025ApJ...980..189G}
}

@ARTICLE{Goicoechea2024,
       author = {{Goicoechea}, J.~R. and {Le Bourlot}, J. and {Black}, J.~H. and {Alarc{\'o}n}, F. and {Bergin}, E.~A. and {Bern{\'e}}, O. and {Bron}, E. and {Canin}, A. and {Chapillon}, E. and {Chown}, R. and {Dartois}, E. and {Gerin}, M. and {Habart}, E. and {Haworth}, T.~J. and {Joblin}, C. and {Kannavou}, O. and {Le Petit}, F. and {Onaka}, T. and {Peeters}, E. and {Pety}, J. and {Roueff}, E. and {Sidhu}, A. and {Schroetter}, I. and {Tabone}, B. and {Tielens}, A.~G.~G.~M. and {Trahin}, B. and {Van De Putte}, D. and {Vicente}, S. and {Zannese}, M.},
        title = "{PDRs4All: X. ALMA and JWST detection of neutral carbon in the externally irradiated disk d203-506: Undepleted gas-phase carbon}",
      journal = {\aap},
         year = 2024,
        month = sep,
       volume = {689},
          eid = {L4},
        pages = {L4},
          doi = {10.1051/0004-6361/202450988},
archivePrefix = {arXiv},
       eprint = {2408.06279},
 primaryClass = {astro-ph.GA},
       adsurl = {https://ui.adsabs.harvard.edu/abs/2024A&A...689L...4G}
}

@dataset{Skiff2014,
       author = {{Skiff}, B.~A.},
        title = "{VizieR Online Data Catalog: Catalogue of Stellar Spectral Classifications (Skiff, 2009- )}",
 howpublished = {VizieR On-line Data Catalog: B/mk.  Originally published in: Lowell Observatory (October 2014)},
         year = 2014,
        month = oct,
          eid = {B/mk},
       adsurl = {https://ui.adsabs.harvard.edu/abs/2014yCat....1.2023S}
}

@ARTICLE{Peeters2024,
       author = {{Peeters}, Els and {Habart}, Emilie and {Bern{\'e}}, Olivier and {Sidhu}, Ameek and {Chown}, Ryan and {Van De Putte}, Dries and {Trahin}, Boris and {Schroetter}, Ilane and {Canin}, Am{\'e}lie and {Alarc{\'o}n}, Felipe and {Schefter}, Bethany and {Khan}, Baria and {Pasquini}, Sofia and {Tielens}, Alexander G.~G.~M. and {Wolfire}, Mark G. and {Dartois}, Emmanuel and {Goicoechea}, Javier R. and {Maragkoudakis}, Alexandros and {Onaka}, Takashi and {Pound}, Marc W. and {Vicente}, S{\'\i}lvia and {Abergel}, Alain and {Bergin}, Edwin A. and {Bernard-Salas}, Jeronimo and {Boersma}, Christiaan and {Bron}, Emeric and {Cami}, Jan and {Cuadrado}, Sara and {Dicken}, Daniel and {Elyajouri}, Meriem and {Fuente}, Asunci{\'o}n and {Gordon}, Karl D. and {Issa}, Lina and {Joblin}, Christine and {Kannavou}, Olga and {Lacinbala}, Ozan and {Languignon}, David and {Le Gal}, Romane and {Meshaka}, Raphael and {Okada}, Yoko and {Robberto}, Massimo and {R{\"o}llig}, Markus and {Schirmer}, Thi{\'e}baut and {Tabone}, Benoit and {Zannese}, Marion and {Aleman}, Isabel and {Allamandola}, Louis and {Auchettl}, Rebecca and {Baratta}, Giuseppe Antonio and {Bejaoui}, Salma and {Bera}, Partha P. and {Black}, John H. and {Boulanger}, Francois and {Bouwman}, Jordy and {Brandl}, Bernhard and {Brechignac}, Philippe and {Br{\"u}nken}, Sandra and {Buragohain}, Mridusmita and {Burkhardt}, Andrew and {Candian}, Alessandra and {Cazaux}, St{\'e}phanie and {Cernicharo}, Jose and {Chabot}, Marin and {Chakraborty}, Shubhadip and {Champion}, Jason and {Colgan}, Sean W.~J. and {Cooke}, Ilsa R. and {Coutens}, Audrey and {Cox}, Nick L.~J. and {Demyk}, Karine and {Meyer}, Jennifer Donovan and {Foschino}, Sacha and {Garc{\'\i}a-Lario}, Pedro and {Gerin}, Maryvonne and {Gottlieb}, Carl A. and {Guillard}, Pierre and {Gusdorf}, Antoine and {Hartigan}, Patrick and {He}, Jinhua and {Herbst}, Eric and {Hornekaer}, Liv and {J{\"a}ger}, Cornelia and {Janot-Pacheco}, Eduardo and {Kaufman}, Michael and {Kendrew}, Sarah and {Kirsanova}, Maria S. and {Klaassen}, Pamela and {Kwok}, Sun and {Labiano}, {\'A}lvaro and {Lai}, Thomas S.-Y. and {Lee}, Timothy J. and {Lefloch}, Bertrand and {Le Petit}, Franck and {Li}, Aigen and {Linz}, Hendrik and {Mackie}, Cameron J. and {Madden}, Suzanne C. and {Mascetti}, Jo{\"e}lle and {McGuire}, Brett A. and {Merino}, Pablo and {Micelotta}, Elisabetta R. and {Misselt}, Karl and {Morse}, Jon A. and {Mulas}, Giacomo and {Neelamkodan}, Naslim and {Ohsawa}, Ryou and {Paladini}, Roberta and {Palumbo}, Maria Elisabetta and {Pathak}, Amit and {Pendleton}, Yvonne J. and {Petrignani}, Annemieke and {Pino}, Thomas and {Puga}, Elena and {Rangwala}, Naseem and {Rapacioli}, Mathias and {Ricca}, Alessandra and {Roman-Duval}, Julia and {Roser}, Joseph and {Roueff}, Evelyne and {Rouill{\'e}}, Ga{\"e}l and {Salama}, Farid and {Sales}, Dinalva A. and {Sandstrom}, Karin and {Sarre}, Peter and {Sciamma-O'Brien}, Ella and {Sellgren}, Kris and {Shenoy}, Sachindev S. and {Teyssier}, David and {Thomas}, Richard D. and {Togi}, Aditya and {Verstraete}, Laurent and {Witt}, Adolf N. and {Wootten}, Alwyn and {Ysard}, Nathalie and {Zettergren}, Henning and {Zhang}, Yong and {Zhang}, Ziwei E. and {Zhen}, Junfeng},
        title = "{PDRs4All: III. JWST's NIR spectroscopic view of the Orion Bar}",
      journal = {\aap},
         year = 2024,
        month = may,
       volume = {685},
          eid = {A74},
        pages = {A74},
          doi = {10.1051/0004-6361/202348244},
archivePrefix = {arXiv},
       eprint = {2310.08720},
 primaryClass = {astro-ph.GA},
       adsurl = {https://ui.adsabs.harvard.edu/abs/2024A&A...685A..74P}
}

@ARTICLE{Berne2023,
       author = {{Bern{\'e}}, Olivier and {Martin-Drumel}, Marie-Aline and {Schroetter}, Ilane and {Goicoechea}, Javier R. and {Jacovella}, Ugo and {Gans}, B{\'e}renger and {Dartois}, Emmanuel and {Coudert}, Laurent H. and {Bergin}, Edwin and {Alarcon}, Felipe and {Cami}, Jan and {Roueff}, Evelyne and {Black}, John H. and {Asvany}, Oskar and {Habart}, Emilie and {Peeters}, Els and {Canin}, Amelie and {Trahin}, Boris and {Joblin}, Christine and {Schlemmer}, Stephan and {Thorwirth}, Sven and {Cernicharo}, Jose and {Gerin}, Maryvonne and {Tielens}, Alexander and {Zannese}, Marion and {Abergel}, Alain and {Bernard-Salas}, Jeronimo and {Boersma}, Christiaan and {Bron}, Emeric and {Chown}, Ryan and {Cuadrado}, Sara and {Dicken}, Daniel and {Elyajouri}, Meriem and {Fuente}, Asunci{\'o}n and {Gordon}, Karl D. and {Issa}, Lina and {Kannavou}, Olga and {Khan}, Baria and {Lacinbala}, Ozan and {Languignon}, David and {Le Gal}, Romane and {Maragkoudakis}, Alexandros and {Meshaka}, Raphael and {Okada}, Yoko and {Onaka}, Takashi and {Pasquini}, Sofia and {Pound}, Marc W. and {Robberto}, Massimo and {R{\"o}llig}, Markus and {Schefter}, Bethany and {Schirmer}, Thi{\'e}baut and {Sidhu}, Ameek and {Tabone}, Benoit and {Van De Putte}, Dries and {Vicente}, S{\'\i}lvia and {Wolfire}, Mark G.},
        title = "{Formation of the methyl cation by photochemistry in a protoplanetary disk}",
      journal = {\nat},
         year = 2023,
        month = sep,
       volume = {621},
       number = {7977},
        pages = {56-59},
          doi = {10.1038/s41586-023-06307-x},
archivePrefix = {arXiv},
       eprint = {2401.03296},
 primaryClass = {astro-ph.GA},
       adsurl = {https://ui.adsabs.harvard.edu/abs/2023Natur.621...56B}
}

@ARTICLE{Berne2024,
       author = {{Bern{\'e}}, Olivier and {Habart}, Emilie and {Peeters}, Els and {Schroetter}, Ilane and {Canin}, Am{\'e}lie and {Sidhu}, Ameek and {Chown}, Ryan and {Bron}, Emeric and {Haworth}, Thomas J. and {Klaassen}, Pamela and {Trahin}, Boris and {Van De Putte}, Dries and {Alarc{\'o}n}, Felipe and {Zannese}, Marion and {Abergel}, Alain and {Bergin}, Edwin A. and {Bernard-Salas}, Jeronimo and {Boersma}, Christiaan and {Cami}, Jan and {Cuadrado}, Sara and {Dartois}, Emmanuel and {Dicken}, Daniel and {Elyajouri}, Meriem and {Fuente}, Asunci{\'o}n and {Goicoechea}, Javier R. and {Gordon}, Karl D. and {Issa}, Lina and {Joblin}, Christine and {Kannavou}, Olga and {Khan}, Baria and {Lacinbala}, Ozan and {Languignon}, David and {Le Gal}, Romane and {Maragkoudakis}, Alexandros and {Meshaka}, Raphael and {Okada}, Yoko and {Onaka}, Takashi and {Pasquini}, Sofia and {Pound}, Marc W. and {Robberto}, Massimo and {R{\"o}llig}, Markus and {Schefter}, Bethany and {Schirmer}, Thi{\'e}baut and {Simmer}, Thomas and {Tabone}, Benoit and {Tielens}, Alexander G.~G.~M. and {Vicente}, S{\'\i}lvia and {Wolfire}, Mark G. and {PDRs4All Team} and {Aleman}, Isabel and {Allamandola}, Louis and {Auchettl}, Rebecca and {Baratta}, Giuseppe Antonio and {Baruteau}, Cl{\'e}ment and {Bejaoui}, Salma and {Bera}, Partha P. and {Black}, John H. and {Boulanger}, Francois and {Bouwman}, Jordy and {Brandl}, Bernhard and {Brechignac}, Philippe and {Br{\"u}nken}, Sandra and {Buragohain}, Mridusmita and {Burkhardt}, Andrew and {Candian}, Alessandra and {Cazaux}, St{\'e}phanie and {Cernicharo}, Jose and {Chabot}, Marin and {Chakraborty}, Shubhadip and {Champion}, Jason and {Colgan}, Sean W.~J. and {Cooke}, Ilsa R. and {Coutens}, Audrey and {Cox}, Nick L.~J. and {Demyk}, Karine and {Meyer}, Jennifer Donovan and {Engrand}, C{\'e}cile and {Foschino}, Sacha and {Garc{\'\i}a-Lario}, Pedro and {Gavilan}, Lisseth and {Gerin}, Maryvonne and {Godard}, Marie and {Gottlieb}, Carl A. and {Guillard}, Pierre and {Gusdorf}, Antoine and {Hartigan}, Patrick and {He}, Jinhua and {Herbst}, Eric and {Hornekaer}, Liv and {J{\"a}ger}, Cornelia and {Janot-Pacheco}, Eduardo and {Kaufman}, Michael and {Kemper}, Francisca and {Kendrew}, Sarah and {Kirsanova}, Maria S. and {Knight}, Collin and {Kwok}, Sun and {Labiano}, {\'A}lvaro and {Lai}, Thomas S.-Y. and {Lee}, Timothy J. and {Lefloch}, Bertrand and {Le Petit}, Franck and {Li}, Aigen and {Linz}, Hendrik and {Mackie}, Cameron J. and {Madden}, Suzanne C. and {Mascetti}, Jo{\"e}lle and {McGuire}, Brett A. and {Merino}, Pablo and {Micelotta}, Elisabetta R. and {Morse}, Jon A. and {Mulas}, Giacomo and {Neelamkodan}, Naslim and {Ohsawa}, Ryou and {Paladini}, Roberta and {Palumbo}, Maria Elisabetta and {Pathak}, Amit and {Pendleton}, Yvonne J. and {Petrignani}, Annemieke and {Pino}, Thomas and {Puga}, Elena and {Rangwala}, Naseem and {Rapacioli}, Mathias and {Ricca}, Alessandra and {Roman-Duval}, Julia and {Roueff}, Evelyne and {Rouill{\'e}}, Ga{\"e}l and {Salama}, Farid and {Sales}, Dinalva A. and {Sandstrom}, Karin and {Sarre}, Peter and {Sciamma-O'Brien}, Ella and {Sellgren}, Kris and {Shannon}, Matthew J. and {Simonnin}, Adrien and {Shenoy}, Sachindev S. and {Teyssier}, David and {Thomas}, Richard D. and {Togi}, Aditya and {Verstraete}, Laurent and {Witt}, Adolf N. and {Wootten}, Alwyn and {Ysard}, Nathalie and {Zettergren}, Henning and {Zhang}, Yong and {Zhang}, Ziwei E. and {Zhen}, Junfeng},
        title = "{A far-ultraviolet-driven photoevaporation flow observed in a protoplanetary disk}",
      journal = {Science},
         year = 2024,
        month = mar,
       volume = {383},
       number = {6686},
        pages = {988-992},
          doi = {10.1126/science.adh2861},
archivePrefix = {arXiv},
       eprint = {2403.00160},
 primaryClass = {astro-ph.GA},
       adsurl = {https://ui.adsabs.harvard.edu/abs/2024Sci...383..988B}
}

@ARTICLE{Vlasblom2026,
       author = {{Vlasblom}, Marissa and {Arabhavi}, Aditya M. and {de Klerk}, Niels and {Kamp}, Inga and {Tabone}, Beno{\^\i}t and {van Dishoeck}, Ewine F.},
        title = "{Detecting Nitrogen Carriers in the Inner Regions of Protoplanetary Disks}",
      journal = {ACS Earth and Space Chemistry},
         year = 2026,
        month = may,
       volume = {10},
       number = {5},
        pages = {1329-1344},
          doi = {10.1021/acsearthspacechem.6c00011},
archivePrefix = {arXiv},
       eprint = {2604.09269},
 primaryClass = {astro-ph.EP},
       adsurl = {https://ui.adsabs.harvard.edu/abs/2026ESC....10.1329V}
}

@ARTICLE{TorresDiaz2024,
       author = {{Torres-D{\'\i}az}, D. and {Basalg{\`e}te}, R. and {Amiaud}, L. and {Philippe}, L. and {Michaut}, X. and {Fillion}, J.-H. and {Lafosse}, A. and {Bertin}, M.},
        title = "{X-ray photo-desorption of NH$_{3}$ and N$_{2}$ from ammonia ices: Mechanisms and yields in protoplanetary disks}",
      journal = {\aap},
         year = 2024,
        month = oct,
       volume = {690},
          eid = {A90},
        pages = {A90},
          doi = {10.1051/0004-6361/202450883},
       adsurl = {https://ui.adsabs.harvard.edu/abs/2024A&A...690A..90T}
}

@ARTICLE{Najita2021,
       author = {{Najita}, Joan R. and {Carr}, John S. and {Brittain}, Sean D. and {Lacy}, John H. and {Richter}, Matthew J. and {Doppmann}, Greg W.},
        title = "{High-resolution Mid-infrared Spectroscopy of GV Tau N: Surface Accretion and Detection of NH$_{3}$ in a Young Protoplanetary Disk}",
      journal = {\apj},
         year = 2021,
        month = feb,
       volume = {908},
       number = {2},
          eid = {171},
        pages = {171},
          doi = {10.3847/1538-4357/abcfc6},
archivePrefix = {arXiv},
       eprint = {2012.00016},
 primaryClass = {astro-ph.SR},
       adsurl = {https://ui.adsabs.harvard.edu/abs/2021ApJ...908..171N}
}

@ARTICLE{Arabhavi2024,
       author = {{Arabhavi}, A.~M. and {Kamp}, I. and {Henning}, Th. and {van Dishoeck}, E.~F. and {Christiaens}, V. and {Gasman}, D. and {Perrin}, A. and {G{\"u}del}, M. and {Tabone}, B. and {Kanwar}, J. and {Waters}, L.~B.~F.~M. and {Pascucci}, I. and {Samland}, M. and {Perotti}, G. and {Bettoni}, G. and {Grant}, S.~L. and {Lagage}, P.~O. and {Ray}, T.~P. and {Vandenbussche}, B. and {Absil}, O. and {Argyriou}, I. and {Barrado}, D. and {Boccaletti}, A. and {Bouwman}, J. and {Caratti o Garatti}, A. and {Glauser}, A.~M. and {Lahuis}, F. and {Mueller}, M. and {Olofsson}, G. and {Pantin}, E. and {Scheithauer}, S. and {Morales-Calder{\'o}n}, M. and {Franceschi}, R. and {Jang}, H. and {Pawellek}, N. and {Rodgers-Lee}, D. and {Schreiber}, J. and {Schwarz}, K. and {Temmink}, M. and {Vlasblom}, M. and {Wright}, G. and {Colina}, L. and {{\"O}stlin}, G.},
        title = "{Abundant hydrocarbons in the disk around a very-low-mass star}",
      journal = {Science},
         year = 2024,
        month = jun,
       volume = {384},
       number = {6700},
        pages = {1086-1090},
          doi = {10.1126/science.adi8147},
archivePrefix = {arXiv},
       eprint = {2406.14293},
 primaryClass = {astro-ph.EP},
       adsurl = {https://ui.adsabs.harvard.edu/abs/2024Sci...384.1086A}
}

@ARTICLE{Tabone2023,
       author = {{Tabone}, B. and {Bettoni}, G. and {van Dishoeck}, E.~F. and {Arabhavi}, A.~M. and {Grant}, S. and {Gasman}, D. and {Henning}, Th. and {Kamp}, I. and {G{\"u}del}, M. and {Lagage}, P.~O. and {Ray}, T. and {Vandenbussche}, B. and {Abergel}, A. and {Absil}, O. and {Argyriou}, I. and {Barrado}, D. and {Boccaletti}, A. and {Bouwman}, J. and {Caratti o Garatti}, A. and {Geers}, V. and {Glauser}, A.~M. and {Justannont}, K. and {Lahuis}, F. and {Mueller}, M. and {Nehm{\'e}}, C. and {Olofsson}, G. and {Pantin}, E. and {Scheithauer}, S. and {Waelkens}, C. and {Waters}, L.~B.~F.~M. and {Black}, J.~H. and {Christiaens}, V. and {Guadarrama}, R. and {Morales-Calder{\'o}n}, M. and {Jang}, H. and {Kanwar}, J. and {Pawellek}, N. and {Perotti}, G. and {Perrin}, A. and {Rodgers-Lee}, D. and {Samland}, M. and {Schreiber}, J. and {Schwarz}, K. and {Colina}, L. and {{\"O}stlin}, G. and {Wright}, G.},
        title = "{A rich hydrocarbon chemistry and high C to O ratio in the inner disk around a very low-mass star}",
      journal = {Nature Astronomy},
         year = 2023,
        month = jul,
       volume = {7},
        pages = {805-814},
          doi = {10.1038/s41550-023-01965-3},
archivePrefix = {arXiv},
       eprint = {2304.05954},
 primaryClass = {astro-ph.EP},
       adsurl = {https://ui.adsabs.harvard.edu/abs/2023NatAs...7..805T}
}

@ARTICLE{Gibb2013,
       author = {{Gibb}, Erika L. and {Horne}, David},
        title = "{Detection of CH$_{4}$ in the GV Tau N Protoplanetary Disk}",
      journal = {\apjl},
         year = 2013,
        month = oct,
       volume = {776},
       number = {2},
          eid = {L28},
        pages = {L28},
          doi = {10.1088/2041-8205/776/2/L28},
       adsurl = {https://ui.adsabs.harvard.edu/abs/2013ApJ...776L..28G}
}

@ARTICLE{McClure2023,
       author = {{McClure}, M.~K. and {Rocha}, W.~R.~M. and {Pontoppidan}, K.~M. and {Crouzet}, N. and {Chu}, L.~E.~U. and {Dartois}, E. and {Lamberts}, T. and {Noble}, J.~A. and {Pendleton}, Y.~J. and {Perotti}, G. and {Qasim}, D. and {Rachid}, M.~G. and {Smith}, Z.~L. and {Sun}, Fengwu and {Beck}, Tracy L. and {Boogert}, A.~C.~A. and {Brown}, W.~A. and {Caselli}, P. and {Charnley}, S.~B. and {Cuppen}, Herma M. and {Dickinson}, H. and {Drozdovskaya}, M.~N. and {Egami}, E. and {Erkal}, J. and {Fraser}, H. and {Garrod}, R.~T. and {Harsono}, D. and {Ioppolo}, S. and {Jim{\'e}nez-Serra}, I. and {Jin}, M. and {J{\o}rgensen}, J.~K. and {Kristensen}, L.~E. and {Lis}, D.~C. and {McCoustra}, M.~R.~S. and {McGuire}, Brett A. and {Melnick}, G.~J. and {{\"O}berg}, Karin I. and {Palumbo}, M.~E. and {Shimonishi}, T. and {Sturm}, J.~A. and {van Dishoeck}, E.~F. and {Linnartz}, H.},
        title = "{An Ice Age JWST inventory of dense molecular cloud ices}",
      journal = {Nature Astronomy},
         year = 2023,
        month = apr,
       volume = {7},
        pages = {431-443},
          doi = {10.1038/s41550-022-01875-w},
archivePrefix = {arXiv},
       eprint = {2301.09140},
 primaryClass = {astro-ph.GA},
       adsurl = {https://ui.adsabs.harvard.edu/abs/2023NatAs...7..431M}
}

@ARTICLE{Ballering2021,
       author = {{Ballering}, Nicholas P. and {Cleeves}, L. Ilsedore and {Anderson}, Dana E.},
        title = "{Simulating Observations of Ices in Protoplanetary Disks}",
      journal = {\apj},
         year = 2021,
        month = oct,
       volume = {920},
       number = {2},
          eid = {115},
        pages = {115},
          doi = {10.3847/1538-4357/ac17ed},
archivePrefix = {arXiv},
       eprint = {2105.12169},
 primaryClass = {astro-ph.EP},
       adsurl = {https://ui.adsabs.harvard.edu/abs/2021ApJ...920..115B}
}

@ARTICLE{Boogert2015,
       author = {{Boogert}, A.~C. Adwin and {Gerakines}, Perry A. and {Whittet}, Douglas C.~B.},
        title = "{Observations of the icy universe.}",
      journal = {\araa},
         year = 2015,
        month = aug,
       volume = {53},
        pages = {541-581},
          doi = {10.1146/annurev-astro-082214-122348},
archivePrefix = {arXiv},
       eprint = {1501.05317},
 primaryClass = {astro-ph.GA},
       adsurl = {https://ui.adsabs.harvard.edu/abs/2015ARA&A..53..541B}
}

@ARTICLE{Oberg2023,
       author = {{{\"O}berg}, Karin I. and {Facchini}, Stefano and {Anderson}, Dana E.},
        title = "{Protoplanetary Disk Chemistry}",
      journal = {\araa},
         year = 2023,
        month = aug,
       volume = {61},
        pages = {287-328},
          doi = {10.1146/annurev-astro-022823-040820},
archivePrefix = {arXiv},
       eprint = {2309.05685},
 primaryClass = {astro-ph.EP},
       adsurl = {https://ui.adsabs.harvard.edu/abs/2023ARA&A..61..287O}
}

@ARTICLE{Oberg2011,
       author = {{{\"O}berg}, Karin I. and {Boogert}, A.~C. Adwin and {Pontoppidan}, Klaus M. and {van den Broek}, Saskia and {van Dishoeck}, Ewine F. and {Bottinelli}, Sandrine and {Blake}, Geoffrey A. and {Evans}, II, Neal J.},
        title = "{The Spitzer Ice Legacy: Ice Evolution from Cores to Protostars}",
      journal = {\apj},
         year = 2011,
        month = oct,
       volume = {740},
       number = {2},
          eid = {109},
        pages = {109},
          doi = {10.1088/0004-637X/740/2/109},
archivePrefix = {arXiv},
       eprint = {1107.5825},
 primaryClass = {astro-ph.GA},
       adsurl = {https://ui.adsabs.harvard.edu/abs/2011ApJ...740..109O}
}

@ARTICLE{PortillaRevelo2025,
       author = {{Portilla-Revelo}, Bayron and {Getman}, Konstantin V. and {Ram{\'\i}rez-Tannus}, Mar{\'\i}a Claudia and {Haworth}, Thomas J. and {Waters}, Rens and {Bik}, Arjan and {Feigelson}, Eric D. and {Kamp}, Inga and {van Terwisga}, Sierk E. and {Frediani}, Jenny and {Henning}, Thomas and {Winter}, Andrew J. and {Roccatagliata}, Veronica and {Preibisch}, Thomas and {Sabbi}, E. and {Zeidler}, Peter and {Kuhn}, Michael A.},
        title = "{XUE: Thermochemical Modeling Suggests a Compact and Gas-depleted Structure for a Distant, Irradiated Protoplanetary Disk}",
      journal = {\apj},
         year = 2025,
        month = may,
       volume = {985},
       number = {1},
          eid = {72},
        pages = {72},
          doi = {10.3847/1538-4357/adc91d},
archivePrefix = {arXiv},
       eprint = {2504.00841},
 primaryClass = {astro-ph.EP},
       adsurl = {https://ui.adsabs.harvard.edu/abs/2025ApJ...985...72P}
}

@ARTICLE{Goicoechea2025,
       author = {{Goicoechea}, Javier R. and {Roncero}, Octavio and {Roueff}, Evelyne and {Black}, John H. and {Schroetter}, Ilane and {Bern{\'e}}, Olivier},
        title = "{H3+ in irradiated protoplanetary disks: Linking far-ultraviolet radiation and water vapor}",
      journal = {\aap},
         year = 2025,
        month = nov,
       volume = {703},
          eid = {A189},
        pages = {A189},
          doi = {10.1051/0004-6361/202555842},
archivePrefix = {arXiv},
       eprint = {2506.05189},
 primaryClass = {astro-ph.GA},
       adsurl = {https://ui.adsabs.harvard.edu/abs/2025A&A...703A.189G}
}

@ARTICLE{Schroetter2025,
       author = {{Schroetter}, I. and {Bern{\'e}}, O. and {Goicoechea}, J.~R. and {Black}, J.~H. and {Roncero}, O. and {Alarcon}, F. and {Amiot}, P. and {Asvany}, O. and {Boersma}, C. and {Br{\"u}nken}, S. and {Cami}, J. and {Coudert}, L. and {Dartois}, E. and {Fuente}, A. and {Gans}, B. and {Gusdorf}, A. and {Jacovella}, U. and {Martin Drumel}, M.~A. and {Onaka}, T. and {Peeters}, E. and {Roueff}, E. and {Tielens}, A.~G.~G.~M. and {Zannese}, M.},
        title = "{PDRs4All: XV. CH radical and H molecular ion in the irradiated protoplanetary disk d203-506}",
      journal = {\aap},
         year = 2025,
        month = jul,
       volume = {699},
          eid = {L13},
        pages = {L13},
          doi = {10.1051/0004-6361/202555841},
archivePrefix = {arXiv},
       eprint = {2506.05186},
 primaryClass = {astro-ph.GA},
       adsurl = {https://ui.adsabs.harvard.edu/abs/2025A&A...699L..13S}
}

@ARTICLE{Zannese2025,
       author = {{Zannese}, M. and {Tabone}, B. and {Habart}, E. and {Dartois}, E. and {Goicoechea}, J.~R. and {Coudert}, L. and {Gans}, B. and {Martin-Drumel}, M.-A. and {Jacovella}, U. and {Faure}, A. and {Godard}, B. and {Tielens}, A.~G.~G.~M. and {Le Gal}, R. and {Black}, J.~H. and {Vicente}, S. and {Bern{\'e}}, O. and {Peeters}, E. and {Van De Putte}, D. and {Chown}, R. and {Sidhu}, A. and {Schroetter}, I. and {Canin}, A. and {Kannavou}, O.},
        title = "{PDRs4All: XI. Detection of infrared CH$^{+}$ and CH$_{3}$$^{+}$ rovibrational emission in the Orion Bar and disk d203-506: Evidence of chemical pumping}",
      journal = {\aap},
         year = 2025,
        month = apr,
       volume = {696},
          eid = {A99},
        pages = {A99},
          doi = {10.1051/0004-6361/202453441},
archivePrefix = {arXiv},
       eprint = {2502.08354},
 primaryClass = {astro-ph.GA},
       adsurl = {https://ui.adsabs.harvard.edu/abs/2025A&A...696A..99Z}
}

@ARTICLE{ODell2017,
       author = {{O'Dell}, C.~R. and {Ferland}, G.~J. and {Peimbert}, M.},
        title = "{Structure and physical conditions in the Huygens region of the Orion nebula}",
      journal = {\mnras},
         year = 2017,
        month = feb,
       volume = {464},
       number = {4},
        pages = {4835-4857},
          doi = {10.1093/mnras/stw2713},
archivePrefix = {arXiv},
       eprint = {1610.06595},
 primaryClass = {astro-ph.GA},
       adsurl = {https://ui.adsabs.harvard.edu/abs/2017MNRAS.464.4835O}
}

@ARTICLE{Banzatti2015,
       author = {{Banzatti}, A. and {Pontoppidan}, K.~M.},
        title = "{An Empirical Sequence of Disk Gap Opening Revealed by Rovibrational CO}",
      journal = {\apj},
         year = 2015,
        month = aug,
       volume = {809},
       number = {2},
          eid = {167},
        pages = {167},
          doi = {10.1088/0004-637X/809/2/167},
archivePrefix = {arXiv},
       eprint = {1507.06650},
 primaryClass = {astro-ph.EP},
       adsurl = {https://ui.adsabs.harvard.edu/abs/2015ApJ...809..167B}
}

@ARTICLE{Horne2012,
       author = {{Horne}, David and {Gibb}, Erika and {Rettig}, Terrence W. and {Brittain}, Sean and {Tilley}, David and {Balsara}, Dinshaw},
        title = "{The Gas/Dust Ratio of Circumstellar Disks: Testing Models of Planetesimal Formation}",
      journal = {\apj},
         year = 2012,
        month = jul,
       volume = {754},
       number = {1},
          eid = {64},
        pages = {64},
          doi = {10.1088/0004-637X/754/1/64},
       adsurl = {https://ui.adsabs.harvard.edu/abs/2012ApJ...754...64H}
}

@ARTICLE{Federman2024,
       author = {{Federman}, Samuel A. and {Megeath}, S. Thomas and {Rubinstein}, Adam E. and {Gutermuth}, Robert and {Narang}, Mayank and {Tyagi}, Himanshu and {Manoj}, P. and {Anglada}, Guillem and {Atnagulov}, Prabhani and {Beuther}, Henrik and {Bourke}, Tyler L. and {Brunken}, Nashanty and {Caratti o Garatti}, Alessio and {Evans}, Neal J. and {Fischer}, William J. and {Furlan}, Elise and {Green}, Joel D. and {Habel}, Nolan and {Hartmann}, Lee and {Karnath}, Nicole and {Klaassen}, Pamela and {Linz}, Hendrik and {Looney}, Leslie W. and {Osorio}, Mayra and {Muzerolle Page}, James and {Nazari}, Pooneh and {Pokhrel}, Riwaj and {Rahatgaonkar}, Rohan and {Rocha}, Will R.~M. and {Sheehan}, Patrick and {Slavicinska}, Katerina and {Stanke}, Thomas and {Stutz}, Amelia M. and {Tobin}, John J. and {Tychoniec}, Lukasz and {Van Dishoeck}, Ewine F. and {Watson}, Dan M. and {Wolk}, Scott and {Yang}, Yao-Lun},
        title = "{Investigating Protostellar Accretion-driven Outflows across the Mass Spectrum: JWST NIRSpec Integral Field Unit 3-5 {\ensuremath{\mu}}m Spectral Mapping of Five Young Protostars}",
      journal = {\apj},
         year = 2024,
        month = may,
       volume = {966},
       number = {1},
          eid = {41},
        pages = {41},
          doi = {10.3847/1538-4357/ad2fa0},
archivePrefix = {arXiv},
       eprint = {2310.03803},
 primaryClass = {astro-ph.SR},
       adsurl = {https://ui.adsabs.harvard.edu/abs/2024ApJ...966...41F}
}

@ARTICLE{Boogert2002,
       author = {{Boogert}, A.~C.~A. and {Hogerheijde}, M.~R. and {Blake}, G.~A.},
        title = "{High-Resolution 4.7 micron Keck/NIRSPEC Spectra of Protostars. I. Ices and Infalling Gas in the Disk of L1489 IRS}",
      journal = {\apj},
         year = 2002,
        month = apr,
       volume = {568},
       number = {2},
        pages = {761-770},
          doi = {10.1086/338969},
archivePrefix = {arXiv},
       eprint = {astro-ph/0112163},
 primaryClass = {astro-ph},
       adsurl = {https://ui.adsabs.harvard.edu/abs/2002ApJ...568..761B}
}

@ARTICLE{Bast2011,
       author = {{Bast}, J.~E. and {Brown}, J.~M. and {Herczeg}, G.~J. and {van Dishoeck}, E.~F. and {Pontoppidan}, K.~M.},
        title = "{Single peaked CO emission line profiles from the inner regions of protoplanetary disks}",
      journal = {\aap},
         year = 2011,
        month = mar,
       volume = {527},
          eid = {A119},
        pages = {A119},
          doi = {10.1051/0004-6361/201015225},
archivePrefix = {arXiv},
       eprint = {1101.1091},
 primaryClass = {astro-ph.SR},
       adsurl = {https://ui.adsabs.harvard.edu/abs/2011A&A...527A.119B}
}

@ARTICLE{Cleveland1979,
       author = {{Cleveland}, William S.},
        title = "{Robust Locally Weighted Regression and Smoothing Scatterplots}",
      journal = {Journal of the American Statistical Association},
         year = 1979,
        month = dec,
       volume = {74},
       number = {368},
        pages = {829-836},
          doi = {10.1080/01621459.1979.10481038}
}

@ARTICLE{Ryabchikova2015,
       author = {{Ryabchikova}, T. and {Piskunov}, N. and {Kurucz}, R.~L. and {Stempels}, H.~C. and {Heiter}, U. and {Pakhomov}, Yu and {Barklem}, P.~S.},
        title = "{A major upgrade of the VALD database}",
      journal = {\physscr},
         year = 2015,
        month = may,
       volume = {90},
       number = {5},
          eid = {054005},
        pages = {054005},
          doi = {10.1088/0031-8949/90/5/054005},
       adsurl = {https://ui.adsabs.harvard.edu/abs/2015PhyS...90e4005R}
}

@ARTICLE{Gordon2022,
       author = {{Gordon}, I.~E. and {Rothman}, L.~S. and {Hargreaves}, R.~J. and {Hashemi}, R. and {Karlovets}, E.~V. and {Skinner}, F.~M. and {Conway}, E.~K. and {Hill}, C. and {Kochanov}, R.~V. and {Tan}, Y. and {Wcis{\l}o}, P. and {Finenko}, A.~A. and {Nelson}, K. and {Bernath}, P.~F. and {Birk}, M. and {Boudon}, V. and {Campargue}, A. and {Chance}, K.~V. and {Coustenis}, A. and {Drouin}, B.~J. and {Flaud}, J.-M. and {Gamache}, R.~R. and {Hodges}, J.~T. and {Jacquemart}, D. and {Mlawer}, E.~J. and {Nikitin}, A.~V. and {Perevalov}, V.~I. and {Rotger}, M. and {Tennyson}, J. and {Toon}, G.~C. and {Tran}, H. and {Tyuterev}, V.~G. and {Adkins}, E.~M. and {Baker}, A. and {Barbe}, A. and {Can{\`e}}, E. and {Cs{\'a}sz{\'a}r}, A.~G. and {Dudaryonok}, A. and {Egorov}, O. and {Fleisher}, A.~J. and {Fleurbaey}, H. and {Foltynowicz}, A. and {Furtenbacher}, T. and {Harrison}, J.~J. and {Hartmann}, J.-M. and {Horneman}, V.-M. and {Huang}, X. and {Karman}, T. and {Karns}, J. and {Kassi}, S. and {Kleiner}, I. and {Kofman}, V. and {Kwabia-Tchana}, F. and {Lavrentieva}, N.~N. and {Lee}, T.~J. and {Long}, D.~A. and {Lukashevskaya}, A.~A. and {Lyulin}, O.~M. and {Makhnev}, V. Yu. and {Matt}, W. and {Massie}, S.~T. and {Melosso}, M. and {Mikhailenko}, S.~N. and {Mondelain}, D. and {M{\"u}ller}, H.~S.~P. and {Naumenko}, O.~V. and {Perrin}, A. and {Polyansky}, O.~L. and {Raddaoui}, E. and {Raston}, P.~L. and {Reed}, Z.~D. and {Rey}, M. and {Richard}, C. and {T{\'o}bi{\'a}s}, R. and {Sadiek}, I. and {Schwenke}, D.~W. and {Starikova}, E. and {Sung}, K. and {Tamassia}, F. and {Tashkun}, S.~A. and {Vander Auwera}, J. and {Vasilenko}, I.~A. and {Vigasin}, A.~A. and {Villanueva}, G.~L. and {Vispoel}, B. and {Wagner}, G. and {Yachmenev}, A. and {Yurchenko}, S.~N.},
        title = "{The HITRAN2020 molecular spectroscopic database}",
      journal = {\jqsrt},
         year = 2022,
        month = jan,
       volume = {277},
          eid = {107949},
        pages = {107949},
          doi = {10.1016/j.jqsrt.2021.107949},
       adsurl = {https://ui.adsabs.harvard.edu/abs/2022JQSRT.27707949G}
}

@ARTICLE{Bergner2024,
       author = {{Bergner}, Jennifer B. and {Sturm}, J.~A. and {Piacentino}, Elettra L. and {McClure}, M.~K. and {{\"O}berg}, Karin I. and {Boogert}, A.~C.~A. and {Dartois}, E. and {Drozdovskaya}, M.~N. and {Fraser}, H.~J. and {Harsono}, Daniel and {Ioppolo}, Sergio and {Law}, Charles J. and {Lis}, Dariusz C. and {McGuire}, Brett A. and {Melnick}, Gary J. and {Noble}, Jennifer A. and {Palumbo}, M.~E. and {Pendleton}, Yvonne J. and {Perotti}, Giulia and {Qasim}, Danna and {Rocha}, W.~R.~M. and {van Dishoeck}, E.~F.},
        title = "{JWST Ice Band Profiles Reveal Mixed Ice Compositions in the HH 48 NE Disk}",
      journal = {\apj},
         year = 2024,
        month = nov,
       volume = {975},
       number = {2},
          eid = {166},
        pages = {166},
          doi = {10.3847/1538-4357/ad79fc},
archivePrefix = {arXiv},
       eprint = {2409.08117},
 primaryClass = {astro-ph.EP},
       adsurl = {https://ui.adsabs.harvard.edu/abs/2024ApJ...975..166B}
}

@ARTICLE{Sturm2023,
       author = {{Sturm}, J.~A. and {McClure}, M.~K. and {Beck}, T.~L. and {Harsono}, D. and {Bergner}, J.~B. and {Dartois}, E. and {Boogert}, A.~C.~A. and {Chiar}, J.~E. and {Cordiner}, M.~A. and {Drozdovskaya}, M.~N. and {Ioppolo}, S. and {Law}, C.~J. and {Linnartz}, H. and {Lis}, D.~C. and {Melnick}, G.~J. and {McGuire}, B.~A. and {Noble}, J.~A. and {{\"O}berg}, K.~I. and {Palumbo}, M.~E. and {Pendleton}, Y.~J. and {Perotti}, G. and {Pontoppidan}, K.~M. and {Qasim}, D. and {Rocha}, W.~R.~M. and {Terada}, H. and {Urso}, R.~G. and {van Dishoeck}, E.~F.},
        title = "{A JWST inventory of protoplanetary disk ices. The edge-on protoplanetary disk HH 48 NE, seen with the Ice Age ERS program}",
      journal = {\aap},
         year = 2023,
        month = nov,
       volume = {679},
          eid = {A138},
        pages = {A138},
          doi = {10.1051/0004-6361/202347512},
archivePrefix = {arXiv},
       eprint = {2309.07817},
 primaryClass = {astro-ph.EP},
       adsurl = {https://ui.adsabs.harvard.edu/abs/2023A&A...679A.138S}
}

@ARTICLE{Gordon2023,
       author = {{Gordon}, Karl D. and {Clayton}, Geoffrey C. and {Decleir}, Marjorie and {Fitzpatrick}, E.~L. and {Massa}, Derck and {Misselt}, Karl A. and {Tollerud}, Erik J.},
        title = "{One Relation for All Wavelengths: The Far-ultraviolet to Mid-infrared Milky Way Spectroscopic R(V)-dependent Dust Extinction Relationship}",
      journal = {\apj},
         year = 2023,
        month = jun,
       volume = {950},
       number = {2},
          eid = {86},
        pages = {86},
          doi = {10.3847/1538-4357/accb59},
archivePrefix = {arXiv},
       eprint = {2304.01991},
 primaryClass = {astro-ph.GA},
       adsurl = {https://ui.adsabs.harvard.edu/abs/2023ApJ...950...86G}
}

@ARTICLE{Caffau2011,
       author = {{Caffau}, E. and {Ludwig}, H.-G. and {Steffen}, M. and {Freytag}, B. and {Bonifacio}, P.},
        title = "{Solar Chemical Abundances Determined with a CO5BOLD 3D Model Atmosphere}",
      journal = {\solphys},
         year = 2011,
        month = feb,
       volume = {268},
       number = {2},
        pages = {255-269},
          doi = {10.1007/s11207-010-9541-4},
archivePrefix = {arXiv},
       eprint = {1003.1190},
 primaryClass = {astro-ph.SR},
       adsurl = {https://ui.adsabs.harvard.edu/abs/2011SoPh..268..255C}
}

@ARTICLE{Allard2012,
       author = {{Allard}, F. and {Homeier}, D. and {Freytag}, B.},
        title = "{Models of very-low-mass stars, brown dwarfs and exoplanets}",
      journal = {Philosophical Transactions of the Royal Society of London Series A},
         year = 2012,
        month = jun,
       volume = {370},
       number = {1968},
        pages = {2765-2777},
          doi = {10.1098/rsta.2011.0269},
archivePrefix = {arXiv},
       eprint = {1112.3591},
 primaryClass = {astro-ph.SR},
       adsurl = {https://ui.adsabs.harvard.edu/abs/2012RSPTA.370.2765A}
}

@ARTICLE{Gehrels1986,
       author = {{Gehrels}, N.},
        title = "{Confidence Limits for Small Numbers of Events in Astrophysical Data}",
      journal = {\apj},
         year = 1986,
        month = apr,
       volume = {303},
        pages = {336},
          doi = {10.1086/164079},
       adsurl = {https://ui.adsabs.harvard.edu/abs/1986ApJ...303..336G}
}

@ARTICLE{Scargle2013,
       author = {{Scargle}, Jeffrey D. and {Norris}, Jay P. and {Jackson}, Brad and {Chiang}, James},
        title = "{Studies in Astronomical Time Series Analysis. VI. Bayesian Block Representations}",
      journal = {\apj},
         year = 2013,
        month = feb,
       volume = {764},
       number = {2},
          eid = {167},
        pages = {167},
          doi = {10.1088/0004-637X/764/2/167},
archivePrefix = {arXiv},
       eprint = {1207.5578},
 primaryClass = {astro-ph.IM},
       adsurl = {https://ui.adsabs.harvard.edu/abs/2013ApJ...764..167S}
}

@ARTICLE{Scargle1998,
       author = {{Scargle}, Jeffrey D.},
        title = "{Studies in Astronomical Time Series Analysis. V. Bayesian Blocks, a New Method to Analyze Structure in Photon Counting Data}",
      journal = {\apj},
         year = 1998,
        month = sep,
       volume = {504},
       number = {1},
        pages = {405-418},
          doi = {10.1086/306064},
archivePrefix = {arXiv},
       eprint = {astro-ph/9711233},
 primaryClass = {astro-ph},
       adsurl = {https://ui.adsabs.harvard.edu/abs/1998ApJ...504..405S}
}

@ARTICLE{Weisskopf2002,
       author = {{Weisskopf}, M.~C. and {Brinkman}, B. and {Canizares}, C. and {Garmire}, G. and {Murray}, S. and {Van Speybroeck}, L.~P.},
        title = "{An Overview of the Performance and Scientific Results from the Chandra X-Ray Observatory}",
      journal = {\pasp},
         year = 2002,
        month = jan,
       volume = {114},
       number = {791},
        pages = {1-24},
          doi = {10.1086/338108},
archivePrefix = {arXiv},
       eprint = {astro-ph/0110308},
 primaryClass = {astro-ph},
       adsurl = {https://ui.adsabs.harvard.edu/abs/2002PASP..114....1W}
}

@INPROCEEDINGS{Garmire2003,
       author = {{Garmire}, Gordon P. and {Bautz}, Mark W. and {Ford}, Peter G. and {Nousek}, John A. and {Ricker}, Jr., George R.},
        title = "{Advanced CCD imaging spectrometer (ACIS) instrument on the Chandra X-ray Observatory}",
    booktitle = {X-Ray and Gamma-Ray Telescopes and Instruments for Astronomy.},
         year = 2003,
       editor = {{Truemper}, Joachim E. and {Tananbaum}, Harvey D.},
       series = {Society of Photo-Optical Instrumentation Engineers (SPIE) Conference Series},
       volume = {4851},
        month = mar,
        pages = {28-44},
          doi = {10.1117/12.461599},
       adsurl = {https://ui.adsabs.harvard.edu/abs/2003SPIE.4851...28G}
}

@BOOK{Tukey1977,
       author = {{Tukey}, John W.},
        title = "{Exploratory data analysis}",
         year = 1977,
       adsurl = {https://ui.adsabs.harvard.edu/abs/1977eda..book.....T}
}

@ARTICLE{Robberto2020,
       author = {{Robberto}, Massimo and {Gennaro}, Mario and {Ubeira Gabellini}, Maria Giulia and {Hillenbrand}, Lynne A. and {Pacifici}, Camilla and {Ubeda}, Leonardo and {Andersen}, Morten and {Barman}, Travis and {Bellini}, Andrea and {Da Rio}, Nicola and {de Mink}, Selma E. and {Lodato}, Giuseppe and {Manara}, Carlo Felice and {Platais}, Imants and {Pueyo}, Laurent and {Strampelli}, Giovanni Maria and {Tan}, Jonathan C. and {Testi}, L.},
        title = "{HST Survey of the Orion Nebula Cluster in the H$_{2}$O 1.4 {\ensuremath{\mu}}m Absorption Band. I. A Census of Substellar and Planetary-mass Objects}",
      journal = {\apj},
         year = 2020,
        month = jun,
       volume = {896},
       number = {1},
          eid = {79},
        pages = {79},
          doi = {10.3847/1538-4357/ab911e},
archivePrefix = {arXiv},
       eprint = {2004.13915},
 primaryClass = {astro-ph.SR},
       adsurl = {https://ui.adsabs.harvard.edu/abs/2020ApJ...896...79R}
}

@ARTICLE{Luhman2024a,
       author = {{Luhman}, K.~L.},
        title = "{Candidates for Substellar Members of the Orion Nebula Cluster from JWST/NIRCam}",
      journal = {\aj},
         year = 2024,
        month = dec,
       volume = {168},
       number = {6},
          eid = {230},
        pages = {230},
          doi = {10.3847/1538-3881/ad812a},
archivePrefix = {arXiv},
       eprint = {2410.10406},
 primaryClass = {astro-ph.GA},
       adsurl = {https://ui.adsabs.harvard.edu/abs/2024AJ....168..230L}
}

@ARTICLE{Jakobsen2022,
       author = {{Jakobsen}, P. and {Ferruit}, P. and {Alves de Oliveira}, C. and {Arribas}, S. and {Bagnasco}, G. and {Barho}, R. and {Beck}, T.~L. and {Birkmann}, S. and {B{\"o}ker}, T. and {Bunker}, A.~J. and {Charlot}, S. and {de Jong}, P. and {de Marchi}, G. and {Ehrenwinkler}, R. and {Falcolini}, M. and {Fels}, R. and {Franx}, M. and {Franz}, D. and {Funke}, M. and {Giardino}, G. and {Gnata}, X. and {Holota}, W. and {Honnen}, K. and {Jensen}, P.~L. and {Jentsch}, M. and {Johnson}, T. and {Jollet}, D. and {Karl}, H. and {Kling}, G. and {K{\"o}hler}, J. and {Kolm}, M.-G. and {Kumari}, N. and {Lander}, M.~E. and {Lemke}, R. and {L{\'o}pez-Caniego}, M. and {L{\"u}tzgendorf}, N. and {Maiolino}, R. and {Manjavacas}, E. and {Marston}, A. and {Maschmann}, M. and {Maurer}, R. and {Messerschmidt}, B. and {Moseley}, S.~H. and {Mosner}, P. and {Mott}, D.~B. and {Muzerolle}, J. and {Pirzkal}, N. and {Pittet}, J.-F. and {Plitzke}, A. and {Posselt}, W. and {Rapp}, B. and {Rauscher}, B.~J. and {Rawle}, T. and {Rix}, H.-W. and {R{\"o}del}, A. and {Rumler}, P. and {Sabbi}, E. and {Salvignol}, J.-C. and {Schmid}, T. and {Sirianni}, M. and {Smith}, C. and {Strada}, P. and {te Plate}, M. and {Valenti}, J. and {Wettemann}, T. and {Wiehe}, T. and {Wiesmayer}, M. and {Willott}, C.~J. and {Wright}, R. and {Zeidler}, P. and {Zincke}, C.},
        title = "{The Near-Infrared Spectrograph (NIRSpec) on the James Webb Space Telescope. I. Overview of the instrument and its capabilities}",
      journal = {\aap},
         year = 2022,
        month = may,
       volume = {661},
          eid = {A80},
        pages = {A80},
          doi = {10.1051/0004-6361/202142663},
archivePrefix = {arXiv},
       eprint = {2202.03305},
 primaryClass = {astro-ph.IM},
       adsurl = {https://ui.adsabs.harvard.edu/abs/2022A&A...661A..80J}
}

@ARTICLE{Rigby2023,
       author = {{Rigby}, Jane and {Perrin}, Marshall and {McElwain}, Michael and {Kimble}, Randy and {Friedman}, Scott and {Lallo}, Matt and {Doyon}, Ren{\'e} and {Feinberg}, Lee and {Ferruit}, Pierre and {Glasse}, Alistair and {Rieke}, Marcia and {Rieke}, George and {Wright}, Gillian and {Willott}, Chris and {Colon}, Knicole and {Milam}, Stefanie and {Neff}, Susan and {Stark}, Christopher and {Valenti}, Jeff and {Abell}, Jim and {Abney}, Faith and {Abul-Huda}, Yasin and {Acton}, D. Scott and {Adams}, Evan and {Adler}, David and {Aguilar}, Jonathan and {Ahmed}, Nasif and {Albert}, Lo{\"\i}c and {Alberts}, Stacey and {Aldridge}, David and {Allen}, Marsha and {Altenburg}, Martin and {{\'A}lvarez-M{\'a}rquez}, Javier and {Alves de Oliveira}, Catarina and {Andersen}, Greg and {Anderson}, Harry and {Anderson}, Sara and {Argyriou}, Ioannis and {Armstrong}, Amber and {Arribas}, Santiago and {Artigau}, Etienne and {Arvai}, Amanda and {Atkinson}, Charles and {Bacon}, Gregory and {Bair}, Thomas and {Banks}, Kimberly and {Barrientes}, Jaclyn and {Barringer}, Bruce and {Bartosik}, Peter and {Bast}, William and {Baudoz}, Pierre and {Beatty}, Thomas and {Bechtold}, Katie and {Beck}, Tracy and {Bergeron}, Eddie and {Bergkoetter}, Matthew and {Bhatawdekar}, Rachana and {Birkmann}, Stephan and {Blazek}, Ronald and {Blome}, Claire and {Boccaletti}, Anthony and {B{\"o}ker}, Torsten and {Boia}, John and {Bonaventura}, Nina and {Bond}, Nicholas and {Bosley}, Kari and {Boucarut}, Ray and {Bourque}, Matthew and {Bouwman}, Jeroen and {Bower}, Gary and {Bowers}, Charles and {Boyer}, Martha and {Bradley}, Larry and {Brady}, Greg and {Braun}, Hannah and {Breda}, David and {Bresnahan}, Pamela and {Bright}, Stacey and {Britt}, Christopher and {Bromenschenkel}, Asa and {Brooks}, Brian and {Brooks}, Keira and {Brown}, Bob and {Brown}, Matthew and {Brown}, Patricia and {Bunker}, Andy and {Burger}, Matthew and {Bushouse}, Howard and {Cale}, Steven and {Cameron}, Alex and {Cameron}, Peter and {Canipe}, Alicia and {Caplinger}, James and {Caputo}, Francis and {Cara}, Mihai and {Carey}, Larkin and {Carniani}, Stefano and {Carrasquilla}, Maria and {Carruthers}, Margaret and {Case}, Michael and {Catherine}, Riggs and {Chance}, Don and {Chapman}, George and {Charlot}, St{\'e}phane and {Charlow}, Brian and {Chayer}, Pierre and {Chen}, Bin and {Cherinka}, Brian and {Chichester}, Sarah and {Chilton}, Zack and {Chonis}, Taylor and {Clampin}, Mark and {Clark}, Charles and {Clark}, Kerry and {Coe}, Dan and {Coleman}, Benee and {Comber}, Brian and {Comeau}, Tom and {Connolly}, Dennis and {Cooper}, James and {Cooper}, Rachel and {Coppock}, Eric and {Correnti}, Matteo and {Cossou}, Christophe and {Coulais}, Alain and {Coyle}, Laura and {Cracraft}, Misty and {Curti}, Mirko and {Cuturic}, Steven and {Davis}, Katherine and {Davis}, Michael and {Dean}, Bruce and {DeLisa}, Amy and {deMeester}, Wim and {Dencheva}, Nadia and {Dencheva}, Nadezhda and {DePasquale}, Joseph and {Deschenes}, Jeremy and {Hunor Detre}, {\"O}rs and {Diaz}, Rosa and {Dicken}, Dan and {DiFelice}, Audrey and {Dillman}, Matthew and {Dixon}, William and {Doggett}, Jesse and {Donaldson}, Tom and {Douglas}, Rob and {DuPrie}, Kimberly and {Dupuis}, Jean and {Durning}, John and {Easmin}, Nilufar and {Eck}, Weston and {Edeani}, Chinwe and {Egami}, Eiichi and {Ehrenwinkler}, Ralf and {Eisenhamer}, Jonathan and {Eisenhower}, Michael and {Elie}, Michelle and {Elliott}, James and {Elliott}, Kyle and {Ellis}, Tracy and {Engesser}, Michael and {Espinoza}, Nestor and {Etienne}, Odessa and {Etxaluze}, Mireya and {Falini}, Patrick and {Feeney}, Matthew and {Ferry}, Malcolm and {Filippazzo}, Joseph and {Fincham}, Brian and {Fix}, Mees and {Flagey}, Nicolas and {Florian}, Michael and {Flynn}, Jim and {Fontanella}, Erin and {Ford}, Terrance and {Forshay}, Peter and {Fox}, Ori and {Franz}, David and {Fu}, Henry and {Fullerton}, Alexander and {Galkin}, Sergey and {Galyer}, Anthony and {Garc{\'\i}a Mar{\'\i}n}, Macarena and {Gardner}, Jonathan P. and {Gardner}, Lisa and {Garland}, Dennis and {Garrett}, Bruce and {Gasman}, Danny and {Gaspar}, Andras and {Gaudreau}, Daniel and {Gauthier}, Peter and {Geers}, Vincent and {Geithner}, Paul and {Gennaro}, Mario and {Giardino}, Giovanna and {Girard}, Julien and {Giuliano}, Mark and {Glassmire}, Kirk and {Glauser}, Adrian},
        title = "{The Science Performance of JWST as Characterized in Commissioning}",
      journal = {\pasp},
         year = 2023,
        month = apr,
       volume = {135},
       number = {1046},
          eid = {048001},
        pages = {048001},
          doi = {10.1088/1538-3873/acb293},
archivePrefix = {arXiv},
       eprint = {2207.05632},
 primaryClass = {astro-ph.IM},
       adsurl = {https://ui.adsabs.harvard.edu/abs/2023PASP..135d8001R}
}

@ARTICLE{Gardner2006,
       author = {{Gardner}, Jonathan P. and {Mather}, John C. and {Clampin}, Mark and {Doyon}, Rene and {Greenhouse}, Matthew A. and {Hammel}, Heidi B. and {Hutchings}, John B. and {Jakobsen}, Peter and {Lilly}, Simon J. and {Long}, Knox S. and {Lunine}, Jonathan I. and {McCaughrean}, Mark J. and {Mountain}, Matt and {Nella}, John and {Rieke}, George H. and {Rieke}, Marcia J. and {Rix}, Hans-Walter and {Smith}, Eric P. and {Sonneborn}, George and {Stiavelli}, Massimo and {Stockman}, H.~S. and {Windhorst}, Rogier A. and {Wright}, Gillian S.},
        title = "{The James Webb Space Telescope}",
      journal = {\ssr},
         year = 2006,
        month = apr,
       volume = {123},
       number = {4},
        pages = {485-606},
          doi = {10.1007/s11214-006-8315-7},
archivePrefix = {arXiv},
       eprint = {astro-ph/0606175},
 primaryClass = {astro-ph},
       adsurl = {https://ui.adsabs.harvard.edu/abs/2006SSRv..123..485G}
}

@ARTICLE{McCaughrean2023,
       author = {{McCaughrean}, M.~J. and {Pearson}, S.~G.},
        title = "{A JWST survey of the Trapezium Cluster \& inner Orion Nebula. I. Observations \& overview}",
      journal = {arXiv e-prints},
         year = 2023,
        month = oct,
          eid = {arXiv:2310.03552},
        pages = {arXiv:2310.03552},
          doi = {10.48550/arXiv.2310.03552},
archivePrefix = {arXiv},
       eprint = {2310.03552},
 primaryClass = {astro-ph.SR},
       adsurl = {https://ui.adsabs.harvard.edu/abs/2023arXiv231003552M}
}

@ARTICLE{Brunn2023,
       author = {{Brunn}, V. and {Marcowith}, A. and {Sauty}, C. and {Padovani}, M. and {Rab}, Ch and {Meskini}, C.},
        title = "{Ionization of inner T Tauri star discs: effects of in situ energetic particles produced by strong magnetic reconnection events}",
      journal = {\mnras},
         year = 2023,
        month = mar,
       volume = {519},
       number = {4},
        pages = {5673-5688},
          doi = {10.1093/mnras/stac3792},
archivePrefix = {arXiv},
       eprint = {2210.10356},
 primaryClass = {astro-ph.SR},
       adsurl = {https://ui.adsabs.harvard.edu/abs/2023MNRAS.519.5673B}
}

@ARTICLE{Kamp2017,
       author = {{Kamp}, I. and {Thi}, W.-F. and {Woitke}, P. and {Rab}, C. and {Bouma}, S. and {M{\'e}nard}, F.},
        title = "{Consistent dust and gas models for protoplanetary disks. II. Chemical networks and rates}",
      journal = {\aap},
         year = 2017,
        month = nov,
       volume = {607},
          eid = {A41},
        pages = {A41},
          doi = {10.1051/0004-6361/201730388},
archivePrefix = {arXiv},
       eprint = {1707.07281},
 primaryClass = {astro-ph.SR},
       adsurl = {https://ui.adsabs.harvard.edu/abs/2017A&A...607A..41K}
}

@ARTICLE{Aresu2011,
       author = {{Aresu}, G. and {Kamp}, I. and {Meijerink}, R. and {Woitke}, P. and {Thi}, W.-F. and {Spaans}, M.},
        title = "{X-ray impact on the protoplanetary disks around T Tauri stars}",
      journal = {\aap},
         year = 2011,
        month = feb,
       volume = {526},
          eid = {A163},
        pages = {A163},
          doi = {10.1051/0004-6361/201015449},
archivePrefix = {arXiv},
       eprint = {1011.5623},
 primaryClass = {astro-ph.SR},
       adsurl = {https://ui.adsabs.harvard.edu/abs/2011A&A...526A.163A}
}

@ARTICLE{Woitke2016,
       author = {{Woitke}, P. and {Min}, M. and {Pinte}, C. and {Thi}, W.-F. and {Kamp}, I. and {Rab}, C. and {Anthonioz}, F. and {Antonellini}, S. and {Baldovin-Saavedra}, C. and {Carmona}, A. and {Dominik}, C. and {Dionatos}, O. and {Greaves}, J. and {G{\"u}del}, M. and {Ilee}, J.~D. and {Liebhart}, A. and {M{\'e}nard}, F. and {Rigon}, L. and {Waters}, L.~B.~F.~M. and {Aresu}, G. and {Meijerink}, R. and {Spaans}, M.},
        title = "{Consistent dust and gas models for protoplanetary disks. I. Disk shape, dust settling, opacities, and PAHs}",
      journal = {\aap},
         year = 2016,
        month = feb,
       volume = {586},
          eid = {A103},
        pages = {A103},
          doi = {10.1051/0004-6361/201526538},
archivePrefix = {arXiv},
       eprint = {1511.03431},
 primaryClass = {astro-ph.EP},
       adsurl = {https://ui.adsabs.harvard.edu/abs/2016A&A...586A.103W}
}

@ARTICLE{Rab2018,
       author = {{Rab}, Ch. and {G{\"u}del}, M. and {Woitke}, P. and {Kamp}, I. and {Thi}, W.-F. and {Min}, M. and {Aresu}, G. and {Meijerink}, R.},
        title = "{X-ray radiative transfer in protoplanetary disks. The role of dust and X-ray background fields}",
      journal = {\aap},
         year = 2018,
        month = jan,
       volume = {609},
          eid = {A91},
        pages = {A91},
          doi = {10.1051/0004-6361/201731443},
archivePrefix = {arXiv},
       eprint = {1711.07249},
 primaryClass = {astro-ph.SR},
       adsurl = {https://ui.adsabs.harvard.edu/abs/2018A&A...609A..91R}
}

@ARTICLE{Thi2013,
       author = {{Thi}, W.~F. and {Kamp}, I. and {Woitke}, P. and {van der Plas}, G. and {Bertelsen}, R. and {Wiesenfeld}, L.},
        title = "{Radiation thermo-chemical models of protoplanetary discs. IV. Modelling CO ro-vibrational emission from Herbig Ae discs}",
      journal = {\aap},
         year = 2013,
        month = mar,
       volume = {551},
          eid = {A49},
        pages = {A49},
          doi = {10.1051/0004-6361/201219210},
archivePrefix = {arXiv},
       eprint = {1210.7654},
 primaryClass = {astro-ph.GA},
       adsurl = {https://ui.adsabs.harvard.edu/abs/2013A&A...551A..49T}
}

@ARTICLE{Thi2011,
       author = {{Thi}, W.-F. and {Woitke}, P. and {Kamp}, I.},
        title = "{Radiation thermo-chemical models of protoplanetary discs - III. Impact of inner rims on spectral energy distributions}",
      journal = {\mnras},
         year = 2011,
        month = apr,
       volume = {412},
       number = {2},
        pages = {711-726},
          doi = {10.1111/j.1365-2966.2010.17741.x},
archivePrefix = {arXiv},
       eprint = {1009.4374},
 primaryClass = {astro-ph.GA},
       adsurl = {https://ui.adsabs.harvard.edu/abs/2011MNRAS.412..711T}
}

@ARTICLE{Kamp2010,
       author = {{Kamp}, I. and {Tilling}, I. and {Woitke}, P. and {Thi}, W.-F. and {Hogerheijde}, M.},
        title = "{Radiation thermo-chemical models of protoplanetary disks. II. Line diagnostics}",
      journal = {\aap},
         year = 2010,
        month = jan,
       volume = {510},
          eid = {A18},
        pages = {A18},
          doi = {10.1051/0004-6361/200913076},
archivePrefix = {arXiv},
       eprint = {0911.1949},
 primaryClass = {astro-ph.SR},
       adsurl = {https://ui.adsabs.harvard.edu/abs/2010A&A...510A..18K}
}

@ARTICLE{Woitke2009,
       author = {{Woitke}, P. and {Kamp}, I. and {Thi}, W.-F.},
        title = "{Radiation thermo-chemical models of protoplanetary disks. I. Hydrostatic disk structure and inner rim}",
      journal = {\aap},
         year = 2009,
        month = jul,
       volume = {501},
       number = {1},
        pages = {383-406},
          doi = {10.1051/0004-6361/200911821},
archivePrefix = {arXiv},
       eprint = {0904.0334},
 primaryClass = {astro-ph.EP},
       adsurl = {https://ui.adsabs.harvard.edu/abs/2009A&A...501..383W}
}

@ARTICLE{Frediani2026,
       author = {{Frediani}, Jenny and {Getman}, Konstantin V. and {Woitke}, Peter and {Portilla-Revelo}, Bayron and {Feigelson}, Eric D. and {Rab}, Christian and {Bik}, Arjan and {Ram{\'\i}rez-Tannus}, Mar{\'\i}a Claudia and {Waters}, Rens and {Henning}, Thomas and {Kamp}, Inga and {Chaparro}, Germ{\'a}n and {Hern{\'a}ndez}, Sebasti{\'a}n and {Rodr{\'\i}guez}, Sim{\'o}n and {Cuartas-Restrepo}, Pablo and {Winter}, Andrew and {Haworth}, Thomas J. and {Preibisch}, Thomas and {van Terwisga}, Sierk E. and {Roccatagliata}, Veronica and {Brandeker}, Alexis},
        title = "{XUE. ProDiMo models of internally and externally irradiated planet-forming disks around 0.3-4.0 solar mass stars (The IRIS project I)}",
      journal = {arXiv e-prints},
         year = 2026,
        month = sep,
          eid = {arXiv:2609.08621},
        pages = {arXiv:2609.08621},
          doi = {10.48550/arXiv.2609.08621},
archivePrefix = {arXiv},
       eprint = {2609.08621},
 primaryClass = {astro-ph.EP},
       adsurl = {https://ui.adsabs.harvard.edu/abs/2026arXiv260908621F}
}

@ARTICLE{PortillaRevelo2026,
       author = {{Portilla-Revelo}, Bayron and {Getman}, Konstantin V. and {K{\'o}sp{\'a}l}, {\'A}gnes and {Semenov}, Dmitry and {Akimkin}, Vitaly and {Zwicky}, Lis and {Forbrich}, Jan and {van Terwisga}, Sierk},
        title = "{The Effect of X-Ray and Accretion Variability on Mid-Infrared Lines in the Young Disk-Bearing Binary DQ Tau}",
      journal = {arXiv e-prints},
         year = 2026,
        month = sep,
          eid = {arXiv:2609.06863},
        pages = {arXiv:2609.06863},
          doi = {10.48550/arXiv.2609.06863},
archivePrefix = {arXiv},
       eprint = {2609.06863},
 primaryClass = {astro-ph.EP},
       adsurl = {https://ui.adsabs.harvard.edu/abs/2026arXiv260906863P}
}

@ARTICLE{Getman2026b,
       author = {{Getman}, Konstantin V. and {Feigelson}, Eric D. and {Garmire}, Gordon P.},
        title = "{Probing Convective Dynamo Activity in Young Suns with the Chandra X-ray Observatory}",
      journal = {Transactions of the International Astronomical Union, Series A},
         year = 2026,
        month = aug,
       volume = {20},
       number = {A32},
        pages = {353-359},
          doi = {10.1017/S1743921324003284},
       adsurl = {https://ui.adsabs.harvard.edu/abs/2026IAUTA..32..353G}
}

@ARTICLE{Getman2026a,
       author = {{Getman}, Konstantin V. and {Feigelson}, Eric D. and {Airapetian}, Vladimir S. and {Garmire}, Gordon P.},
        title = "{X-Ray Evolution of Young Stars: Early Dimming and Coronal Softening in Solar-mass Stars with Implications for Planetary Atmospheres}",
      journal = {\apj},
         year = 2026,
        month = apr,
       volume = {1001},
       number = {2},
          eid = {133},
        pages = {133},
          doi = {10.3847/1538-4357/ae2e00},
archivePrefix = {arXiv},
       eprint = {2512.12055},
 primaryClass = {astro-ph.SR},
       adsurl = {https://ui.adsabs.harvard.edu/abs/2026ApJ..1001..133G}
}

@ARTICLE{Getman2026c,
       author = {{Getman}, Konstantin V. and
                 {Ram{\'i}rez-Tannus}, Mar{\'i}a Claudia and
                 {Frediani}, Jenny and
                 {Luhman}, Kevin L. and
                 {Feigelson}, Eric D. and
                 {Portilla-Revelo}, Bayron and
                 {Rab}, Christian and
                 {Brunn}, Valentin and
                 {Forbrich}, Jan and
                 {Haworth}, Thomas J. and
                 {Waters}, Rens and
                 {Bik}, Arjan and
                 {Law}, Charles J. and
                 {Winter}, Andrew J. and
                 {Airapetian}, Vladimir S. and
                 {Waggoner}, Abygail R. and
                 {Kuhn}, Michael A. and
                 {Cleeves}, L. Ilsedore and
                 {Dzib}, Sergio A. and
                 {Padovani}, Marco},
        title = "{Tracing Accretion and Inner-Disk Gas Evolution in the Orion Nebula via Br$\alpha$ and CO}",
      journal = {\apj},
         year = 2026,
         note = {submitted}
}

@ARTICLE{Getman2025,
       author = {{Getman}, Konstantin V. and {Kochukhov}, Oleg and {Ninan}, Joe P. and {Feigelson}, Eric D. and {Airapetian}, Vladimir S. and {Waggoner}, Abygail R. and {Cleeves}, L. Ilsedore and {Forbrich}, Jan and {Dzib}, Sergio A. and {Law}, Charles J. and {Rab}, Christian and {Krolikowski}, Daniel M.},
        title = "{Multi-observatory Study of Young Stellar Energetic Flares (MORYSEF): No Evidence for Abnormally Strong Stellar Magnetic Fields after Powerful X-Ray Flares}",
      journal = {\apj},
         year = 2025,
        month = feb,
       volume = {980},
       number = {1},
          eid = {57},
        pages = {57},
          doi = {10.3847/1538-4357/ada55c},
archivePrefix = {arXiv},
       eprint = {2501.01321},
 primaryClass = {astro-ph.SR},
       adsurl = {https://ui.adsabs.harvard.edu/abs/2025ApJ...980...57G}
}

@ARTICLE{Getman2024,
       author = {{Getman}, Konstantin V. and {Feigelson}, Eric D. and {Waggoner}, Abygail R. and {Cleeves}, L. Ilsedore and {Forbrich}, Jan and {Ninan}, Joe P. and {Kochukhov}, Oleg and {Airapetian}, Vladimir S. and {Dzib}, Sergio A. and {Law}, Charles J. and {Rab}, Christian},
        title = "{Multi-Observatory Research of Young Stellar Energetic Flares (MORYSEF): X-Ray-flare-related Phenomena and Multi-epoch Behavior}",
      journal = {\apj},
         year = 2024,
        month = dec,
       volume = {976},
       number = {2},
          eid = {195},
        pages = {195},
          doi = {10.3847/1538-4357/ad8562},
archivePrefix = {arXiv},
       eprint = {2410.05126},
 primaryClass = {astro-ph.SR},
       adsurl = {https://ui.adsabs.harvard.edu/abs/2024ApJ...976..195G}
}

@ARTICLE{GetmanFeigelson2021,
       author = {{Getman}, Konstantin V. and {Feigelson}, Eric D.},
        title = "{X-Ray Superflares from Pre-main-sequence Stars: Flare Energetics and Frequency}",
      journal = {\apj},
         year = 2021,
        month = jul,
       volume = {916},
       number = {1},
          eid = {32},
        pages = {32},
          doi = {10.3847/1538-4357/ac00be},
archivePrefix = {arXiv},
       eprint = {2105.04768},
 primaryClass = {astro-ph.SR},
       adsurl = {https://ui.adsabs.harvard.edu/abs/2021ApJ...916...32G}
}

@ARTICLE{Ilgner2006,
       author = {{Ilgner}, M. and {Nelson}, R.~P.},
        title = "{On the ionisation fraction in protoplanetary disks. III. The effect of X-ray flares on gas-phase chemistry}",
      journal = {\aap},
         year = 2006,
        month = aug,
       volume = {455},
       number = {2},
        pages = {731-740},
          doi = {10.1051/0004-6361:20065308},
archivePrefix = {arXiv},
       eprint = {astro-ph/0605290},
 primaryClass = {astro-ph},
       adsurl = {https://ui.adsabs.harvard.edu/abs/2006A&A...455..731I}
}

@ARTICLE{Washinoue2024,
       author = {{Washinoue}, Haruka and {Takasao}, Shinsuke and {Furuya}, Kenji},
        title = "{Effect of Time-varying X-Ray Emission from Stellar Flares on the Ionization of Protoplanetary Disks}",
      journal = {\apj},
         year = 2024,
        month = nov,
       volume = {976},
       number = {1},
          eid = {25},
        pages = {25},
          doi = {10.3847/1538-4357/ad7fdf},
archivePrefix = {arXiv},
       eprint = {2406.04946},
 primaryClass = {astro-ph.SR},
       adsurl = {https://ui.adsabs.harvard.edu/abs/2024ApJ...976...25W}
}

@ARTICLE{Brunn2024,
       author = {{Brunn}, V. and {Rab}, Ch and {Marcowith}, A. and {Sauty}, C. and {Padovani}, M. and {Meskini}, C.},
        title = "{Impacts of energetic particles from T Tauri flares on inner protoplanetary discs}",
      journal = {\mnras},
         year = 2024,
        month = jun,
       volume = {530},
       number = {4},
        pages = {3669-3687},
          doi = {10.1093/mnras/stae1105},
archivePrefix = {arXiv},
       eprint = {2404.16459},
 primaryClass = {astro-ph.HE},
       adsurl = {https://ui.adsabs.harvard.edu/abs/2024MNRAS.530.3669B}
}

@ARTICLE{Waggoner2022,
       author = {{Waggoner}, Abygail R. and {Cleeves}, L. Ilsedore},
        title = "{Classification of X-Ray Flare-driven Chemical Variability in Protoplanetary Disks}",
      journal = {\apj},
         year = 2022,
        month = mar,
       volume = {928},
       number = {1},
          eid = {46},
        pages = {46},
          doi = {10.3847/1538-4357/ac549f},
archivePrefix = {arXiv},
       eprint = {2202.06962},
 primaryClass = {astro-ph.SR},
       adsurl = {https://ui.adsabs.harvard.edu/abs/2022ApJ...928...46W}
}

@ARTICLE{Rab2017,
       author = {{Rab}, Ch. and {G{\"u}del}, M. and {Padovani}, M. and {Kamp}, I. and {Thi}, W.-F. and {Woitke}, P. and {Aresu}, G.},
        title = "{Stellar energetic particle ionization in protoplanetary disks around T Tauri stars}",
      journal = {\aap},
         year = 2017,
        month = jul,
       volume = {603},
          eid = {A96},
        pages = {A96},
          doi = {10.1051/0004-6361/201630241},
archivePrefix = {arXiv},
       eprint = {1702.07522},
 primaryClass = {astro-ph.SR},
       adsurl = {https://ui.adsabs.harvard.edu/abs/2017A&A...603A..96R}
}

@ARTICLE{Hu2022,
       author = {{Hu}, Junxiang and {Airapetian}, Vladimir S. and {Li}, Gang and {Zank}, Gary and {Jin}, Meng},
        title = "{Extreme energetic particle events by superflare-asssociated CMEs from solar-like stars}",
      journal = {Science Advances},
         year = 2022,
        month = mar,
       volume = {8},
       number = {12},
          eid = {eabi9743},
        pages = {eabi9743},
          doi = {10.1126/sciadv.abi9743},
       adsurl = {https://ui.adsabs.harvard.edu/abs/2022SciA....8I9743H}
}

@INPROCEEDINGS{Airapetian2023,
       author = {{Airapetian}, Vladimir and {Jin}, Meng and {Hu}, Junxiang and {Lynch}, Benjamin and {Green}, James and {France}, Kevin and {Paudel}, Rishi and {Danchi}, William and {Chen}, Thomas and {Rau}, Gioia},
        title = "{Expanding The Frontiers: From The Solar Corona and The Wind to Evolving Space Weather of Planet Hosting Stars}",
    booktitle = {Bulletin of the American Astronomical Society},
         year = 2023,
       volume = {55},
        month = jul,
          eid = {005},
        pages = {005},
          doi = {10.3847/25c2cfeb.41e10693},
       adsurl = {https://ui.adsabs.harvard.edu/abs/2023BAAS...55c.005A}
}

@ARTICLE{Hazra2022,
       author = {{Hazra}, Gopal and {Vidotto}, Aline A. and {Carolan}, Stephen and {Villarreal D'Angelo}, Carolina and {Manchester}, Ward},
        title = "{The impact of coronal mass ejections and flares on the atmosphere of the hot Jupiter HD189733b}",
      journal = {\mnras},
         year = 2022,
        month = feb,
       volume = {509},
       number = {4},
        pages = {5858-5871},
          doi = {10.1093/mnras/stab3271},
archivePrefix = {arXiv},
       eprint = {2111.04531},
 primaryClass = {astro-ph.EP},
       adsurl = {https://ui.adsabs.harvard.edu/abs/2022MNRAS.509.5858H}
}

@ARTICLE{AlvaradoGomez2022,
       author = {{Alvarado-G{\'o}mez}, Juli{\'a}n D. and {Cohen}, Ofer and {Drake}, Jeremy J. and {Fraschetti}, Federico and {Poppenhaeger}, Katja and {Garraffo}, Cecilia and {Chebly}, Judy and {Ilin}, Ekaterina and {Harbach}, Laura and {Kochukhov}, Oleg},
        title = "{Simulating the Space Weather in the AU Mic System: Stellar Winds and Extreme Coronal Mass Ejections}",
      journal = {\apj},
         year = 2022,
        month = apr,
       volume = {928},
       number = {2},
          eid = {147},
        pages = {147},
          doi = {10.3847/1538-4357/ac54b8},
archivePrefix = {arXiv},
       eprint = {2202.07949},
 primaryClass = {astro-ph.SR},
       adsurl = {https://ui.adsabs.harvard.edu/abs/2022ApJ...928..147A}
}

@ARTICLE{Airapetian2020,
       author = {{Airapetian}, V.~S. and {Barnes}, R. and {Cohen}, O. and {Collinson}, G.~A. and {Danchi}, W.~C. and {Dong}, C.~F. and {Del Genio}, A.~D. and {France}, K. and {Garcia-Sage}, K. and {Glocer}, A. and {Gopalswamy}, N. and {Grenfell}, J.~L. and {Gronoff}, G. and {G{\"u}del}, M. and {Herbst}, K. and {Henning}, W.~G. and {Jackman}, C.~H. and {Jin}, M. and {Johnstone}, C.~P. and {Kaltenegger}, L. and {Kay}, C.~D. and {Kobayashi}, K. and {Kuang}, W. and {Li}, G. and {Lynch}, B.~J. and {L{\"u}ftinger}, T. and {Luhmann}, J.~G. and {Maehara}, H. and {Mlynczak}, M.~G. and {Notsu}, Y. and {Osten}, R.~A. and {Ramirez}, R.~M. and {Rugheimer}, S. and {Scheucher}, M. and {Schlieder}, J.~E. and {Shibata}, K. and {Sousa-Silva}, C. and {Stamenkovi{\'c}}, V. and {Strangeway}, R.~J. and {Usmanov}, A.~V. and {Vergados}, P. and {Verkhoglyadova}, O.~P. and {Vidotto}, A.~A. and {Voytek}, M. and {Way}, M.~J. and {Zank}, G.~P. and {Yamashiki}, Y.},
        title = "{Impact of space weather on climate and habitability of terrestrial-type exoplanets}",
      journal = {International Journal of Astrobiology},
         year = 2020,
        month = apr,
       volume = {19},
       number = {2},
        pages = {136-194},
          doi = {10.1017/S1473550419000132},
archivePrefix = {arXiv},
       eprint = {1905.05093},
 primaryClass = {astro-ph.EP},
       adsurl = {https://ui.adsabs.harvard.edu/abs/2020IJAsB..19..136A}
}

@ARTICLE{Gorti2009,
       author = {{Gorti}, U. and {Hollenbach}, D.},
        title = "{Photoevaporation of Circumstellar Disks By Far-Ultraviolet, Extreme-Ultraviolet and X-Ray Radiation from the Central Star}",
      journal = {\apj},
         year = 2009,
        month = jan,
       volume = {690},
       number = {2},
        pages = {1539-1552},
          doi = {10.1088/0004-637X/690/2/1539},
archivePrefix = {arXiv},
       eprint = {0809.1494},
 primaryClass = {astro-ph},
       adsurl = {https://ui.adsabs.harvard.edu/abs/2009ApJ...690.1539G}
}

@ARTICLE{Waggoner2023,
       author = {{Waggoner}, Abygail R. and {Cleeves}, L. Ilsedore and {Loomis}, Ryan A. and {Aikawa}, Yuri and {Bae}, Jaehan and {Bergner}, Jennifer B. and {Booth}, Alice S. and {Calahan}, Jenny K. and {Cataldi}, Gianni and {Law}, Charles J. and {Le Gal}, Romane and {Long}, Feng and {{\"O}berg}, Karin I. and {Teague}, Richard and {Wilner}, David J.},
        title = "{MAPS: Constraining Serendipitous Time Variability in Protoplanetary Disk Molecular Ion Emission}",
      journal = {\apj},
         year = 2023,
        month = oct,
       volume = {956},
       number = {2},
          eid = {103},
        pages = {103},
          doi = {10.3847/1538-4357/acf315},
archivePrefix = {arXiv},
       eprint = {2308.11699},
 primaryClass = {astro-ph.EP},
       adsurl = {https://ui.adsabs.harvard.edu/abs/2023ApJ...956..103W}
}

@ARTICLE{Espaillat2023,
       author = {{Espaillat}, C.~C. and {Thanathibodee}, T. and {Pittman}, C.~V. and {Sturm}, J.~A. and {McClure}, M.~K. and {Calvet}, N. and {Walter}, F.~M. and {Franco-Hern{\'a}ndez}, R. and {Muzerolle Page}, J.},
        title = "{JWST Detects Neon Line Variability in a Protoplanetary Disk}",
      journal = {\apjl},
         year = 2023,
        month = nov,
       volume = {958},
       number = {1},
          eid = {L4},
        pages = {L4},
          doi = {10.3847/2041-8213/ad023d},
archivePrefix = {arXiv},
       eprint = {2311.07739},
 primaryClass = {astro-ph.SR},
       adsurl = {https://ui.adsabs.harvard.edu/abs/2023ApJ...958L...4E}
}

@ARTICLE{Cleeves2017,
       author = {{Cleeves}, L. Ilsedore and {Bergin}, Edwin A. and {{\"O}berg}, Karin I. and {Andrews}, Sean and {Wilner}, David and {Loomis}, Ryan},
        title = "{Variable H$^{13}$CO$^{+}$ Emission in the IM Lup Disk: X-Ray Driven Time-dependent Chemistry?}",
      journal = {\apjl},
         year = 2017,
        month = jul,
       volume = {843},
       number = {1},
          eid = {L3},
        pages = {L3},
          doi = {10.3847/2041-8213/aa76e2},
archivePrefix = {arXiv},
       eprint = {1706.00833},
 primaryClass = {astro-ph.SR},
       adsurl = {https://ui.adsabs.harvard.edu/abs/2017ApJ...843L...3C}
}

@ARTICLE{Johansen2011,
       author = {{Johansen}, A. and {Klahr}, H. and {Henning}, Th.},
        title = "{High-resolution simulations of planetesimal formation in turbulent protoplanetary discs}",
      journal = {\aap},
         year = 2011,
        month = may,
       volume = {529},
          eid = {A62},
        pages = {A62},
          doi = {10.1051/0004-6361/201015979},
archivePrefix = {arXiv},
       eprint = {1010.4757},
 primaryClass = {astro-ph.EP},
       adsurl = {https://ui.adsabs.harvard.edu/abs/2011A&A...529A..62J}
}

@ARTICLE{Lim2024,
       author = {{Lim}, Jeonghoon and {Simon}, Jacob B. and {Li}, Rixin and {Armitage}, Philip J. and {Carrera}, Daniel and {Lyra}, Wladimir and {Rea}, David G. and {Yang}, Chao-Chin and {Youdin}, Andrew N.},
        title = "{Streaming Instability and Turbulence: Conditions for Planetesimal Formation}",
      journal = {\apj},
         year = 2024,
        month = jul,
       volume = {969},
       number = {2},
          eid = {130},
        pages = {130},
          doi = {10.3847/1538-4357/ad47a2},
archivePrefix = {arXiv},
       eprint = {2312.12508},
 primaryClass = {astro-ph.EP},
       adsurl = {https://ui.adsabs.harvard.edu/abs/2024ApJ...969..130L}
}

@ARTICLE{Espaillat2013,
       author = {{Espaillat}, C. and {Ingleby}, L. and {Furlan}, E. and {McClure}, M. and {Spatzier}, A. and {Nieusma}, J. and {Calvet}, N. and {Bergin}, E. and {Hartmann}, L. and {Miller}, J.~M. and {Muzerolle}, J.},
        title = "{Tracing High-energy Radiation from T Tauri Stars Using Mid-infrared Neon Emission from Disks}",
      journal = {\apj},
         year = 2013,
        month = jan,
       volume = {762},
       number = {1},
          eid = {62},
        pages = {62},
          doi = {10.1088/0004-637X/762/1/62},
archivePrefix = {arXiv},
       eprint = {1211.2335},
 primaryClass = {astro-ph.SR},
       adsurl = {https://ui.adsabs.harvard.edu/abs/2013ApJ...762...62E}
}

@INPROCEEDINGS{Glassgold2000,
       author = {{Glassgold}, A.~E. and {Feigelson}, E.~D. and {Montmerle}, T.},
        title = "{Effects of Energetic Radiation in Young Stellar Objects}",
    booktitle = {Protostars and Planets IV},
         year = 2000,
       editor = {{Mannings}, V. and {Boss}, A.~P. and {Russell}, S.~S.},
        month = may,
        pages = {429},
       adsurl = {https://ui.adsabs.harvard.edu/abs/2000prpl.conf..429G}
}

@ARTICLE{Glassgold2007,
       author = {{Glassgold}, Alfred E. and {Najita}, Joan R. and {Igea}, Javier},
        title = "{Neon Fine-Structure Line Emission by X-Ray Irradiated Protoplanetary Disks}",
      journal = {\apj},
         year = 2007,
        month = feb,
       volume = {656},
       number = {1},
        pages = {515-523},
          doi = {10.1086/510013},
archivePrefix = {arXiv},
       eprint = {astro-ph/0611094},
 primaryClass = {astro-ph},
       adsurl = {https://ui.adsabs.harvard.edu/abs/2007ApJ...656..515G}
}

@ARTICLE{Owen2019,
       author = {{Owen}, James E.},
        title = "{Atmospheric Escape and the Evolution of Close-In Exoplanets}",
      journal = {Annual Review of Earth and Planetary Sciences},
         year = 2019,
        month = may,
       volume = {47},
        pages = {67-90},
          doi = {10.1146/annurev-earth-053018-060246},
archivePrefix = {arXiv},
       eprint = {1807.07609},
 primaryClass = {astro-ph.EP},
       adsurl = {https://ui.adsabs.harvard.edu/abs/2019AREPS..47...67O}
}

@ARTICLE{Ercolano2021,
       author = {{Ercolano}, Barbara and {Picogna}, Giovanni and {Monsch}, Kristina and {Drake}, Jeremy J. and {Preibisch}, Thomas},
        title = "{The dispersal of protoplanetary discs - II: photoevaporation models with observationally derived irradiating spectra}",
      journal = {\mnras},
         year = 2021,
        month = dec,
       volume = {508},
       number = {2},
        pages = {1675-1685},
          doi = {10.1093/mnras/stab2590},
archivePrefix = {arXiv},
       eprint = {2109.04113},
 primaryClass = {astro-ph.EP},
       adsurl = {https://ui.adsabs.harvard.edu/abs/2021MNRAS.508.1675E}
}

@ARTICLE{Woitke2024,
       author = {{Woitke}, P. and {Thi}, W.-F. and {Arabhavi}, A.~M. and {Kamp}, I. and {K{\'o}sp{\'a}l}, {\'A}. and {{\'A}brah{\'a}m}, P.},
        title = "{2D disc modelling of the JWST line spectrum of EX Lupi}",
      journal = {\aap},
         year = 2024,
        month = mar,
       volume = {683},
          eid = {A219},
        pages = {A219},
          doi = {10.1051/0004-6361/202347730},
archivePrefix = {arXiv},
       eprint = {2311.18321},
 primaryClass = {astro-ph.EP},
       adsurl = {https://ui.adsabs.harvard.edu/abs/2024A&A...683A.219W}
}

@MANUAL{RCoreTeam2026,
       author = {{R Core Team}},
        title = "{R: A Language and Environment for Statistical Computing}",
 organization = {R Foundation for Statistical Computing},
      address = {Vienna, Austria},
         year = 2026,
          url = {https://www.R-project.org/}
}

@INPROCEEDINGS{Fruscione2006,
       author = {{Fruscione}, Antonella and {McDowell}, Jonathan C. and {Allen}, Glenn E. and {Brickhouse}, Nancy S. and {Burke}, Douglas J. and {Davis}, John E. and {Durham}, Nick and {Elvis}, Martin and {Galle}, Elizabeth C. and {Harris}, Daniel E. and {Huenemoerder}, David P. and {Houck}, John C. and {Ishibashi}, Bish and {Karovska}, Margarita and {Nicastro}, Fabrizio and {Noble}, Michael S. and {Nowak}, Michael A. and {Primini}, Frank A. and {Siemiginowska}, Aneta and {Smith}, Randall K. and {Wise}, Michael},
        title = "{CIAO: Chandra's data analysis system}",
    booktitle = {Observatory Operations: Strategies, Processes, and Systems},
         year = 2006,
       editor = {{Silva}, David R. and {Doxsey}, Rodger E.},
       series = {Society of Photo-Optical Instrumentation Engineers (SPIE) Conference Series},
       volume = {6270},
        month = jun,
          eid = {62701V},
        pages = {62701V},
          doi = {10.1117/12.671760},
       adsurl = {https://ui.adsabs.harvard.edu/abs/2006SPIE.6270E..1VF}
}

@MISC{Bushouse2025,
       author = {{Bushouse}, Howard and {Eisenhamer}, Jonathan and
                 {Dencheva}, Nadia and {Davies}, James and
                 {Greenfield}, Perry and {Morrison}, Jane and
                 {Hodge}, Phil and {Simon}, Bernie and
                 {Grumm}, David and {Droettboom}, Michael and
                 {Slavich}, Edward and {Sosey}, Megan and
                 {Pauly}, Tyler and {Miller}, Todd and
                 {Jedrzejewski}, Robert and {Hack}, Warren and
                 {Davis}, David and {Crawford}, Steven and
                 {Law}, David and {Gordon}, Karl and
                 {Regan}, Michael and {Cara}, Mihai and
                 {MacDonald}, Ken and {Bradley}, Larry and
                 {Shanahan}, Clare and {Jamieson}, William and
                 {Teodoro}, Mairan and {Williams}, Thomas and
                 {Pena-Guerrero}, Maria and {Graham}, Brett and
                 {Molter}, Edward and {Brandt}, Timothy and
                 {Hayes}, Christian and {Cooper}, Rachel and
                 {Clarke}, Melanie and {Filippazzo}, Joseph},
        title = "{JWST Calibration Pipeline}",
         year = 2025,
        month = nov,
      version = {1.20.2},
          doi = {10.5281/zenodo.17515973},
    publisher = {Zenodo}
}

@software{Broos2012,
       author = {{Broos}, Patrick and {Townsley}, Leisa and {Getman}, Konstantin and {Bauer}, Franz},
        title = "{AE: ACIS Extract}",
 howpublished = {Astrophysics Source Code Library, record ascl:1203.001},
         year = 2012,
        month = mar,
          eid = {ascl:1203.001},
archivePrefix = {ascl},
       eprint = {1203.001},
       adsurl = {https://ui.adsabs.harvard.edu/abs/2012ascl.soft03001B}
}

@ARTICLE{Broos2010,
       author = {{Broos}, Patrick S. and {Townsley}, Leisa K. and {Feigelson}, Eric D. and {Getman}, Konstantin V. and {Bauer}, Franz E. and {Garmire}, Gordon P.},
        title = "{Innovations in the Analysis of Chandra-ACIS Observations}",
      journal = {\apj},
         year = 2010,
        month = may,
       volume = {714},
       number = {2},
        pages = {1582-1605},
          doi = {10.1088/0004-637X/714/2/1582},
archivePrefix = {arXiv},
       eprint = {1003.2397},
 primaryClass = {astro-ph.HE},
       adsurl = {https://ui.adsabs.harvard.edu/abs/2010ApJ...714.1582B}
}

@ARTICLE{Getman2022,
       author = {{Getman}, Konstantin V. and {Feigelson}, Eric D. and {Garmire}, Gordon P. and {Broos}, Patrick S. and {Kuhn}, Michael A. and {Preibisch}, Thomas and {Airapetian}, Vladimir S.},
        title = "{Evolution of X-Ray Activity in <25 Myr Old Pre-main Sequence Stars}",
      journal = {\apj},
         year = 2022,
        month = aug,
       volume = {935},
       number = {1},
          eid = {43},
        pages = {43},
          doi = {10.3847/1538-4357/ac7c69},
archivePrefix = {arXiv},
       eprint = {2203.02047},
 primaryClass = {astro-ph.SR},
       adsurl = {https://ui.adsabs.harvard.edu/abs/2022ApJ...935...43G}
}

@INPROCEEDINGS{Alexander2014,
       author = {{Alexander}, R. and {Pascucci}, I. and {Andrews}, S. and {Armitage}, P. and {Cieza}, L.},
        title = "{The Dispersal of Protoplanetary Disks}",
    booktitle = {Protostars and Planets VI},
         year = 2014,
       editor = {{Beuther}, Henrik and {Klessen}, Ralf S. and {Dullemond}, Cornelis P. and {Henning}, Thomas},
        month = jan,
        pages = {475-496},
          doi = {10.2458/azu_uapress_9780816531240-ch021},
archivePrefix = {arXiv},
       eprint = {1311.1819},
 primaryClass = {astro-ph.EP},
       adsurl = {https://ui.adsabs.harvard.edu/abs/2014prpl.conf..475A}
}

@ARTICLE{Getman2018,
       author = {{Getman}, K.~V. and {Feigelson}, E.~D. and {Kuhn}, M.~A. and {Bate}, M.~R. and {Broos}, P.~S. and {Garmire}, G.~P.},
        title = "{Intracluster age gradients in numerous young stellar clusters}",
      journal = {\mnras},
         year = 2018,
        month = may,
       volume = {476},
       number = {1},
        pages = {1213-1223},
          doi = {10.1093/mnras/sty302},
archivePrefix = {arXiv},
       eprint = {1804.05077},
 primaryClass = {astro-ph.GA},
       adsurl = {https://ui.adsabs.harvard.edu/abs/2018MNRAS.476.1213G}
}

@ARTICLE{Getman2014b,
       author = {{Getman}, Konstantin V. and {Feigelson}, Eric D. and {Kuhn}, Michael A.},
        title = "{Core-Halo Age Gradients and Star Formation in the Orion Nebula and NGC 2024 Young Stellar Clusters}",
      journal = {\apj},
         year = 2014,
        month = jun,
       volume = {787},
       number = {2},
          eid = {109},
        pages = {109},
          doi = {10.1088/0004-637X/787/2/109},
archivePrefix = {arXiv},
       eprint = {1403.2742},
 primaryClass = {astro-ph.SR},
       adsurl = {https://ui.adsabs.harvard.edu/abs/2014ApJ...787..109G}
}

@ARTICLE{Winter2022,
       author = {{Winter}, Andrew J. and {Haworth}, Thomas J.},
        title = "{The external photoevaporation of planet-forming discs}",
      journal = {European Physical Journal Plus},
         year = 2022,
        month = oct,
       volume = {137},
       number = {10},
          eid = {1132},
        pages = {1132},
          doi = {10.1140/epjp/s13360-022-03314-1},
archivePrefix = {arXiv},
       eprint = {2206.11910},
 primaryClass = {astro-ph.EP},
       adsurl = {https://ui.adsabs.harvard.edu/abs/2022EPJP..137.1132W}
}

@ARTICLE{Getman2005,
       author = {{Getman}, K.~V. and {Flaccomio}, E. and {Broos}, P.~S. and {Grosso}, N. and {Tsujimoto}, M. and {Townsley}, L. and {Garmire}, G.~P. and {Kastner}, J. and {Li}, J. and {Harnden}, Jr., F.~R. and {Wolk}, S. and {Murray}, S.~S. and {Lada}, C.~J. and {Muench}, A.~A. and {McCaughrean}, M.~J. and {Meeus}, G. and {Damiani}, F. and {Micela}, G. and {Sciortino}, S. and {Bally}, J. and {Hillenbrand}, L.~A. and {Herbst}, W. and {Preibisch}, T. and {Feigelson}, E.~D.},
        title = "{Chandra Orion Ultradeep Project: Observations and Source Lists}",
      journal = {\apjs},
         year = 2005,
        month = oct,
       volume = {160},
       number = {2},
        pages = {319-352},
          doi = {10.1086/432092},
archivePrefix = {arXiv},
       eprint = {astro-ph/0410136},
 primaryClass = {astro-ph},
       adsurl = {https://ui.adsabs.harvard.edu/abs/2005ApJS..160..319G}
}

@ARTICLE{BenjaminiHochberg1995,
       author = {{Benjamini}, Yoav and {Hochberg}, Yosef},
        title = "{Controlling the False Discovery Rate: A Practical and Powerful Approach to Multiple Testing}",
      journal = {Journal of the Royal Statistical Society: Series B (Methodological)},
         year = 1995,
        month = jan,
       volume = {57},
       number = {1},
        pages = {289--300},
          doi = {10.1111/j.2517-6161.1995.tb02031.x},
          url = {https://rss.onlinelibrary.wiley.com/doi/10.1111/j.2517-6161.1995.tb02031.x}
}

@ARTICLE{Miller2001,
       author = {{Miller}, Christopher J. and {Genovese}, Christopher and {Nichol}, Robert C. and {Wasserman}, Larry and {Connolly}, Andrew and {Reichart}, Daniel and {Hopkins}, Andrew and {Schneider}, Jeff and {Moore}, Andrew},
        title = "{Controlling the False-Discovery Rate in Astrophysical Data Analysis}",
      journal = {\aj},
         year = 2001,
        month = dec,
       volume = {122},
       number = {6},
        pages = {3492-3505},
          doi = {10.1086/324109},
archivePrefix = {arXiv},
       eprint = {astro-ph/0107034},
 primaryClass = {astro-ph},
       adsurl = {https://ui.adsabs.harvard.edu/abs/2001AJ....122.3492M}
}

@ARTICLE{Zeileis2005,
       author = {{Zeileis}, Achim and {Grothendieck}, Gabor},
        title = "{zoo: S3 Infrastructure for Regular and Irregular Time Series}",
      journal = {Journal of Statistical Software},
         year = 2005,
        month = may,
       volume = {14},
       number = {6},
        pages = {1--27},
          doi = {10.18637/jss.v014.i06},
          url = {https://www.jstatsoft.org/article/view/v014i06}
}

@software{Pontoppidan2026,
       author = {{Pontoppidan}, Klaus M. and {Salyk}, Colette and {Banzatti}, Andrea},
        title = "{ctool: Continuum tool}",
 howpublished = {Astrophysics Source Code Library, record ascl:2601.002},
         year = 2026,
        month = jan,
          eid = {ascl:2601.002},
archivePrefix = {ascl},
       eprint = {2601.002},
       adsurl = {https://ui.adsabs.harvard.edu/abs/2026ascl.soft01002P}
}

@ARTICLE{Megeath2012,
       author = {{Megeath}, S.~T. and {Gutermuth}, R. and {Muzerolle}, J. and {Kryukova}, E. and {Flaherty}, K. and {Hora}, J.~L. and {Allen}, L.~E. and {Hartmann}, L. and {Myers}, P.~C. and {Pipher}, J.~L. and {Stauffer}, J. and {Young}, E.~T. and {Fazio}, G.~G.},
        title = "{The Spitzer Space Telescope Survey of the Orion A and B Molecular Clouds. I. A Census of Dusty Young Stellar Objects and a Study of Their Mid-infrared Variability}",
      journal = {\aj},
         year = 2012,
        month = dec,
       volume = {144},
       number = {6},
          eid = {192},
        pages = {192},
          doi = {10.1088/0004-6256/144/6/192},
archivePrefix = {arXiv},
       eprint = {1209.3826},
 primaryClass = {astro-ph.GA},
       adsurl = {https://ui.adsabs.harvard.edu/abs/2012AJ....144..192M}
}

@ARTICLE{Arulanantham2025,
       author = {{Arulanantham}, Nicole and {Salyk}, Colette and {Pontoppidan}, Klaus and {Banzatti}, Andrea and {Zhang}, Ke and {{\"O}berg}, Karin and {Long}, Feng and {Carr}, John and {Najita}, Joan and {Pascucci}, Ilaria and {Colmenares}, Mar{\'\i}a Jos{\'e} and {Xie}, Chengyan and {Huang}, Jane and {Green}, Joel and {Andrews}, Sean M. and {Blake}, Geoffrey A. and {Bergin}, Edwin A. and {Pinilla}, Paola and {Vioque}, Miguel and {Dahl}, Emma and {Raul}, Eshan and {Krijt}, Sebastiaan and {The Jdiscs Collaboration}},
        title = "{The JDISC Survey: Linking the Physics and Chemistry of Inner and Outer Protoplanetary Disk Zones}",
      journal = {\aj},
         year = 2025,
        month = aug,
       volume = {170},
       number = {2},
          eid = {67},
        pages = {67},
          doi = {10.3847/1538-3881/addd01},
archivePrefix = {arXiv},
       eprint = {2505.07562},
 primaryClass = {astro-ph.SR},
       adsurl = {https://ui.adsabs.harvard.edu/abs/2025AJ....170...67A}
}

@ARTICLE{Brown2013,
       author = {{Brown}, J.~M. and {Pontoppidan}, K.~M. and {van Dishoeck}, E.~F. and {Herczeg}, G.~J. and {Blake}, G.~A. and {Smette}, A.},
        title = "{VLT-CRIRES Survey of Rovibrational CO Emission from Protoplanetary Disks}",
      journal = {\apj},
         year = 2013,
        month = jun,
       volume = {770},
       number = {2},
          eid = {94},
        pages = {94},
          doi = {10.1088/0004-637X/770/2/94},
archivePrefix = {arXiv},
       eprint = {1304.4961},
 primaryClass = {astro-ph.SR},
       adsurl = {https://ui.adsabs.harvard.edu/abs/2013ApJ...770...94B}
}
\bibliographystyle{aasjournalv7.1}



\end{document}